\documentclass[12pt]{article}
\usepackage{epsf,axodraw}
\newcommand{\lsim}{\stackrel{<}{_\sim}}
\newcommand{\gsim}{\stackrel{>}{_\sim}}
\newcommand{\pslash}{\hspace{-3pt}\not\!p}
\newcommand{\qslash}{\hspace{-3pt}\not\!q}
\newcommand{\kslash}{\hspace{-3pt}\not\!k}
\newcommand{\lslash}{\hspace{-3pt}\not\hspace{-1pt}l}

\def\gsim{\ \rlap{\raise 3pt \hbox{$>$}}{\lower 3pt \hbox{$\sim$}}\ }
\def\lsim{\ \rlap{\raise 3pt \hbox{$<$}}{\lower 3pt \hbox{$\sim$}}\ }
\newcommand\epjc[3]{{Eur.\ Phys.\ J. }{\bf C#1} (#2) #3}

\newcommand\npb[3]{{Nucl.\ Phys.\ }{\bf B#1} (#2) #3}
\newcommand\npps[3]{{Nucl.\ Phys.\ (Proc.\ Suppl.)\ }{\bf B#1} (#2) #3}
\newcommand\plb[3]{{Phys.\ Lett.\ }{\bf B#1} (#2) #3}
\newcommand\prd[3]{{Phys.\ Rev.\ }{\bf D#1} (#2) #3}
\newcommand\prep[3]{{Phys.\ Rep.\ }{\bf#1} (#2) #3}
\newcommand\prl[3]{{Phys.\ Rev.\ Lett.\ }{\bf#1} (#2) #3}
\newcommand\rmp[3]{{Rev.\ Mod.\ Phys.\ }{\bf#1} (#2) #3}
\newcommand\zpc[3]{{Z.\ Phys.\ }{\bf C#1} (#2) #3}

\newcommand\sjnp[3]{{Sov.\ J.\ Nucl.\ Phys.\ }{\bf#1} (#2) #3}

\newcommand\yf[3]{{Yad.\ Fiz.\ }{\bf#1} (#2) #3}

\newcommand\ibid[3]{{\em ibid.\ }{\bf#1} (#2) #3}
\newcommand\jpg[3]{{J.\ Phys.\ }{\bf G#1} (#2) #3}
\newcommand\ijmpa[3]{{Int.\ J.\ Mod.\ Phys.\ }{\bf A#1} (#2) #3}
\newcommand\arnps[3]{{Ann.\ Rev.\ Nucl.\ Part.\ Sci.\ }{\bf #1}
 (#2) #3}

\newcommand{\hepph}[1]{{hep-ph/#1}}
\newcommand{\heplat}[1]{{hep-lat/#1}}
\newcommand{\hepex}[1]{{hep-ex/#1}}

\begin{document}

\thispagestyle{empty}
\begin{flushright}
CERN-TH/2000-159\\
CLNS~00/1675\\
PITHA~00/06\\
SHEP~00/06\\[0.15cm]
hep-ph/0006124\\
June 13, 2000
\end{flushright}

\vspace*{0.8cm}
\centerline{\Large\bf 
QCD factorization for exclusive non-leptonic}
\vspace*{0.2cm}
\boldmath 
\centerline{\Large\bf 
$B$-meson decays: General arguments}
\unboldmath
\vspace*{0.2cm}
\centerline{\Large\bf 
and the case of heavy--light final states}

\vspace*{1.5cm}
\centerline{{\sc M. Beneke${}^{a}$}, {\sc G. Buchalla${}^b$},
{\sc M. Neubert${}^c$} and {\sc C.T. Sachrajda${}^d$}}

\vspace*{1cm}
\centerline{\sl ${}^a$Institut f\"ur Theoretische Physik E, RWTH Aachen}
\centerline{\sl D - 52056 Aachen, Germany}
\vspace*{0.2cm}
\centerline{\sl ${}^b$Theory Division, CERN, CH-1211 Geneva 23,
                Switzerland}
\vspace*{0.2cm}
\centerline{\sl ${}^c$Newman Laboratory of Nuclear Studies}
\centerline{\sl Cornell University, Ithaca, NY 14853,  USA}
\vspace*{0.2cm}
\centerline{\sl ${}^d$Department of Physics and Astronomy, University
  of Southampton}
\centerline{\sl Southampton SO17 1BJ, UK}

\vspace*{0.9cm}
\centerline{\bf Abstract}
\vspace*{0.3cm}

\noindent We provide a rigorous basis for factorization for a large 
class of non-leptonic two-body $B$-meson decays in the heavy-quark 
limit. The resulting factorization formula incorporates elements of the 
naive factorization approach and the hard-scattering approach, but 
allows us to compute systematically radiative (``non-factorizable'') 
corrections to naive factorization for decays such as $B\to D\pi$ and 
$B\to \pi \pi$. We first discuss the factorization formula from a 
general point of view. We then consider factorization for decays into 
heavy-light final states (such as $B\to D\pi$) in more detail, 
including a proof of the factorization formula at two-loop order. 
Explicit results for the leading QCD corrections to factorization are 
presented and compared to existing measurements of branching fractions 
and final-state interaction phases.

\noindent 
\vfill

\newpage
\pagenumbering{arabic}


\section{Introduction}

Non-leptonic, two-body $B$-meson decays, although simple as far as the 
underlying weak decay of the $b$ quark is concerned, are complicated on 
account of strong-interaction effects. If these effects 
could be computed, this 
would enhance tremendously our ability to uncover the origin of CP violation 
in weak interactions from data on a variety of such decays being
collected at the $B$ factories.

In this paper we begin a systematic analysis of weak 
heavy-meson decays into two energetic mesons based on the factorization 
properties of decay amplitudes in quantum chromodynamics (QCD). 
(Some of the results have already been presented in \cite{BBNS99}.) 
As in the classic analysis of semi-leptonic $B\to D$ transitions 
\cite{IW89,VS87}, our arguments make extensive use of the fact that the $b$ 
quark is heavy compared to the intrinsic scale of strong 
interactions. This allows us to deduce that non-leptonic decay amplitudes 
in the heavy-quark limit have a simple structure. The arguments to 
reach this conclusion, however, are quite different from those used for 
semi-leptonic decays, since for non-leptonic decays a 
large momentum is transferred to at least one of the final-state 
mesons. The results of this work justify naive factorization of 
four fermion operators for many, but not all, non-leptonic decays and 
imply that corrections termed ``non-factorizable'', 
which up to now have been thought to 
be intractable, can be calculated rigorously, if the 
mass of the weakly decaying quark is large enough. This leads to 
a large number of predictions for CP-violating $B$ decays in the 
heavy-quark limit, for which measurements will soon become available.  

Weak decays of heavy mesons involve three fundamental scales, the weak 
interaction scale $M_W$, the $b$-quark mass $m_b$, and the QCD  
scale $\Lambda_{\rm QCD}$, which are strongly ordered: 
$M_W\gg m_b\gg \Lambda_{\rm QCD}$. The underlying weak decay being 
computable, all theoretical work concerns strong-interaction corrections. 
The strong-interaction effects which involve virtualities above the 
scale $m_b$ are well understood. 
They renormalize the coefficients of local operators ${\cal O}_i$ in 
the weak effective Hamiltonian \cite{BBL}, 
so that the amplitude for the decay 
$B\to M_1 M_2$ is given by
\begin{equation}
\label{effham}
{\cal A}(B\to M_1 M_2) = \frac{G_F}{\sqrt{2}} \sum_i 
\lambda_i\,C_i(\mu)\,
\langle M_1 M_2 |{\cal O}_i|B\rangle(\mu),
\end{equation}
where $G_F$ is the Fermi constant. Each term in the sum is 
the product of a Cabibbo-Kobayashi-Maskawa (CKM) factor $\lambda_i$, 
a coefficient function $C_i(\mu)$, which incorporates strong-interaction 
effects above the scale $\mu\sim m_b$, and a matrix 
element of an operator ${\cal O}_i$. The most difficult 
theoretical problem is to compute these matrix elements or, 
at least, to reduce them to simpler non-perturbative objects. 

There exist a variety of treatments of this problem, on many of which 
we will comment later, which rely on assumptions of some sort. Here we 
identify two somewhat contrary lines of approach. (A more
comprehensive discussion of the literature on non-leptonic $B$ 
decays is given in a separate section of this paper.)

The first approach, which we shall call ``naive factorization'', 
replaces the matrix element of a four-fermion operator in a heavy-quark 
decay by the product of the matrix elements of two currents  
\cite{FS78,CaMa78}, for example,
\begin{equation}
\label{fac1}
\langle \pi^+\pi^- |(\bar{u} b)_{\rm V-A}(\bar{d} u)_{\rm V-A}|\bar{B}_d
\rangle \to \langle \pi^-|(\bar{d} u)_{\rm V-A}|0\rangle\,\langle 
\pi^+|(\bar{u} b)_{\rm V-A}|\bar{B}_d\rangle. 
\end{equation}
This assumes that the exchange of ``non-factorizable'' 
gluons between the $\pi^-$ and the 
$(\bar{B_d}\,\pi^+)$ system can be neglected, if the virtuality of the 
gluons is below $\mu\sim m_b$. The non-leptonic decay amplitude reduces 
to the product of a form factor and a decay constant. This assumption 
is in general not justified, except in the limit of a large number of 
colours in some cases. It deprives the amplitude of any physical 
mechanism that could account for rescattering in the final state  
and for the generation of a strong phase shift between different 
amplitudes. ``Non-factorizable'' radiative corrections must also exist, 
because the scale dependence of the two sides of 
(\ref{fac1}) is different. Since ``non-factorizable'' corrections at 
scales larger than $\mu$ are taken into account in deriving the 
effective weak Hamiltonian, it appears rather arbitrary to leave 
them out below the scale $\mu$. 

The correct scale dependence can be restored by computing the transition 
matrix element for an inclusive or partonic final state and by absorbing the 
correction into effective scale-independent coefficients. However,
without a systematic approach to computing the hadronic matrix elements, 
this sidelines the real question of how to improve the parametric 
accuracy of the naive factorization approach.

Various generalizations of the naive factorization approach have 
been proposed, which include new parameters that account for 
non-factorizable corrections. In the most general form, 
these generalizations have nothing to do with the original 
``factorization'' ansatz, but amount to a general parameterization 
of the matrix elements, including those of penguin operators. Such 
general parameterizations are exact, but at the price of introducing 
many unknown parameters and eliminating any theoretical input on 
strong-interaction dynamics. Making such a parameterization useful 
requires certain assumptions that relate these 
parameters.

The second method used to study non-leptonic decays is the hard-scattering 
approach. Here the assumption is that the decay is 
dominated by hard gluon exchange. The decay amplitude is then expressed 
as a convolution of a hard-scattering factor with light-cone wave 
functions of the participating mesons, for example,
\begin{equation}
\label{hsa1}
\langle \pi^+\pi^- |(\bar{u} b)_{\rm V-A}(\bar{d} u)_{\rm V-A}|\bar{B}_d
\rangle \to \int_0^1\!d\xi du dv \,\Phi_B(\xi)\,\Phi_{\pi}(u)\,
\Phi_{\pi}(v)\,T(\xi,u,v;m_b).
\end{equation} 
This is analogous to more 
familiar applications of this method to hard exclusive reactions 
involving only light hadrons \cite{LB80,EfRa80}. 

For many hard exclusive processes the hard-scattering contribution 
represents the leading term in 
an expansion in $\Lambda_{\rm QCD}/Q$, where $Q$ denotes the hard scale. 
However, the short-distance dominance of hard exclusive processes 
is not enforced kinematically and relies crucially on the properties 
of hadronic wave functions. There is an important difference between 
light mesons and heavy mesons regarding these wave functions, because 
the light quark in a heavy meson at rest naturally has a small momentum of 
order $\Lambda_{\rm QCD}$, 
while for fast light mesons a configuration with a soft quark
is suppressed by the meson's wave function. As a consequence the soft 
(or Feynman) mechanism is power suppressed for hard exclusive processes 
involving light mesons, but it is of leading power, and in fact larger 
than the hard-scattering contribution by a factor 
$1/\alpha_s(\sqrt{m_b\Lambda_{\rm QCD}})$, 
for heavy-meson decays. (The arguments that lead to this conclusion 
will be reviewed below.)

A standard analysis of higher-order corrections to the hard-scattering 
amplitude $T$ in (\ref{hsa1}) shows  
that the configuration in which the final-state 
meson picks up the soft spectator quark of the heavy meson as a 
soft quark is suppressed by a Sudakov form factor, if the meson has 
large momentum. This suggests that the hard-scattering term may 
become dominant even for heavy-meson decays, if the heavy-quark 
mass is very large. However, calculation of the $B\to \pi$ form 
factors in the QCD sum rule approach \cite{KRWY97,BBB98} indicates
that the soft 
contribution dominates for $b$ quarks with $m_b\approx 5\,$GeV. 
Even if Sudakov suppression were effective, arguing away the soft 
contribution in this way is not completely satisfactory; a 
factorization formula that separates soft and hard contributions on 
the basis of power counting alone is more desirable.

It is clear from this discussion that a satisfactory treatment 
should take into account soft contributions (and hence provide the 
correct asymptotic limit -- if we ignore Sudakov suppression factors), 
but also allow us to compute corrections to the
naive factorization result in a systematic way 
(and hence result in a scheme- and 
scale-independent expression up to corrections of higher order in 
the strong coupling $\alpha_s$). 

It is not at all obvious that 
such a treatment would result in a predictive framework. We will show 
that this does indeed happen for most non-leptonic two-body $B$ decays. 
Our main conclusion is that ``non-factorizable'' corrections are 
dominated by hard gluon exchange, while the soft effects that survive 
in the heavy-quark limit are confined to the $(B M_1)$ system, where 
$M_1$ denotes the meson that picks up the spectator quark in 
the $B$ meson. This result 
is expressed as a factorization formula, which is valid up to 
corrections suppressed by $\Lambda_{\rm QCD}/m_b$. At leading 
power non-perturbative contributions are parameterized by the physical 
form factors for the $B\to M_1$ transition and 
leading-twist light-cone distribution amplitudes of the mesons. 
Hard perturbative corrections can be computed 
systematically in a way similar to the hard-scattering approach. 
On the other hand, because the $B\to M_1$ transition is parameterized by 
a form factor, we recover the result 
of naive factorization at lowest order in $\alpha_s$. 
An important implication of the factorization 
formula is that strong rescattering phases are either perturbative or 
power suppressed in $m_b$. It is worth emphasizing that the 
decoupling of $M_2$ occurs in the presence of soft interactions 
in the $(B M_1)$ system. In other words, while strong-interaction effects in 
the $B\to M_1$ transition are 
not confined to small transverse distances, the other meson $M_2$ 
is predominantly produced as a compact object with small transverse 
extension. The decoupling of soft effects then follows from  
``colour transparency''. The colour-transparency argument for 
exclusive $B$ decays has already been noted in the literature 
\cite{Bj89,DG91}, but it has never been developed into a factorization 
formula that could be used to obtain quantitative predictions.

The approach described in this paper is general and applies to decays into 
a heavy and a light meson (such as $B\to D\pi$) as well as to decays into 
two light mesons (such as $B\to \pi\pi$, $B\to \pi K$, etc.). Factorization 
does not hold, however, for decays such as $B\to \pi D$ and $B\to D D_s$, 
in which the meson that does {\em not\/} pick up the spectator quark in 
the $B$ meson is heavy. For the special case of the ratio 
$\Gamma(B\to D^*\pi)/\Gamma(B\to D\pi)$ Politzer and Wise   
evaluated ``non-factorizable'' 
one-loop corrections several years ago \cite{PW91}. 
Their result agrees with the result obtained from the general factorization 
formula proposed here.

The outline of the paper is as follows: in Sect.~\ref{factform} we state 
the factorization formula in its general form and define the various 
elements of the formula, in particular the light-cone distribution 
amplitudes. In Sect.~\ref{arguments} we collect the arguments that 
lead to the factorization formula. We show how light-cone distribution 
amplitudes enter, discuss the heavy-quark scaling of the $B\to\pi$ 
form factor and the cancellation of soft and collinear effects. We 
also address the issue of multi-particle Fock states and annihilation 
topologies, which are power suppressed in $\Lambda_{\rm QCD}/m_b$. The 
arguments of this section are appropriate for decays into a heavy and a 
light meson, as well as, with some modifications, to decays into two 
light mesons. However, we will keep the 
discussion qualitative and leave technical details to later sections. 
In Sect.~\ref{oneloop} we discuss the cancellation of long-distance 
singularities 
at one-loop order, and present the calculation of the hard-scattering 
functions at this order for decays into a heavy and a light meson. 
Sect.~\ref{allorders} extends the proof of the cancellation of 
singularities to two-loop order and provides arguments for factorization 
to all orders. In Sect.~\ref{bdpi} we consider the phenomenology of 
decays into a heavy and a light meson on the basis of the factorization 
formula. We examine to what extent a charm meson should 
be considered as heavy or light and discuss various tests of the theoretical 
framework. A critical review of and comparison with other approaches 
to exclusive non-leptonic decays is given in Sect.~\ref{sec:comparison}.
Sect.~\ref{conclusion} contains our conclusion.

Except for the general discussion, we restrict this paper to the 
proof of factorization and the phenomenology for decays into a 
heavy and a light meson. The more elaborate technical arguments needed to 
establish the factorization formula for decays into two light mesons, 
together with an adequate discussion of the heavy-quark limit in this 
case, will be given in a subsequent paper.


\section{Statement of the factorization formula}
\label{factform}

In this section we summarize the main result of this paper, the 
factorization formula for non-leptonic 
$B$ decays. We introduce relevant terminology and definitions. 

\subsection{The idea of factorization}
\label{idea}

In the context of non-leptonic decays the term ``factorization'' is 
usually applied to the approximation of the matrix element of 
a four fermion operator by the product of a form factor and a decay 
constant, see (\ref{fac1}). Corrections to this approximation are called 
``non-factorizable''. We will refer to this approximation as 
``naive factorization'' and use quotes on ``non-factorizable'' to avoid 
confusion with the meaning of factorization in the context of 
hard processes in QCD.

In the latter context 
factorization refers to the separation of long-distance contributions
to the process from a short-distance part that depends only 
on the large scale $m_b$. The short-distance part can be computed in 
an expansion in the strong coupling $\alpha_s(m_b)$. The long-distance 
contributions must be computed non-perturbatively or determined 
experimentally. 
The advantage is that these non-perturbative parameters are often 
simpler in structure than the original quantity, or they are process 
independent. For example, factorization 
applied to hard processes in inclusive hadron-hadron collisions 
requires only 
parton distributions as non-perturbative inputs. Parton distributions 
are much simpler objects than the original matrix element with two 
hadrons in the initial state.  On the other hand, factorization 
applied to the $B\to D$ form 
factor leads to a non-perturbative object (the ``Isgur-Wise function'') 
which is still a function of the momentum transfer. However, the benefit 
here is that symmetries 
relate this function to other form factors. 
In the case of non-leptonic $B$ decays, 
the simplification is primarily of the first kind (simpler structure). 
We call those 
effects non-factorizable (without quotes) which depend on 
the long-distance properties of the $B$ meson and both final-state mesons 
combined. 

The factorization properties of non-leptonic decay amplitudes depend on 
the two-meson final state. We call a meson ``light'', if its mass $m$ 
remains finite in the heavy-quark limit. A meson is called ``heavy'', if 
we assume that its mass scales with $m_b$ in the heavy-quark limit, 
such that $m/m_b$ stays fixed. In principle, we could still have 
$m\gg \Lambda_{\rm QCD}$ for a light meson. Charm mesons could be 
considered as light in this sense. However, unless otherwise mentioned, 
we assume that $m$ is of order $\Lambda_{\rm QCD}$ for a light meson, 
and we consider charm mesons as heavy. We also restrict the term ``heavy 
mesons'' to mesons of a heavy and a light quark and do not include 
onia of two heavy quarks. The difference is that heavy and light mesons 
have large transverse extension of order $1/\Lambda_{\rm QCD}$, 
while the transverse size of onia becomes small in the heavy-quark limit.

Although not necessary, it is useful to describe non-leptonic decays in the 
$B$-meson rest frame. In this paper quantities which are not Lorentz 
invariant will always refer to this frame. In evaluating 
the scaling 
behaviour of the decay amplitudes we assume that the energies of both 
final-state mesons 
scale like $m_b$ in the heavy-quark limit. We do not consider 
explicitly the so-called small velocity limit for heavy mesons in 
which $m_b\sim m$ while $m_b-m\gg \Lambda_{\rm QCD}$ stays fixed 
in the heavy-quark limit, which implies 
$m/m_b\to 1$. Although our results remain 
valid in this limit, the assumption that $m/m_b$ stays fixed simplifies 
the discussion, because we do not have to distinguish the scales 
$m_b$ and $m_b-m$.

\subsection{The factorization formula}

\begin{figure}[t]
   \vspace{-3cm}
   \epsfysize=23cm
   \epsfxsize=16cm
   \centerline{\epsffile{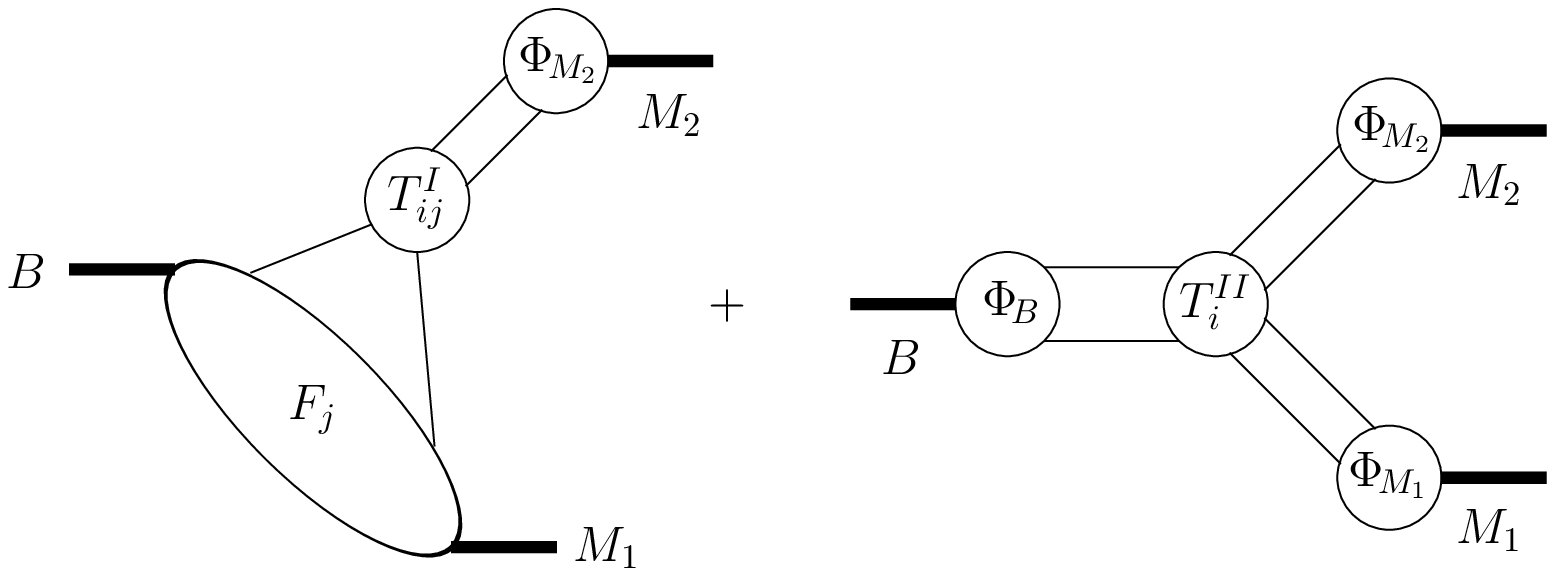}}
   \vspace*{-15.5cm}
\caption[dummy]{\label{fig1}\small Graphical representation of 
the factorization formula. Only one of the two form-factor terms in 
(\ref{fff}) is shown for simplicity.}
\end{figure}

We consider weak decays $B\to M_1 M_2$ in the heavy-quark limit and 
differentiate between decays into final states containing a heavy and a 
light meson or two light mesons. Up 
to power corrections of order $\Lambda_{\rm QCD}/m_b$ the transition 
matrix element of an operator ${\cal O}_i$ in the weak effective 
Hamiltonian is given by
\begin{eqnarray}
\label{fff}
\langle M_1 M_2|{\cal O}_i|\bar{B}\rangle &=& 
\sum_j F_j^{B\to M_1}(m_2^2)\,\int_0^1 du\,T_{ij}^I(u)\,\Phi_{M_2}(u) 
\,\,+\,\,(M_1\leftrightarrow M_2)\nonumber\\
&&\hspace*{-2cm}
+\,\int_0^1 d\xi du dv \,T_i^{II}(\xi,u,v)\,
\Phi_B(\xi)\,\Phi_{M_1}(v)\,\Phi_{M_2}(u) \nonumber\\
&&\mbox{if $M_1$ and $M_2$ are both light,} \\
\langle M_1 M_2|{\cal O}_i|\bar{B}\rangle &=& 
\sum_j F_j^{B\to M_1}(m_2^2)\,\int_0^1 du\,T_{ij}^I(u)\,\Phi_{M_2}(u) 
\nonumber\\
&&\mbox{if $M_1$ is heavy and $M_2$ is light.} \nonumber
\end{eqnarray} 
Here $F_j^{B\to M_{1,2}}(m_{2,1}^2)$ denotes a $B\to M_{1,2}$ form factor, 
and $\Phi_X(u)$ is the light-cone distribution amplitude for the 
quark-antiquark Fock state of meson $X$. These non-perturbative 
quantities will be defined precisely in the next subsection. 
$T_{ij}^I(u)$ and $T_i^{II}(\xi,u,v)$ are hard-scattering functions, 
which are perturbatively calculable. The hard-scattering kernels and 
light-cone distribution amplitudes depend on a factorization scale 
and scheme, which is suppressed in the notation of (\ref{fff}).
Finally, $m_{1,2}$ denote the light meson masses.  
Eq.~(\ref{fff}) is represented graphically in 
Fig.~\ref{fig1}. (The second line of the first equation in (\ref{fff}) is 
somewhat simplified and may require including an integration over 
transverse momentum in the $B$ meson starting from order 
$\alpha_s^2$, see the remarks after (\ref{setperpzero}).)

As it stands, the first equation in (\ref{fff}) applies to decays into two 
light mesons, for which the spectator quark in the $B$ meson (in the 
following 
simply referred to as the ``spectator quark'') can go to either of the
final-state mesons. An example is the decay 
$B^-\to \pi^0 K^-$. If the spectator quark can go only to one of the 
final-state mesons, as for example in $\bar B_d\to \pi^+ K^-$, 
we call this meson $M_1$ and the second form-factor term 
on the right-hand side of (\ref{fff}) is absent. 

The factorization formula simplifies when the spectator quark goes 
to a heavy meson (second equation in (\ref{fff})), 
such as in $\bar{B}_d\to D^+ \pi^-$. In this case 
the third term on the right-hand side of (\ref{fff}), which accounts for 
hard interactions with the spectator quark, can be dropped because it is 
power suppressed in the heavy-quark limit. In the opposite situation that 
the spectator quark goes to a light meson but the other meson is heavy, 
factorization does not hold because the heavy meson is neither 
fast nor small and cannot be factorized from the $B\to M_1$ transition. 
However, such amplitudes are again power suppressed in the heavy-quark limit 
relative to amplitudes in which the spectator quark goes to a heavy meson 
while the other meson is light. (These statements will be justified in 
detail in Sect.~\ref{arguments}.) We also note that factorization 
{\em does\/} hold, at least formally, if the emission particle $M_2$ is an 
onium. Finally, notice that annihilation topologies do not appear in the 
factorization formula. They do not contribute at leading order in 
the heavy-quark expansion.

Any hard interaction costs a power of $\alpha_s$. As 
a consequence the third term in (\ref{fff}) is absent at order 
$\alpha_s^0$. Since at this order the functions $T^I_{ij}(u)$ are 
independent of $u$, the convolution integral results in a meson decay 
constant and we see that (\ref{fff}) reproduces naive 
factorization. The factorization formula allows us to compute 
radiative corrections to this result to all orders in $\alpha_s$. 
Further corrections are suppressed by powers of $\Lambda_{\rm QCD}/m_b$ in 
the heavy-quark limit.

The significance and usefulness of the factorization formula 
stems from the fact that the non-perturbative quantities which appear 
on the right-hand side of (\ref{fff}) 
are much simpler than the original non-leptonic 
matrix element on the left-hand side. This is because they either 
reflect universal properties of a single meson state (light-cone 
distribution amplitudes) or refer only to a $B\to\mbox{meson}$ transition 
matrix element of a local current (form factors). While it is extremely 
difficult, if not impossible \cite{MT90}, to compute the original 
matrix element $\langle M_1 M_2|{\cal O}_i|\bar{B}\rangle$ in lattice 
QCD, form factors and light-cone distribution amplitudes are 
already being computed in this way, although with significant systematic 
errors at present. 
Alternatively, form factors can be obtained using data on semi-leptonic 
decays, and light-cone distribution amplitudes by comparison with 
other hard exclusive processes.

Adopting the terminology introduced earlier, Eq.~(\ref{fff}) implies that 
there exist no non-factorizable effects (in the sense of QCD 
factorization) at leading order in the 
heavy-quark expansion. Since the form factors and light-cone 
distribution amplitudes are real, all final-state interactions 
and the strong phases generated by them are part of the calculable 
hard-scattering functions. This and the absence of non-factorizable 
corrections is unlikely to be true beyond leading order in the heavy-quark 
expansion, because there 
exist soft gluon effects that connect $M_1$ and $M_2$, which are 
suppressed by one power of $\Lambda_{\rm QCD}/m_b$.

\subsection{Definition of non-perturbative parameters}

\subsubsection{Form factors}

The form factors $F_j^{B\to M}(q^2)$ in (\ref{fff}) arise in the 
decomposition of matrix elements of the form 
\begin{equation}
\langle M(p')|\bar{q}\Gamma b|\bar{B}(p)\rangle,
\end{equation}
where $\Gamma$ can be any irreducible Dirac matrix 
that appears after contraction of the hard subgraph to a local 
vertex with respect to the $B\to M$ transition. For the purpose of 
discussion we will often refer to the matrix element 
of the vector current, which is conventionally parameterized by 
two scalar form factors:
\begin{equation}
\langle P(p')|\bar{q}\gamma^\mu b|\bar{B}(p)\rangle = F_+^{B\to P}(q^2)
\,(p^\mu+{p'}^\mu) + \Big[F_0^{B\to P}(q^2)-F_+^{B\to P}(q^2)\Big]\,
\frac{m_B^2-m_P^2}{q^2}\,q^\mu ,
\end{equation}
where $q=p-p'$.
The pseudoscalar meson is denoted by $P$, $m_P$ is its mass and 
$m_B$ the mass of the $B$ meson. For $q^2=0$ the two form factors 
coincide, $F_+^{B\to P}(0)=F_0^{B\to P}(0)$. The scaling of 
$F_+^{B\to P}(0)$ with $m_b$ will be discussed in Sect.~\ref{arguments}.

Note that we write (\ref{fff}) in terms of physical form factors. 
In principle, Fig.~\ref{fig1} could be looked upon in two different ways. 
In the first way, we suppose that the region represented by `$F$' accounts 
only for the soft contributions to the $B\to M_1$ form factor. The hard 
contributions to the form factor can be considered as part of $T^{I}_{ij}$
or as part of the 
second diagram, i.e.\ as part of the hard-scattering factor $T^{II}_i$. 
Performing this split-up requires that one understands the factorization 
of hard and soft contributions to the form factor. If $M_1$ is 
heavy, this amounts to matching the form factor onto a form factor
defined in heavy-quark effective theory. However, for a light meson 
$M_1$, the factorization of hard and soft contributions to the form factor 
is not yet completely understood. We bypass this 
problem by interpreting `$F$' as the physical form factor, including 
hard and soft contributions. The hard contributions to the form factor 
should then be omitted from the hard-scattering kernel $T^{II}_i$ 
and a subtraction has to be performed in $T^{I}_{ij}$ beginning at 
two-loop order (see also Sect.~\ref{allorders}). The relevant diagrams are 
easily identified. An additional advantage of using physical form factors 
is that they are directly related to  measurable quantities, or 
to the form factors obtained from lattice QCD or QCD sum rules.

\subsubsection{Light-cone distribution amplitudes of light mesons}

Light-cone distribution amplitudes play the same role for hard 
exclusive processes that parton distributions play for inclusive 
processes. As in the latter case, the leading-twist distribution 
amplitudes, which are the ones we need at leading power in the 
$1/m_b$ expansion, are given by two-particle operators with a certain 
helicity structure. The helicity structure is determined by the angular 
momentum of the meson and the fact that the spinor of an energetic 
quark has only two large components.

The leading-twist light-cone distribution amplitudes for pseudoscalar mesons 
($P$), longitudinally polarized vector mesons ($V_{||}$), and transversely 
polarized vector mesons ($V_\perp$) with flavour content $(\bar{q} q')$ are 
\begin{eqnarray}
\label{distamps}
\langle P(q)|\bar{q}(y)_{\alpha} q'(x)_{\beta}|0\rangle\Big|_{(x-y)^2=0}
&=& \frac{i f_P}{4}\,(\not\!q\gamma_5)_{\beta
\alpha}\,\int_0^1 du\,e^{i(\bar{u} qx+u qy)}\,\Phi_P(u,\mu), 
\nonumber\\
\langle V_{||}(q)|\bar{q}(y)_{\alpha} q'(x)_{\beta}|0\rangle\Big|_{(x-y)^2=0}
&=& -\frac{i f_V}{4}\!\!\not\!q_{\beta
\alpha}\,\int_0^1 du\,e^{i(\bar{u} qx+u qy)}\,\Phi_{||}(u,\mu),
\\
\langle V_\perp(q)|\bar{q}(y)_{\alpha} q'(x)_{\beta}|0\rangle\Big|_{(x-y)^2=0}
&=& -\frac{i f_T(\mu)}{8}\,[\not\!\varepsilon^*_\perp,
\not\!q\,]_{\beta
\alpha}\,\int_0^1 du\,e^{i(\bar{u} qx+u qy)}\,\Phi_\perp(u,\mu).
\nonumber
\end{eqnarray}
The equality sign is to be understood as ``equal up to higher-twist 
terms'', and it is also understood that the operator on the 
left-hand side is a colour singlet. We use the 
``bar''-notation throughout this paper, i.e.\ $\bar{v}\equiv 1-v$ for 
any longitudinal momentum fraction variable. The parameter 
$\mu$ is the renormalization scale of the 
light-cone operators on the left-hand side. The distribution amplitudes 
are normalized as $\int_0^1 du\,\Phi_{X}(u,\mu)=1$ $(X=P,||,\perp)$. 
One defines the asymptotic distribution amplitude as the limit in 
which the renormalization scale is sent to infinity. All three distribution 
amplitudes introduced above have the same asymptotic form
\begin{equation}
\label{asform}
\Phi_{X}(u,\mu)\stackrel{\mu\to\infty}{=} 6 u\bar{u}.
\end{equation}
The decay constants appearing in (\ref{distamps}) refer to the normalization 
in which $f_\pi=131\,$MeV. ($f_T(\mu)$ is scale dependent, because 
it is related to the matrix element of the non-conserved tensor 
current.) $\varepsilon_\perp^\mu$ is the polarization vector 
of a transversely polarized vector meson. For longitudinally polarized 
vector mesons we can identify $\varepsilon_{||}^\mu=q^\mu/m_V$, where $m_V$ 
is the vector-meson mass. Corrections to this are suppressed by two powers 
of $m_V/q^0\sim \Lambda_{\rm QCD}/m_b$. We have used this fact to eliminate 
the polarization vector of $V_{||}$ in (\ref{distamps}). There is a 
path-ordered exponential that connects the two quark fields at 
different positions and makes the light-cone operators gauge invariant. 
In (\ref{distamps}) we have suppressed this standard factor.

The use of light-cone distribution amplitudes in non-leptonic $B$ 
decays requires justification, which we will provide in 
Sect.~\ref{arguments}. The decay amplitude is then 
calculated as follows: assign momentum $uq$ to the quark in the 
outgoing (light) meson with momentum $q$ and assign momentum 
$\bar{u} q$ to the antiquark. Write down the on-shell amplitude 
in momentum space with outgoing quark and antiquark of momentum $uq$ and 
$\bar{u}q$, respectively, and perform the replacement
\begin{equation}
\label{curproj}
\bar{u}_{\alpha a}(uq)\Gamma(u,\ldots)_{\alpha\beta,ab,\ldots}
v_{\beta b}(\bar{u}q) \,\longrightarrow\,
\frac{i f_P}{4 N_c}\int_0^1 du\,\Phi_P(u)\,(\not\!q\gamma_5)_
{\beta\alpha}\Gamma(u,\ldots)_{\alpha\beta,aa,\ldots}
\end{equation}
for pseudoscalars and, with obvious modifications, for vector mesons.
(Here $N_c=3$ refers to the number of colours.)

Even when working with light-cone distribution amplitudes (light-cone 
wave functions integrated over transverse momentum) it is not always 
justified to perform the collinear approximation on the external 
quark (antiquark) lines right away. One may have to keep the transverse 
components of the quark momentum $k$ and be allowed to 
put $k=u q$ only after some operations on the amplitude have been 
carried out. However, these subtleties do not concern calculations 
that use only leading-twist light-cone distributions.

\boldmath
\subsubsection{Light-cone distribution amplitudes of $B$ mesons}
\unboldmath

It is intuitive that light-cone distribution amplitudes 
for light mesons appear in non-leptonic decays. The relevance of light-cone 
distribution amplitudes of $B$ mesons is less clear, because the 
spectator quark in the $B$ meson is not energetic in the $B$-meson rest 
frame. Hence if we assign momentum $l$ to the spectator quark, 
all components of $l$ are of order $\Lambda_{\rm QCD}$. 

The $B$-meson light-cone distribution amplitude appears only in the third 
term on the right-hand side of (\ref{fff}), the hard spectator interaction 
term. As discussed above, this term is of leading power only for 
decays into two light mesons or decays into a light meson and an 
onium. One finds, at least at order $\alpha_s$, that the hard spectator 
interaction amplitude depends only on $p'\!\cdot l$ at leading 
order in $1/m_b$, where 
$p'$ is the momentum of the light meson that picks up the spectator 
quark. We introduce light-cone components 
\begin{equation}
v_{\pm}=\frac{v^0\pm v^3}{\sqrt{2}} 
\end{equation}
for any vector $v$. If we choose $p'$ such that only $p'_-$ is non-zero, 
the hard spectator amplitude $A(l,\ldots)$ 
depends only on $l_+$.  The decay amplitude 
for the general two-particle Fock state of the $B$ meson is given by 
the integral over the full Bethe-Salpeter wave function
\begin{equation}
\Psi_B(z,p) = \langle 0|\bar{q}_{\alpha}(z)[\ldots]b_{\beta}(0)
|\bar{B}_d(p)
\rangle = \int\frac{d^4 l}{(2\pi)^4}\,e^{-i lz}\,\hat{\Psi}_B
(l,p)
\end{equation}
with the partonic decay amplitude. (The dots denote the path-ordered 
exponential required to make the matrix element gauge invariant.) 
We then approximate 
\begin{equation}
\label{setperpzero}
\int\frac{d^4 l}{(2\pi)^4}\,A(l,\ldots)\,\hat{\Psi}_B(l,p)
= \int dl_+ \,A(l_+,\ldots)\,\int\frac{d^2 l_\perp dl_-}{(2\pi)^4}
\,\hat{\Psi}_B(l,p), 
\end{equation}
which is valid up to power corrections. 
Since the wave function can be integrated over $l_\perp$ and $l_-$ it 
follows that we need $\Psi(z,p)$ only for light-like $z$, i.e.\ for
$z_+=z_\perp=0$. We used this property to express the hard spectator 
interaction in (\ref{fff}) in terms of an integral over longitudinal 
momentum fraction $\xi$. It is possible that this cannot be 
justified in higher orders in perturbation theory. 
If not, the hard spectator interaction 
has to be generalized to include an integration over the 
transverse momentum. We leave this issue to subsequent work on 
factorization in decays into two light mesons. (An example for which 
this generalization is necessary is the decay $B^+\to\gamma e^+\nu$, 
which has recently been discussed in \cite{KPY00}.) The qualitative
discussion for light-light final states in this paper is not 
affected by this potential complication.

For the most general decomposition of the light-cone distribution 
amplitude at leading order in $1/m_b$, we make use of the fact that in 
the $B$-meson rest frame only the upper two components of the $b$-quark 
spinor are large. However, since the spectator quark is neither energetic 
nor heavy, no further restriction on the components of the spectator-quark 
spinor exists. We then find that the $B$ meson is described by two scalar 
wave functions at leading power, which we can choose as 
\begin{eqnarray}
\label{bwave}
\langle 0|\bar{q}_{\alpha}(z)[\ldots]b_{\beta}(0)|\bar{B}_d(p)
\rangle\Big|_{z_+=z_\perp=0} &=&
\nonumber\\ 
&&\hspace*{-5cm}
-\frac{i f_B}{4}\,[(\not\!p+m_b)\gamma_5]_{\beta\gamma}\,
\int_0^1 d\xi\,e^{-i\xi p_+ z_-} \left[\Phi_{B1}(\xi)+
\not\! n_- \Phi_{B2}(\xi)\right]_{\gamma\alpha},
\end{eqnarray}
where $n_-=(1,0,0,-1)$, and the normalization conditions are 
\begin{equation}
\label{normwave}
\int_0^1 d\xi\,\Phi_{B1}(\xi)=1,\qquad
\int_0^1 d\xi\,\Phi_{B2}(\xi)=0.
\end{equation}
The light spectator carries longitudinal momentum fraction 
$\xi\equiv l_+/p_+$. At leading power in $1/m_b$, we can neglect the 
difference between the $b$-quark mass and the $B$-meson mass. We 
emphasize that (\ref{bwave}) gives the most general decomposition 
of the leading-power light-cone distribution amplitude only if 
the transverse momentum of the spectator quark $l_\perp$ can be 
neglected in the hard-scattering amplitude at leading power in 
an expansion in $1/m_b$. If this is not the case, the $B$ meson 
is still described by two scalar wave functions at leading power; however, 
the right-hand side of (\ref{bwave}) has to be modified.

Contrary to the distribution amplitudes of light mesons, the $B$-meson 
distribution amplitudes are poorly known, even theoretically. At scales 
much larger than $m_b$, the $B$ meson is like a light meson and the 
distribution amplitude should approach a symmetric form. At scales of 
order $m_b$ and smaller, one expects the distribution amplitudes to 
be very asymmetric with $\xi$ of order $\Lambda_{\rm QCD}/m_b$. 

We will use the decomposition (\ref{bwave}) for the qualitative 
discussion of factorization in Sect.~\ref{arguments}. This will be 
sufficient since the remainder of the paper, which provides technical 
arguments for factorization, is restricted to decays into heavy-light 
final states, for which the hard spectator interaction, which requires 
the $B$-meson wave function, is absent. For a technical proof of 
factorization for decays into two light mesons the definition (\ref{bwave}) 
is not satisfactory for several reasons. The $B$-meson wave functions 
are defined in full QCD and contain an implicit dependence on $m_b$ 
that should be made explicit. This concerns logarithms 
of $m_b$ which have to be summed in order to define the heavy-quark 
limit properly. This can be done by matching the distribution amplitudes 
on distribution amplitudes defined 
in heavy-quark effective theory, although this is not mandatory 
at leading power in $1/m_b$. The distribution amplitudes in heavy-quark 
effective theory are expressed more naturally in terms of $l_+$ 
rather than 
the variable $\xi$, which is $m_b$ dependent. Logarithmic effects in 
$m_b$ should then be absorbed into the strong coupling and into the 
$B$-meson distribution amplitude, or summed in other ways. 
In this respect it is worth noting 
that the evolution of the $B$-meson distribution amplitude at scales below 
$m_b$ is driven by soft singularities rather than by collinear ones. The 
singularity structure implies that the integral over $l_+$ actually extends 
to infinity, because the energy of the heavy quark is infinite in the 
soft limit. In other words, even if the ``primordial'' $B$-meson 
distribution contains only momenta of order $\Lambda_{\rm QCD}$, evolution 
generates a tail that extends to infinite momenta. 
A complete definition of the $B$-meson wave function therefore contains a 
cut-off $\mu$ such that $l_+<\mu$ in addition to the cut-off 
in transverse momentum related to collinear singularities. 
We will return to these issues in a subsequent paper devoted to 
factorization for decays into two light mesons. 

\section{Arguments for factorization}
\label{arguments}

In this section we provide the basic power-counting arguments that lead 
to the factorized structure of (\ref{fff}). We shall do so by analyzing 
qualitatively the hard, soft and collinear contributions of the simplest 
diagrams in each class of contributions.

The plan of this section is as follows. We begin by spelling out the 
kinematic properties and dynamical assumptions from which power counting 
in $1/m_b$ and the relevance of light-cone distributions follow. We then 
discuss the heavy-quark scaling for the $B$-meson form factors and review 
the argument why the soft contribution is not suppressed. The analysis of 
``non-factorizable'' diagrams for decays into a heavy and a light meson 
and into two light mesons is presented subsequently. This includes a 
discussion of power suppression of the contributions from annihilation 
diagrams. The following subsection is devoted to the implications of the 
factorization formula for final-state interactions. Next we discuss decays 
in which the emission particle is a heavy-light meson, for which 
factorization (even naive factorization) does not hold, and decays in 
which the emission particle is a heavy quarkonium, for which factorization
holds in the formal heavy-quark limit. We then discuss in more detail the 
power suppression of the contributions from non-leading Fock states 
(higher-twist light-cone distribution amplitudes) of the mesons. The 
section concludes by mentioning some limitations of the QCD factorization 
approach.

\subsection{Preliminaries and power counting}

\begin{figure}[t]
   \vspace{-2.2cm}
   \epsfysize=23cm
   \epsfxsize=16cm
   \centerline{\epsffile{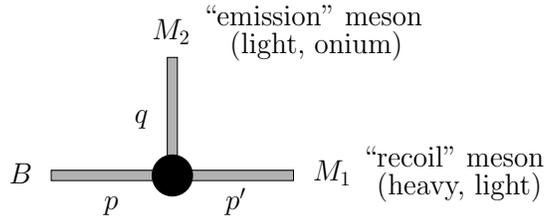}}
   \vspace*{-17.7cm}
\caption[dummy]{\label{fig2}\small Kinematics and notation.}
\end{figure}

In this section we label the meson which picks up the spectator quark 
by $M_1$ and assign momentum $p'$ to it. If $M_1$ is light, we choose 
the coordinate axis 
so that only $p'_-$ is large, i.e.\ of order $m_b$. The other meson is 
labeled $M_2$ with momentum $q$. Unless otherwise stated, $M_2$ will 
be assumed to be light, and only 
$q_+$ is of order $m_b$. When meson masses are 
neglected, $p'=m_B n_-/2$, $q=m_B n_+/2$ with $n_\pm=(1,0,0,\pm 1)$. 
See Fig.~\ref{fig2} 
for notation and further terminology. In subsequent diagrams 
lines directed upwards will always belong to $M_2$ as in Fig.~\ref{fig2}. 

The simplest diagrams that we can draw for a non-leptonic decay 
amplitude assign a quark and antiquark to each meson. We choose the 
quark and antiquark momentum in $M_2$ as 
\begin{equation}
\label{moms}
k_q = u q+k_\perp +\frac{\vec{k}_\perp^{\,2}}{2 u m_B}\,n_-,
\qquad
k_{\bar q} = \bar{u} q-k_\perp +\frac{\vec{k}_\perp^{\,2}}{2 \bar{u} m_B}\,n_-.
\end{equation}
Note that 
$q\not= k_q+k_{\bar{q}}$, but the off-shellness $(k_q+k_{\bar{q}})^2$ is 
of the same order as the light meson mass, which we can neglect 
at leading power in $1/m_b$. A similar decomposition is used 
for $M_1$ in terms of $v$, $p'$ and $k_\perp'$. 
Let $l$ denote the momentum of the spectator 
quark. The decay amplitude is then a function 
\begin{equation}
A(p',q;l;u,k_\perp;v,k'_\perp)
\end{equation}
convoluted with meson wave functions.

We start by considering the case for which 
$M_1$ is heavy. To prove (\ref{fff}) in this case one has to show 
that: 
\begin{itemize}
\item[1)] There is no leading (in $1/m_b$) contribution from the endpoint 
regions $u\sim \Lambda_{\rm QCD}/m_b$ and $
\bar{u}\sim \Lambda_{\rm QCD}/m_b$.
\item[2)] One can set $k_\perp=0$ in the amplitude (more generally, expand 
the amplitude in $k_\perp$) after collinear subtractions, which can be 
absorbed into the wave function of $M_2$. This, together with 1), 
guarantees that the amplitude is legitimately 
expressed in terms of the light-cone distribution amplitude of $M_2$.
\item[3)] The leading contribution comes from $\bar{v}\sim 
\Lambda_{\rm QCD}/m_b$, which guarantees the absence of a hard spectator 
interaction term.
\item[4)] After subtraction of infrared contributions corresponding to 
the light-cone distribution amplitude and the form factor, the leading 
contributions to the amplitude come only from internal lines with 
virtuality that scales with $m_b$.
\item[5)] Non-valence Fock states are non-leading.
\end{itemize}
If $M_1$ is light the same statements apply, except that there is now a 
leading contribution from large momentum transfer to the spectator 
quark, so that $\bar{v}$ can be of order 1. In order to verify the structure 
of the third term in the first equation in 
(\ref{fff}), one then has to show that for any 
hard spectator interaction the amplitude depends only on $l_+$, and 
that one can set 
$k'_\perp=0$ in addition to $k_\perp=0$ 
after collinear subtractions appropriate to the wave functions 
of $M_1$, $M_2$ and $B$.

The requirement that after subtractions virtualities should be large is 
obvious to guarantee the infrared finiteness of the hard-scattering 
functions $T^I_{ij}$ and $T^{II}_i$. Let us comment on setting 
transverse momenta in the wave functions to zero and on endpoint 
contributions.

Neglecting transverse momenta requires that we count them as order 
$\Lambda_{\rm QCD}$ when comparing terms of different magnitude in the 
scattering amplitude. This conforms to our intuition, and the assumption 
of the parton model, that intrinsic transverse momenta are limited to 
hadronic scales. However, in QCD transverse momenta are not limited, but 
logarithmically distributed up to the hard scale. The important point is 
that those contributions that violate the starting assumption of limited 
transverse momentum can be 
absorbed into the universal meson light-cone distribution amplitudes. 
The statement that 
transverse momenta can be counted of order $\Lambda_{\rm QCD}$ is 
to be understood after these subtractions have been performed.

The second comment concerns ``endpoint contributions'' in the 
convolution integrals over longitudinal momentum fractions. These 
contributions are dangerous, because we may be able to demonstrate 
the infrared safety of the hard-scattering amplitude under assumption 
of generic $u$ and independent of the shape of the meson 
distribution amplitude, but for $u\to 0$ or $u\to 1$ a propagator that was 
assumed to be off-shell approaches the mass-shell. If such a
contribution is of leading power, we 
do not expect the perturbative calculation of the hard-scattering 
function to be reliable.

Estimating endpoint contributions requires knowledge of the endpoint 
behaviour of the light-cone distribution amplitude. Since the 
distribution amplitude enters the factorization formula at a 
renormalization scale of order $m_b$, we can use the asymptotic 
form of the wave function to estimate the endpoint contribution. 
(More generally, we only have to assume that the distribution amplitude 
at a given scale has the same endpoint behaviour as the asymptotic 
distribution amplitude. This is generally the case, unless there 
is a conspiracy of terms in the Gegenbauer expansion of the 
distribution amplitude. If such a conspiracy existed at some scale, it would 
be immediately destroyed by evolving the distribution amplitude 
to a slightly different scale.) 
Using (\ref{asform}) we count a light meson distribution amplitude as 
order $\Lambda_{\rm QCD}/m_b$ in the endpoint region (defined as the region
where $u$ or $\bar{u}$ is 
of order $\Lambda_{\rm QCD}/m_b$, such that the quark or antiquark momentum 
is of order $\Lambda_{\rm QCD}$) and order $1$ away from the endpoint 
($X=P,||,\perp$):
\begin{equation}
\label{powerpi}
\Phi_{X}(u) \sim \left\{
\begin{array}{cl}
1 \,; & \quad \mbox{generic $u$,} \\[0.1cm]
\Lambda_{\rm QCD}/m_b \,; & \quad u,\,\bar{u} \sim \Lambda_{\rm QCD}/m_b.
\end{array}
\right.
\end{equation}
Note that the endpoint region has size of order $\Lambda_{\rm QCD}/m_b$ 
so that the endpoint suppression is $\sim(\Lambda_{\rm QCD}/m_b)^2$. 
This suppression has to be weighted against potential enhancements 
of the partonic amplitude when one of the propagators approaches 
the mass shell. 

The counting for $B$ mesons, or heavy mesons in general, is different. 
Given the normalization condition (\ref{normwave}), 
we count
\begin{equation}
\label{powerb}
\Phi_{B1}(\xi) \sim \left\{
\begin{array}{cl}
m_b/\Lambda_{\rm QCD} \,; & \quad\xi \sim \Lambda_{\rm QCD}/m_b, \\[0.1cm]
0 \,; & \quad\xi\sim 1.
\end{array}
\right.
\end{equation}
The zero probability for a light spectator with momentum of order 
$m_b$ must be understood as a boundary condition for the wave function 
renormalized at a scale much below $m_b$. There is a small probability 
for hard fluctuations that transfer large momentum to the spectator as 
discussed above. This ``hard tail'' is generated by evolution of 
the wave function from a hadronic scale to a scale of order $m_b$. 
If we assume that the initial distribution at the hadronic scale 
falls sufficiently rapidly for $\xi\gg \Lambda_{\rm QCD}/m_b$, this 
remains true after evolution. We shall assume a sufficiently 
fast fall-off, so that, for the purposes of power counting, the probability 
that the $l_+$ ($=\xi p_+$) component of the spectator quark's momentum is  
of order $m_b$ can be set to zero. 
If $M_1$ is a heavy meson, the same counting that 
applies to the $B$ meson is valid also for $M_1$. Despite the fact
that $M_1$ has momentum of order $m_b$, we do not need to distinguish the 
$B$- and $M_1$-meson rest frames for the purpose of power counting, because 
the two frames are not connected by a parametrically large boost (i.e.\ the 
Lorentz factor of the boost 
is of order 1 and not of order $m_b/\Lambda_{\rm QCD}$). In 
other words, the components of the spectator quark in $M_1$ are still 
of order $\Lambda_{\rm QCD}$. 

\boldmath
\subsection{The $B\to M_1$ form factor}
\unboldmath
\label{formfactor}

\begin{figure}[t]
   \vspace{-2.7cm}
   \epsfysize=27cm
   \epsfxsize=18cm
   \centerline{\epsffile{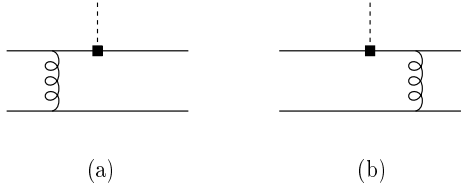}}
   \vspace*{-21.5cm}
\caption[dummy]{\label{fig3}\small Leading contributions to the 
$B\to M_1$ form factor in the hard-scattering approach. The dashed line 
represents the weak current. The two lines to the left belong to 
the $B$ meson, the ones to the right to the recoil meson $M_1$.}
\end{figure}

We now consider the form factor for the $B\to M_1$ transition and 
demonstrate that it receives a leading contribution from soft 
gluon exchange. This implies that a non-leptonic decay cannot be
treated completely in the hard-scattering picture, and therefore 
that the form 
factor should enter the factorization formula as a non-perturbative 
quantity, as in (\ref{fff}). We begin the argument with the diagrams 
shown in Fig.~\ref{fig3}, which would be leading if the 
$B\to M_1$ transition could be considered as a hard process. We shall 
also establish how the form factors scale with the mass of the heavy 
quark.

\boldmath
\subsubsection{$M_1$ heavy ($B\to D$ transitions)}
\unboldmath

The case when the final-state meson is heavy, for example a $D$ meson, 
is particularly simple. When 
the gluon exchanged in Fig.~\ref{fig3} is hard, the final spectator 
quark has momentum of order $m_b$. According to the counting rule 
(\ref{powerb}) this configuration has no overlap with the $D$-meson 
wave function. On the other hand, there is no suppression for 
soft gluons in Fig.~\ref{fig3}. It follows that the dominant behaviour 
of the $B\to D$ form 
factor in the heavy-quark limit is given by soft processes.

To answer the question how the form factor scales in the heavy-quark 
limit, we note that since the form factor is dominated 
by soft processes we can exploit the 
heavy-quark symmetries. (The discussion in this section aims only at counting 
powers of $m_b$, that is we ignore logarithmic effects in $m_b$ 
which arise from hard corrections to the $b\to c$ vertex.) Heavy-quark 
symmetry implies that the form factor 
scales like a constant, since it is equal to one at zero
velocity transfer and is independent of $m_b$ as long as the Lorentz
boost that connects the $B$ and $D$ rest frames is independent of 
$m_b$. The same conclusion also follows from the power-counting rules for 
light-cone wave functions. To see 
this, we represent the form factor by an overlap integral of 
light-cone wave functions (not integrated over transverse momentum),
\begin{equation}
\label{overlap1}
F_{+,0}^{B\to D}(0) \sim 
\int\frac{d\xi d^2k_\perp}{16\pi^3}\,\Psi_{B}(\xi,k_\perp)
\,\Psi_{D}(\xi'(\xi),k_\perp),
\end{equation}
where $\xi'(\xi)$ is fixed by kinematics and we have set $q^2=0$. The 
probability of finding the $B$ meson in its valence Fock state is 
of order 1 in the heavy-quark limit, i.e.\ 
\begin{equation}
\label{normkt}
\int\frac{d\xi d^2k_\perp}{16\pi^3}\,|\Psi_{B,D}(\xi,k_\perp)|^2\sim 1.
\end{equation}
Counting $k_\perp\sim \Lambda_{\rm QCD}$ and 
$d\xi\sim \Lambda_{\rm QCD}/m_b$, we deduce that 
$\Psi_B(\xi,k_\perp) \sim m_b^{1/2}/\Lambda_{\rm QCD}^{3/2}$. 
An alternative way to arrive at this result uses the relation 
\begin{equation}
\Phi_B(\xi) \sim \frac{1}{f_B}\,\,\int_{k_\perp<\mu} \!\!d^2k_\perp 
\Psi_{B}(\xi,k_\perp),
\end{equation}
together with $f_B\sim \Lambda_{\rm QCD}^{3/2}/m_b^{1/2}$ and the 
normalization condition for $\Phi_B(\xi)$. From (\ref{overlap1}), 
we then obtain the scaling law 
\begin{eqnarray}
F_{+,0}^{B\to D}(0) \sim 1,
\label{scalingpi}
\end{eqnarray}
in agreement with our earlier power-counting estimate. The
representation (\ref{overlap1}) of the form factor as an overlap of 
wave functions for the two-particle Fock components of the heavy-meson 
wave function is not rigorous, because there is no reason 
to assume that the contribution from higher Fock states with 
additional soft gluons is suppressed. The consistency with the 
estimate based on heavy-quark symmetry shows that these additional 
contributions are not larger than the two-particle contribution.

\boldmath
\subsubsection{$M_1$ light ($B\to\pi$ transitions)}
\unboldmath

The case of the heavy-light form factor is more complicated. When the 
exchanged gluon in Fig.~\ref{fig3} is soft, one of the quark 
constituents of $M_1$ is soft (for the purpose of illustration, 
for the remainder of this subsection we will take $M_1$ to be a pion). 
This configuration is suppressed by the endpoint behaviour of the 
pion distribution amplitude given by (\ref{powerpi}). In addition 
we now also have  
a hard contribution, for which there is no wave-function suppression.

We begin with the hard contribution. By assumption both quarks that 
form the pion have longitudinal momenta of order $m_b$, so that the 
virtuality of the exchanged gluon is of order $m_b \Lambda_{\rm QCD}$. 
These gluons can be treated perturbatively in the heavy-quark limit. 
The calculation of the diagrams shown in Fig.~\ref{fig3}, setting 
$q^2=0$ as an example, results in
\begin{eqnarray}
\label{hardbpi}
F_{+,0}^{B\to \pi}(0) &=& 
\frac{\pi\alpha_s C_F}{N_c}\,\frac{f_\pi f_B}{
m_b^2}\int_0^1 d\xi du\,[\Phi_{B1}(\xi)-2\Phi_{B2}(\xi)]\,
\Phi_\pi(u)\,\frac{1}{\xi \bar{u}^2}
\nonumber\\
&&\hspace*{-1.5cm}\,+\,\mbox{terms with }1/\bar{u}.
\end{eqnarray}
(For the calculation of the $1/\bar{u}$ term the $B$-meson 
distribution amplitude has to be generalized as indicated after 
(\ref{bwave}). However, the precise expression for the $1/\bar{u}$ 
term is not necessary for the subsequent discussion.) With 
$\bar{u}\sim 1$, $\xi\sim \Lambda_{\rm QCD}/m_b$ and the scaling 
behaviours of the distribution amplitudes discussed earlier, we obtain
\begin{equation}
F_{+,0;\,\rm hard}^{B\to \pi}(0) \sim \alpha_s(\sqrt{m_b\Lambda_{\rm QCD}})\,
\left(\frac{\Lambda_{\rm QCD}}{m_b}\right)^{\!3/2}.
\end{equation}
To our knowledge, this scaling behaviour was first derived in 
\cite{CZ90}. 

However, the computation of the hard contribution is not self-consistent. 
With $\Phi_{\pi}(u)\propto u\bar{u}$, the integral in (\ref{hardbpi}) 
diverges logarithmically for $u\to 1$ \cite{SHB90,BuDo91}. In this limit the 
momentum of the exchanged gluon approaches zero. If we interpret 
$\alpha_s$ times this logarithmic divergence as a constant of order 
1, we obtain an estimate for the soft contribution to the form factor:
\begin{equation}
\label{softbpi}
F_{+,0;\,\rm soft}^{B\to \pi}(0) \sim
\left(\frac{\Lambda_{\rm QCD}}{m_b}\right)^{\!3/2}.
\end{equation}
There is an alternative way to arrive at this result. As in 
(\ref{overlap1}) we represent the soft contribution to the form factor 
by an overlap integral of wave 
functions not integrated over 
transverse momentum. The difference is that for $u(\xi)\sim 
\Lambda_{\rm QCD}/m_b$ the wave function for the pion scales 
as $\Psi_\pi(u(\xi),k_\perp)\sim 1/m_b$. Eq.~(\ref{softbpi}) then 
follows from (\ref{overlap1}). We therefore conclude that the hard and 
soft contributions to the heavy-light form factor have the same 
scaling behaviour in the heavy-quark limit. The hard contribution is 
suppressed by one power of $\alpha_s$. This is why 
the standard approach to hard, exclusive processes \cite{LB80,EfRa80} 
is not applicable to heavy-light form factors, 
as noticed already in \cite{CZ90}. Note that 
both ways of arriving at this conclusion make use of the fact that 
the pion's light-cone distribution amplitude vanishes linearly near 
the endpoints $u=0$ or 1. Since the applicability of the Fock-state 
expansion is doubtful for endpoint regions, relying on the endpoint 
behaviour of the two-particle wave function makes the power-counting 
estimate (\ref{softbpi}) appear less solid than the estimate 
for the $B\to D$ form factor. 

The dominance of the soft contribution has been a major motivation 
for applying light-cone QCD sum rules to the calculation of the 
$B\to \pi$ form factor \cite{CZ90}. In this framework, the leading 
contribution is again given by a diagram that corresponds to the 
soft overlap term. The first order radiative correction 
\cite{KRWY97,BBB98} contains both hard and soft contributions, in 
accordance with the above discussion. Furthermore, the heavy-quark 
scaling is also consistent with the one observed above \cite{BBB98}. 
However, it should be noted that the heavy-quark scaling law in the 
framework of QCD sum rules also relies on the endpoint  behaviour 
of the pion wave function and therefore does not provide an 
independent verification of the scaling behaviour.

The upshot of this discussion is that the heavy-to-light form factor 
is not fully calculable in perturbative QCD (using light-cone distribution 
amplitudes), because the form factor is dominated by a soft endpoint 
contribution. At this point, it is worth recalling that we have 
neglected logarithmic effects in $m_b$. Summing such logarithms 
results in a Sudakov form factor that suppresses the kinematic 
configuration when almost all momentum in the $b\to u$ transition is 
transferred to the $u$ quark, i.e.\ it suppresses the singularity 
at $u=1$ in (\ref{hardbpi}). (A similar situation occurs for the 
pion form factor at next-to-leading power and is discussed in 
\cite{GT82}.) If the soft contribution were suppressed  
sufficiently by the Sudakov form factor, as would be the case in the limit 
of an asymptotically large bottom quark mass, the heavy-to-light 
form factor would be calculable perturbatively in terms of 
light-cone distribution amplitudes. By the arguments provided later 
in this section, the entire non-leptonic decay amplitude into 
two light mesons (but not a heavy and a light meson) could then be 
brought into the form of the second line of the first equation in 
(\ref{fff}). This 
observation is the starting point for the hard-scattering approach 
to non-leptonic decays as discussed further in Sect.~\ref{sec:comparison}. 
However, it appears unlikely to us that Sudakov suppression  
makes the soft contribution negligible for $m_b\approx 5\,$GeV. 
For this reason we prefer to keep the factorization formula in the 
more general form of (\ref{fff}). The important point is that 
factorization is still valid under less restrictive assumptions 
that admit a soft contribution to the heavy-light form factor. Nothing 
is lost keeping this more general form, as the form factor, which then 
appears as an additional non-perturbative input, can be obtained 
experimentally or from other methods. 

We shall see later that the first and second line in the factorization 
formula (\ref{fff}) are of the same order in the heavy-quark limit, but 
that the second line is suppressed by one power of $\alpha_s$. This 
conclusion depends on the assumed endpoint behaviour of the light-cone 
distribution amplitude of a light meson and on neglecting a 
potential Sudakov suppression of the endpoint contribution. Let us 
mention what changes if these assumptions are not valid. If Sudakov 
suppression is effective, or if the light-cone distribution amplitude 
vanishes more rapidly than linearly near the endpoint, then the hard 
contribution to the form factor is leading and both terms in the 
factorization formula are of the same order in the heavy-quark limit 
and in $\alpha_s$. If, on the other hand, the light meson distribution 
amplitude vanishes less rapidly than linearly near the endpoint, or 
if soft effects are larger than indicated by the endpoint behaviour 
of the two-particle wave function, then the first line in (\ref{fff}) 
becomes more important. (If the wave function does not vanish at the 
endpoint, the factorization formula breaks down.) In the following 
we shall assume the canonical endpoint behaviour provided by the 
asymptotic wave function. We shall also restrict ourselves to 
power counting and neglect possible Sudakov form factors.

To conclude this discussion of the form factor, we also mention that 
a complete treatment of logarithms of $m_b$ goes far beyond the Sudakov 
form factor mentioned above. The factor $1/\bar{u}^2$ that causes the 
divergence of the integral in (\ref{hardbpi}) comes only from the 
first diagram of Fig.~\ref{fig3}. In \cite{ASY94} this term is absorbed 
into a redefinition of the $B$-meson wave function by an eikonal phase. 
If this could be done to all orders, this would remove the need to include 
a soft contribution to the form factor at leading power in $1/m_b$, at 
least in perturbation theory. It is then shown in \cite{ASY94} 
that there is another 
Sudakov form factor that suppresses the contribution from the 
small-$\xi$ region to the integral in 
(\ref{hardbpi}). A complete discussion of logarithms 
of $m_b$ to all orders in perturbation theory has, to our knowledge, 
never been given. We shall not pursue this in this paper, too, since 
we use the form factor as an input.

\subsection{Non-leptonic decay amplitudes}
\label{nlamp}

We now turn to the qualitative discussion of the lowest-order 
and one-gluon exchange diagrams that could contribute to the 
hard-scattering kernels $T^I_{ij}(u)$ and $T^{II}_i(\xi,u,v)$ in 
(\ref{fff}). In the figures which follow, with the 
exception of the annihilation diagrams, the two lines directed 
upwards represent $M_2$, which we shall assume to be a $\pi^-$ 
for definiteness. The two 
lines on the left represent the $\bar{B}_d$, the lower line being the 
light $\bar d$ spectator. The two lines directed to the right represent 
$M_1$, taken to be a $D^+$ or $\pi^+$ for definiteness. 
The black square marks the weak decay vertex for 
$b\to q \bar{u} d$ where $q=c,u$ (see also Fig.~\ref{fig2}).

\subsubsection{Lowest-order diagram}
\begin{figure}[t]
   \vspace{-4cm}
   \epsfysize=27cm
   \epsfxsize=18cm
   \centerline{\epsffile{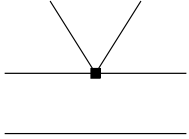}}
   \vspace*{-20.5cm}
\caption[dummy]{\label{fig4}\small Leading-order contribution to 
the hard-scattering kernel $T^I_{ij}(u)$. The weak decay of the 
$b$ quark through a four-fermion operator (of current-current or penguin 
type) is represented by the black square. See text and 
Fig.~\ref{fig2} for further notation.}
\end{figure}

There is a single diagram with no hard gluon interactions shown 
in Fig.~\ref{fig4}. According to (\ref{powerb}) 
the spectator quark is soft, and since it does not undergo a hard 
interaction it is absorbed as a soft quark by the recoiling meson. 
This is evidently a contribution to the left-hand diagram of  
Fig.~\ref{fig1}, involving the $B\to D$ ($B\to \pi$) form factor 
in the case of $\bar{B}_d\to D^+\pi^-$ ($\bar{B}_d\to \pi^+\pi^-$). 
The hard subprocess in Fig.~\ref{fig4} is just given by the insertion 
of a four-fermion operator and hence it   
does not depend on the longitudinal momentum fraction $u$ of the two 
quarks that form the emitted $\pi^-$. Consequently, the 
lowest-order contribution to $T_{ij}^I(u)$ in (\ref{fff}) is 
independent of $u$, and the $u$-integral 
reduces to the normalization condition for the pion wave function. The 
result is, not surprisingly, that the factorization formula (\ref{fff}) 
reproduces the result of naive factorization, if we neglect gluon 
exchange. 

Note that the physical picture underlying this lowest-order process is 
that the spectator quark (which is part of the $B\to D$ or 
$B\to \pi$ form factor) is soft. If this is the case, the hard-scattering 
approach misses the leading contribution to the non-leptonic decay 
amplitude. 

Putting together all factors relevant to 
power counting we find that, in the heavy-quark limit, the decay 
amplitude scales as 
\begin{equation}
\label{abd}
{\cal A}(\bar{B}_d\to D^+ \pi^-) \sim G_F m_b^2 \,F^{B\to D}(0) \,f_\pi 
\sim G_F m_b^2\,\Lambda_{\rm QCD}
\end{equation}
for a decay into a heavy-light final state (in which the spectator quark 
is absorbed by the heavy meson), and 
\begin{equation}
\label{abpi}
{\cal A}(\bar{B}_d\to \pi^+ \pi^-) \sim G_F m_b^2 \,F^{B\to \pi}(0)\,f_\pi 
\sim G_F m_b^{1/2}\,\Lambda^{5/2}_{\rm QCD}
\end{equation}
for a decay into two light mesons. 
Other contributions must be compared with these scaling rules.

\subsubsection{Factorizable diagrams}
\begin{figure}[t]
   \vspace{-3.3cm}
   \epsfysize=27cm
   \epsfxsize=18cm
   \centerline{\epsffile{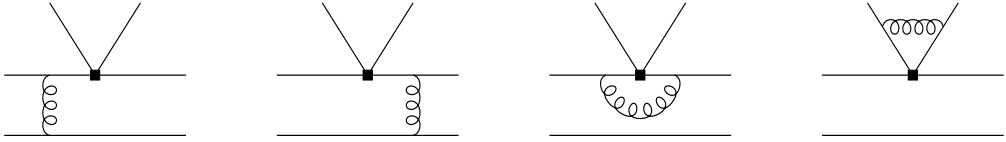}}
   \vspace*{-21.1cm}
\caption[dummy]{\label{fig5}\small Diagrams at order $\alpha_s$ that 
need not be calculated.}
\end{figure}

In order to justify naive factorization (Fig.~\ref{fig4}) as the leading 
term in an expansion in $\alpha_s$ and $\Lambda_{\rm QCD}/m_b$, we 
must show that radiative corrections are either suppressed in one 
of these two parameters, or already contained in the definition of 
the form factor and the decay constant of $M_2$ (the pion).

Consider the diagrams shown in Fig.~\ref{fig5}. The first three diagrams 
are part of the form factor and do not contribute to the hard-scattering 
kernels. Since the first and third diagrams contain leading
contributions from the region in
which the gluon is soft (as discussed for the first diagram in the 
previous subsection), they should not be considered as corrections to 
Fig.~\ref{fig4}. This is of no consequence since these soft contributions 
are absorbed into the physical form factor. The diagrams also have 
hard contributions, which we could isolate and compute. For instance, 
the hard contributions in the third diagram  
are those that go into the short-distance coefficient when the 
physical form factor is matched onto the heavy-to-heavy or heavy-to-light 
form factor in heavy-quark effective theory. However, we do not 
perform this matching here, neither do we attempt to construct a 
factorization formula for the heavy-light form factor itself. Rather, 
the form factor that appears in (\ref{fff}) is the form factor in full QCD, 
which is also the one directly measured in experiments.

The fourth diagram in Fig.~\ref{fig5} is also factorizable. 
In general, this diagram would split 
into a hard contribution and a contribution to the evolution of the pion 
distribution amplitude. However, as the leading-order diagram 
(Fig.~\ref{fig4}) involves only 
the normalization integral of the pion distribution amplitude, 
the sum of the fourth diagram in Fig.~\ref{fig5} 
and the wave-function renormalization of the quarks in the emitted pion 
vanishes. In other words, these diagrams renormalize the  $(\bar{u}d)$ 
light-quark $V-A$ current, which however is conserved.

\subsubsection{``Non-factorizable'' vertex corrections}
\begin{figure}[t]
   \vspace{-4cm}
   \epsfysize=27cm
   \epsfxsize=18cm
   \centerline{\epsffile{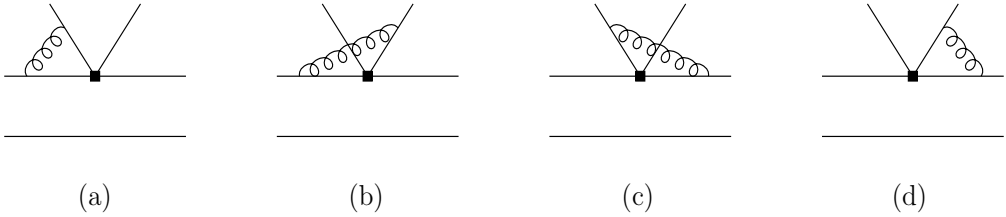}}
   \vspace*{-19.7cm}
\caption[dummy]{\label{fig6}\small ``Non-factorizable'' vertex 
corrections.}
\end{figure}

We now begin the analysis of  ``non-factorizable'' diagrams, i.e.\ diagrams
containing gluon exchanges 
that do not belong to the form factor for the $B\to M_1$ transition 
or the decay constant of $M_2$. At order $\alpha_s$ these diagrams 
can be divided into four groups: vertex corrections, penguin diagrams, 
hard spectator interactions and annihilation diagrams. We discuss 
these four cases in turn. 

The vertex corrections shown in Fig.~\ref{fig6} violate the naive 
factorization ansatz (\ref{fac1}). One of the key points of this 
paper is that these diagrams are calculable nonetheless. Let us 
summarize the argument here. The explicit evaluation of these diagrams 
can be found in Sect.~\ref{oneloop}.
A generalization of the argument to higher orders
is given in Sect.~\ref{allorders}.

The statement is that these diagrams form an order-$\alpha_s$ 
correction to the hard-scattering kernel $T^I_{ij}(u)$. To demonstrate 
this, we have to show that: (a) The transverse momentum of the 
quarks that form $M_2$ can be neglected at leading power, i.e.\ 
the two momenta in (\ref{moms}) can be approximated by $u q$ and 
$\bar{u} q$, respectively. This guarantees that only a convolution 
in the longitudinal momentum fraction $u$ appears in the factorization 
formula. (b) The contribution from the soft-gluon
region and gluons collinear to the direction of $M_2$ and $M_1$ 
(if $M_1$ is light) is power suppressed. In practice this means that 
the sum of these diagrams cannot contain any infrared divergences at 
leading power in $1/m_b$.

Neither of the two conditions holds true for any of the four diagrams 
individually, as each of them separately is collinearly and infrared 
divergent. As will be shown in detail later, the infrared divergences 
cancel when one sums over the gluon attachments to the two quarks 
comprising the emission pion ((a+b), (c+d) in Fig.~\ref{fig6}). 
This cancellation is a technical 
manifestation of Bjorken's colour-transparency argument \cite{Bj89}: 
soft gluon interactions with the emitted colour-singlet 
$\bar{u}d$ pair are suppressed, 
because they interact only with the colour dipole moment of the compact 
light-quark pair. Collinear divergences cancel after summing over gluon 
attachments to the $b$ and $c$ (or $u$) quark line ((a+c), (b+d) in 
Fig.~\ref{fig6}); in the light-cone gauge, 
collinear divergences are absent altogether. Thus the sum 
of the four diagrams 
(a-d) involves only hard gluon exchange at leading power. Because 
the hard gluons transfer large momentum to the quarks that form the 
emission pion, the hard-scattering factor now results in a non-trivial 
convolution with the pion distribution amplitude. 
``Non-factorizable'' contributions are therefore non-universal, i.e.\ they 
depend on what type of meson $M_2$ is. 

Note that the colour-transparency argument, and hence the cancellation 
of soft gluon effects, applies only if the 
$\bar{u}d$ pair is compact. This is not the case if the emitted 
pion is formed in a very asymmetric configuration, in which one of 
the quarks carries almost all of the pion's momentum. Since the 
probability for forming a pion in such an endpoint configuration 
is of order $(\Lambda_{\rm QCD}/m_b)^2$, they could become important 
only if the hard-scattering amplitude favoured the production 
of these asymmetric pairs, i.e.\ if $T^I_{ij}\sim 1/u^2$ for $u\to 0$ 
(or  $T^I_{ij}\sim 1/\bar{u}^2$ for $u\to 1$). However, such strong
endpoint singularities in the hard-scattering amplitude do not occur.

\subsubsection{Penguin diagrams}
\begin{figure}[t]
   \vspace{-4.6cm}
   \epsfysize=27cm
   \epsfxsize=18cm
   \centerline{\epsffile{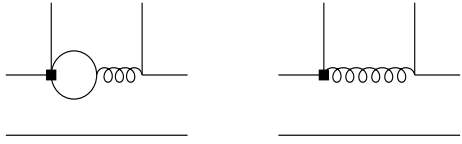}}
   \vspace*{-19.9cm}
\caption[dummy]{\label{fig7}\small Diagram with a ``penguin'' 
contraction. The second diagram represents a contribution from the 
chromomagnetic dipole operator in the weak effective Hamiltonian.}
\end{figure}

The penguin diagram (first diagram in Fig.~\ref{fig7}) exists for 
$\bar{B}_d\to \pi^+\pi^-$ but not for $\bar{B}_d\to D^+\pi^-$. We need 
to show again that, at leading order in $1/m_b$, all internal lines in 
this diagram are hard. 

Consider first the two final-state quarks into which the gluon splits.  
The quark that goes into the recoil $\pi^+$ to the right must always be 
energetic to make an energetic pion, because the 
$\bar{d}$ spectator quark is soft. The configuration in which the other 
quark is soft is suppressed by the endpoint behaviour of the 
light-cone distribution amplitude of the $\pi^-$. We conclude that 
the gluon splits into two energetic quarks that fly in opposite 
directions, and that the gluon has large virtuality $\sim\bar{u} m_b^2$, 
where $\bar{u}$ is the longitudinal momentum fraction of the antiquark 
in the $\pi^-$. In principle, one of the quarks in the 
quark loop can still be soft, if the loop momentum is soft and the 
gluon momentum flows asymmetrically through the loop. But this 
configuration is suppressed by two powers of $\Lambda_{\rm QCD}/m_b$ relative 
to the configuration where both quarks 
carry large momentum of order $m_b$, as follows from the structure of 
a vacuum polarization diagram. As a result the penguin diagram  
contributes to the hard-scattering kernel $T_{ij}^I(u)$ at order 
$\alpha_s$, just as the vertex diagrams do. The same argument shows 
that the chromomagnetic dipole diagram (second diagram in 
Fig.~\ref{fig7}) is also a calculable correction to the hard-scattering 
kernel. An explicit calculation of these diagrams can be found in 
\cite{BBNS99}.

Note that this argument provides a rigorous justification for the 
Bander-Silver\-man-Soni (BSS) mechanism \cite{BSS79} to generate 
strong-interaction phases perturbatively by means of the rescattering 
phase of the penguin loop. In particular, the gluon virtuality 
$k^2=\bar{u} m_b^2$, which is usually treated as a phenomenological 
parameter, is unambiguously determined by the kinematics of the 
decay process together with the weighting of $\bar u$
implied by the pion wave function. 
At the same time it should be noted that the BSS 
mechanism does not provide a complete description of final-state 
interactions even in the heavy-quark limit, as the vertex 
diagrams (c,d) of Fig.~\ref{fig6} also generate imaginary parts, 
which are of the same order as those of the penguin diagram. A more 
detailed discussion of final-state interaction phases will be 
presented in Sect.~\ref{fsi}.

\subsubsection{Hard spectator interaction}
\label{subsec:hardspec}
\begin{figure}[t]
   \vspace{-3.5cm}
   \epsfysize=27cm
   \epsfxsize=18cm
   \centerline{\epsffile{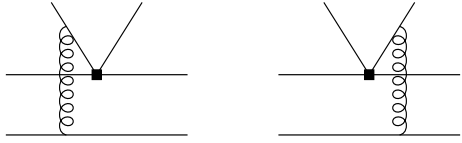}}
   \vspace*{-21.2cm}
\caption[dummy]{\label{fig8}\small ``Non-factorizable'' spectator interactions.}
\end{figure}

Up to this point, we have not obtained a contribution to the second line 
of (\ref{fff}), i.e.\ to the hard-scattering term in Fig.~\ref{fig1} (as 
opposed to the form-factor term). The diagrams shown in Fig.~\ref{fig8} 
cannot be associated with the form-factor term. These diagrams would 
impede factorization if there existed a soft contribution at 
leading power. While such terms are present in 
each of the two diagrams separately, to leading power they cancel 
in the sum over the two gluon attachments to the $\bar{u}d$ pair 
by the same colour-transparency 
argument that was applied to the ``non-factorizable'' vertex 
corrections. For decays into two light mesons there is a further 
suppression of soft gluon exchange because of the endpoint 
suppression of the light-cone distribution amplitude for the recoiling 
meson $M_1$. 

We consider first the decay into a heavy and a light meson 
($\bar{B}_d\to D^+\pi^-$) in more detail. We still have to show that 
after the soft cancellation the remaining soft contribution is 
power suppressed relative to the leading-order contribution 
(\ref{abd}). A straightforward calculation leads to the following 
(simplified) result for the sum of the two diagrams:
\begin{eqnarray}
\label{specd}
{\cal A}(\bar{B}_d\to D^+ \pi^-)_{({\rm Fig.~\ref{fig8}})} 
&\sim& G_F \,f_{\pi} f_D f_B\,
\alpha_s\,
\int_0^1 \frac{d\xi}{\xi}\,\Phi_{B1}(\xi)
\int_0^1 \frac{d\eta}{\eta}\,\Phi_{D1}(\eta)
\int_0^1 \frac{d u}{u}\,\Phi_{\pi}(u)
\nonumber\\
&& \hspace*{-3cm} \sim G_F \,\alpha_s\,m_b\,\Lambda_{\rm QCD}^{2}.
\end{eqnarray}
This is indeed power suppressed relative to (\ref{abd}). Note that 
the gluon virtuality is of order $\xi\eta m_b^2\sim \Lambda_{\rm QCD}^2$ 
and so, strictly speaking, the calculation in terms of light-cone distribution 
amplitudes cannot be justified. Nevertheless, we use (\ref{specd}) to 
estimate the size of the soft contribution, as we did for the heavy-light 
form factor. On the contrary, when the gluon is hard, it transfers 
large momentum to the spectator quark. According to our power-counting 
rule (\ref{powerb}), such a configuration has no overlap with either 
the $B$- or the $D$-meson wave function. We therefore conclude that 
the hard spectator interaction does not contribute to heavy-light final 
states at leading power in the heavy-quark expansion. The factorization 
formula (\ref{fff}) then assumes a simpler form, with the second 
line omitted, as discussed earlier.

For decays into two light mesons ($\bar{B}_d\to \pi^+\pi^-$) the explicit 
expression for the sum of the two diagrams is similar to the one 
above \cite{BBNS99}:
\begin{eqnarray}
\label{specpi}
{\cal A}(\bar{B}_d\to \pi^+ \pi^-)_{({\rm Fig.~\ref{fig8}})} 
&\sim& G_F \,f_{\pi}^2 f_B\,
\alpha_s\,\int_0^1 \frac{d\xi}{\xi}\,\Phi_{B1}(\xi)
\int_0^1 \frac{d v}{\bar{v}}\,\Phi_{\pi}(v)
\int_0^1 \frac{d u}{u}\,\Phi_{\pi}(u)
\nonumber\\
&& \hspace*{-3cm} \sim 
\left\{
\begin{array}{ll}
G_F \,\alpha_s\,m_b^{1/2}\,\Lambda_{\rm QCD}^{5/2} \,; & 
\mbox{hard gluon},\\[0.1cm]
G_F \,\alpha_s\,m_b^{-1/2}\,\Lambda_{\rm QCD}^{7/2} \,; & 
\mbox{soft gluon}.
\end{array}
\right.
\end{eqnarray}
The soft contribution is suppressed as discussed above, but 
the hard contribution is of the same order as (\ref{abpi}), with an 
additional factor of $\alpha_s$. (The hard gluon has momentum of order 
$m_b$, but its virtuality is only of order $m_b\Lambda_{\rm QCD}$, 
similar to the hard contribution to the $B\to \pi$ form factor.) 
Eq.~(\ref{specpi}) results in 
a contribution to the second hard-scattering kernel, $T^{II}_i(\xi,u,v)$, 
in (\ref{fff}). In the heavy-quark limit, the hard spectator interaction 
is of the same order as the vertex corrections and penguin contributions 
to the first hard-scattering kernel. (See, however, the comments 
at the end of Sect.~\ref{formfactor} concerning a modification of 
this statement in the presence of Sudakov form factors.)

\subsubsection{Annihilation topologies}
\label{subsec:annihilation}
\begin{figure}[t]
   \vspace{-5.1cm}
   \epsfysize=27cm
   \epsfxsize=18cm
   \centerline{\epsffile{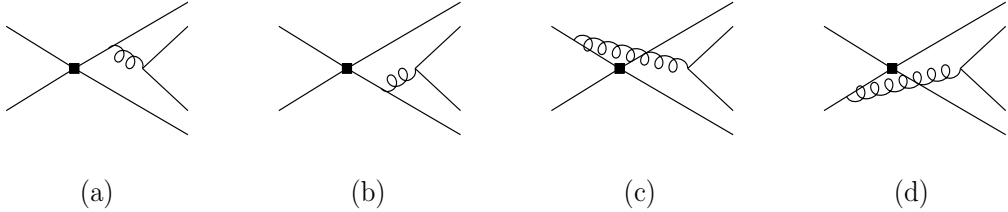}}
   \vspace*{-18.5cm}
\caption[dummy]{\label{fig9}\small Annihilation diagrams.}
\end{figure}

Our final concern in this subsection are the annihilation diagrams 
(Fig.~\ref{fig9}) which contribute to $\bar{B}_d\to D^+ \pi^-$ and 
$\bar{B}_d\to \pi^+ \pi^-$. The hard part of these diagrams would 
amount to another contribution to the second hard-scattering kernel, 
$T^{II}_i(\xi,u,v)$. The soft part, if unsuppressed, would violate 
factorization. However, we shall show now that the hard part as well 
as the soft part are suppressed by at least one power of 
$\Lambda_{\rm QCD}/m_b$.

\vspace*{0.4cm}
\noindent
{\em Light-light final states ($\bar{B}_d\to \pi^+ \pi^-$)}

\vspace*{0.2cm}
\noindent
We begin with the two diagrams (a,b). Suppose first that all four 
light quarks in the final state are energetic. Then the virtuality 
of the gluon is of order $m_b^2$. If we now let one of the quarks be 
soft, the gluon virtuality can decrease to $m_b \Lambda_{\rm QCD}$ and 
the amplitude is then enhanced by a factor $m_b/\Lambda_{\rm QCD}$. 
(In particular cases, the virtuality of the internal quark line 
can also become small. However, closer inspection shows 
that in this case the numerator also becomes small and there is no 
further enhancement of the amplitude.) This enhancement is over-compensated 
by a suppression with two powers of $\Lambda_{\rm QCD}/m_b$, where 
one power arises from the endpoint behaviour of the pion distribution 
amplitude and another from the small region of phase space 
considered. The configuration where two final-state quarks are soft 
is even further suppressed. It follows that the leading contribution 
to (a,b) arises when all four quarks are energetic. Since the 
integral over the $B$-meson wave function simply gives the 
normalization integral, it is easy to see that the diagrams scale 
at most as 
\begin{equation}
\label{ann1}
G_F\,f_\pi^2 f_B\alpha_s\sim G_F\,\alpha_s m_b^{-1/2} 
\Lambda_{\rm QCD}^{7/2},
\end{equation}
which is one power of $\Lambda_{\rm QCD}/m_b$ smaller than 
(\ref{abpi}). (In fact, current conservation implies that the result 
is proportional to the difference of quark masses at the annihilation 
vertex. Hence the sum of (a) and (b) vanishes 
for $\bar{B}_d\to \pi^+ \pi^-$.)

The hard part of diagrams (c,d) (all four quarks energetic) 
obviously also scales as (\ref{ann1}). A difference to (a,b) arises 
when some of the quarks are soft. For emission of the 
light $q\bar{q}$ pair from the $B$-meson 
spectator quark (diagram (d)) it may happen that the endpoint 
contribution from a single soft final-state quark is not 
suppressed relative to the hard part, because the amplitude is 
enhanced by a large internal gluon and quark propagator. 
(But since the gluon virtuality is still of order $m_b\Lambda_{\rm QCD}$,  
there remains a factor $\alpha_s(\sqrt{m_b\Lambda_{\rm QCD}})$). 
However, since the hard part is power suppressed relative to (\ref{abpi}), 
power suppression continues to hold for the entire graph.

\vspace*{0.4cm}
\noindent
{\em Heavy-light final states ($\bar{B}_d\to D^+ \pi^-$)}
\vspace*{0.2cm}

\noindent
The power counting is different for  
$\bar{B}_d\to D^+\pi^-$, because the light quark that 
goes into the $D$ meson must always be soft according to (\ref{powerb}), 
and hence the virtuality of the gluon is never larger than 
$m_b\Lambda_{\rm QCD}$. Nevertheless, we obtain power suppression 
also in this case. The argument is as follows. We can write 
the annihilation amplitude as 
\begin{equation}
\label{ann2}
{\cal A}(\bar{B}_d\to D^+ \pi^-)_{({\rm Fig.~\ref{fig9}})} 
\sim G_F \,f_{\pi} f_D f_B\,
\alpha_s\,\int_0^1 d\xi d\eta d u
\,\Phi_{B1}(\xi)\,\Phi_{D1}(\eta)\,\Phi_{\pi}(u)\,
T^{\rm ann.}(\xi,\eta,u),
\end{equation}
where the dimensionless function $T^{\rm ann.}(\xi,\eta,u)$ 
is a product of propagators and vertices. The product of decay 
constants scales as $\Lambda_{\rm QCD}^4/m_b$. Since 
$d\xi\,\Phi_{B1}(\xi)$ scales as 1 and so does $d\eta\,\Phi_{D1}(\eta)$, 
while $du\,\Phi_\pi(u)$ is never larger than 1, the amplitude can 
only compete with the leading-order result (\ref{abd}) if 
$T^{\rm ann.}(\xi,\eta,u)$ can be made of order 
$(m_b/\Lambda_{\rm QCD})^3$ or larger. Since $T^{\rm ann.}(\xi,\eta,u)$ 
contains only two propagators, this can be achieved only  
if both quarks the gluon splits into are soft, in which case 
$T^{\rm ann.}(\xi,\eta,u)\sim (m_b/\Lambda_{\rm QCD})^4$. But then 
$du\,\Phi_\pi(u)\sim (\Lambda_{\rm QCD}/m_b)^2$ so that this 
contribution is power suppressed.

\subsubsection{Summary}

To summarize the discussion up to this point: for the decay into a light 
emitted and a heavy recoiling meson (such as our example $\bar{B}_d\to 
D^+ \pi^-$) 
the second factorization formula in (\ref{fff}) holds. The hard-scattering 
kernel $T^I_{ij}(u)$ is computed in lowest order 
from the diagram shown in Fig.~\ref{fig4}, and at order $\alpha_s$ from the 
vertex diagrams in Fig.~\ref{fig6}. For decays 
into two light mesons, the more complicated first formula in (\ref{fff}) 
applies. Then, in addition to the vertex diagrams, there are 
penguin contributions (Fig.~\ref{fig7}) to the kernel $T^I_{ij}(u)$, and 
there is a non-vanishing hard-scattering term in (\ref{fff}). The kernel 
$T_i^{II}(\xi,u,v)$ 
is computed from the diagrams shown in Fig.~\ref{fig8}. In both cases, 
naive factorization 
follows when one neglects all corrections of order 
$\Lambda_{\rm QCD}/m_b$ {\em and\/}  
of order $\alpha_s$. Eq.~(\ref{fff}) allows us to compute systematically 
corrections to higher order in $\alpha_s$, but still neglects power 
corrections of order $\Lambda_{\rm QCD}/m_b$.

Some of the loop diagrams entering the calculation of the hard-scattering 
kernels have imaginary parts which contribute to the 
strong rescattering phases. It follows from our discussion that these 
imaginary parts are of order $\alpha_s$ or 
$\Lambda_{\rm QCD}/m_b$. This demonstrates that 
strong phases vanish in the heavy-quark limit (unless the real parts of the 
amplitudes are also suppressed). Since this statement 
goes against the folklore that prevails from the present understanding 
of this issue, we shall return to this point in Sect.~\ref{fsi}.

In a common terminology, the decays which we have treated explicitly so far 
are called ``class-I'' decays. The distinction of ``class-I'', ``class-II'' 
and ``class-III'' decays refers to colour factors and charge combinatorics 
arising in naive factorization. It is clear that this distinction is not 
relevant to QCD factorization in the sense of (\ref{fff}), which relies 
on the hardness and virtuality of partons. This means that the factorization 
formula applies to any decay into two light mesons, irrespective of whether 
the decay is class-I, class-II or dominated by penguin operators. 
Factorization also works for all decays into heavy-light final states, 
in which the light spectator quark in the $B$ meson is absorbed by the 
heavy final-state particle (class-I). Factorization does {\em not\/} work for 
a heavy-light final state, when the spectator quark is picked up by the 
light meson (class-II), for example $\bar{B}_d\to \pi^0 D^0$. 
We will return to this point in Sect.~\ref{hl}. 

Our discussion has so far been based on the leading two-particle 
valence-quark Fock state of the mesons. To complete the discussion 
we shall argue in Sect.~\ref{otherfock} that the contributions to the 
decay amplitude from higher Fock components 
of the meson wave functions are power suppressed. In 
Sect.~\ref{limitations} we will discuss some of the limitations of the 
applicability of the factorization 
formula in practice, recalling that the physical mass of the 
$b$ quark is not asymptotically large.

\subsection{Remarks on final-state interactions}
\label{fsi}

Since the subject of final-state interactions, and of 
strong-interaction phases in particular, is of paramount importance 
for the interpretation of CP-violating observables, we discuss 
here in some more detail the implications of QCD factorization for 
this issue.

Final-state interactions are usually discussed in terms of intermediate 
hadronic states. This is suggested by the unitarity relation (taking 
$B\to \pi\pi$ for definiteness) 
\begin{equation}
\label{unitarity}
\mbox{Im}\,{\cal A}_{B\to \pi\pi} \sim \sum_n 
{\cal A}_{B\to n}{\cal A}_{n\to \pi\pi}^*,
\end{equation}
where $n$ runs over all hadronic intermediate states. We can also 
interpret the sum in (\ref{unitarity}) as extending over   
intermediate states of partons. The partonic interpretation is justified 
by the dominance of hard rescattering in the heavy-quark limit. In 
this limit the number of physical intermediate states is arbitrarily 
large. We may then argue on the grounds of parton-hadron duality 
that their average is described well enough (up to $\Lambda_{\rm QCD}/
m_b$ corrections, say) by a partonic calculation. This is the picture 
implied by (\ref{fff}). The hadronic language is in principle 
exact. However, the large number of intermediate states makes 
it intractable to observe systematic cancellations, which 
usually occur in an inclusive sum over hadronic intermediate states.

A particular contribution to the right-hand side of (\ref{unitarity}) 
is elastic rescattering ($n=\pi\pi$). The energy dependence of the 
total elastic $\pi\pi$-scattering cross section is governed by 
soft pomeron behaviour. Hence the strong-interaction phase 
of the $B\to \pi\pi$ amplitude due to elastic rescattering alone  
increases slowly in the heavy-quark limit \cite{DGPS96}. On 
general grounds, it is rather improbable that elastic rescattering 
gives an appropriate representation of the imaginary part of 
the decay amplitude in the heavy-quark limit. This 
expectation is also borne out in the framework of Regge behaviour, as 
discussed in \cite{DGPS96}, where the importance (in fact, dominance) of 
inelastic rescattering is emphasized. However, the approach pursued 
in \cite{DGPS96} leaves open the possibility of soft rescattering phases 
that do not vanish in the heavy-quark limit, as well as the possibility 
of systematic cancellations, for which the Regge approach does not provide 
an appropriate theoretical framework.

Eq.~(\ref{fff}) implies that such systematic cancellations 
do occur in the sum over all intermediate states $n$. It is worth 
recalling that similar cancellations are not uncommon for hard 
processes. Consider the example of $e^+ e^-\to\,$hadrons at large 
energy $q$. While the production of any hadronic final state 
occurs on a time scale of order $1/\Lambda_{\rm QCD}$ (and would 
lead to infrared divergences if we attempted to describe it using 
perturbation theory), the inclusive cross section given by the sum 
over all hadronic final states is described very well by a 
$q\bar{q}$ pair that lives over a short time scale of order $1/q$. In 
close analogy, while each particular hadronic intermediate state 
$n$ in (\ref{unitarity}) cannot be described partonically, the 
sum over all intermediate states is accurately represented by 
a $q\bar{q}$ fluctuation of small transverse size of order $1/m_b$. 
Because the $q\bar{q}$ pair is small, the physical picture of 
rescattering is very different from elastic $\pi\pi$ scattering.

In perturbation theory, the pomeron is associated with two-gluon 
exchange. The analysis of two-loop contributions to the non-leptonic 
decay amplitude in Sect.~\ref{allorders} shows that the soft 
and collinear cancellations that guarantee the partonic interpretation 
of rescattering extend to two-gluon exchange. (Strictly speaking, 
the analysis of Sect.~\ref{allorders} applies only to decays into a 
heavy and a light meson. However, the cancellation in the soft-soft 
region, which is relevant to the present discussion, goes through 
unmodified if both final-state mesons are light.) Hence, the soft 
final-state interactions are again subleading as required by the validity of 
(\ref{fff}). As far as the hard rescattering contributions are concerned, 
two-gluon exchange plus ladder graphs between 
a compact $q\bar{q}$ pair with energy of order $m_b$ and transverse 
size of order $1/m_b$ and the other pion does not lead to large 
logarithms, and hence there is no possibility to construct the (hard) pomeron. 
Note the difference with elastic vector-meson production through a virtual 
photon, which also involves a compact $q\bar{q}$ pair. However, 
in this case one considers $s\gg Q^2$, where $\sqrt{s}$ 
is the photon-proton center-of-mass 
energy and $Q$ the virtuality of the photon. This implies that the 
$q\bar{q}$ fluctuation is born long before it hits the proton. It 
is this difference of time scales, non-existent in non-leptonic 
$B$ decays, that permits pomeron exchange in 
elastic vector-meson production in $\gamma^* p$ collisions. 

It follows from (\ref{fff}) that the leading strong-interaction phase is 
of order $\alpha_s$ in the heavy-quark limit. (More precisely, the imaginary 
part of the decay amplitude is of order $\alpha_s$, so rescattering phases 
are small unless the real part, which starts at order $\alpha_s^0$, 
is suppressed.) The same statement holds 
for rescattering in general. For instance, 
according to the duality argument, a 
penguin contraction with a charm loop represents 
the sum over all intermediate states of the form 
$D\bar{D}$, $J/\Psi\rho$, etc. that rescatter into two pions. 

As is clear from the discussion, parton-hadron duality is crucial for 
the validity of (\ref{fff}) beyond perturbative factorization. Proving 
quantitatively to what accuracy we can expect duality to hold is, as yet, an   
unsolved problem in QCD. In the absence of 
a solution, it is worth noting that the 
same (often implicit) assumption is fundamental to 
many successful QCD predictions in jet physics and hadron-hadron collisions. 
In particular, the duality assumption that the sum over all hadronic states 
in (\ref{unitarity}) is calculable in terms of partons (given the 
dominance of hard scattering) is the same assumption that forms the 
basis for the application of the operator product expansion to 
{\em inclusive\/} non-leptonic heavy-quark decays \cite{Bigi:1992su}.

\boldmath
\subsection{Non-leptonic decays when $M_2$ is not light}
\unboldmath
\label{hl}

The analysis of non-leptonic decay amplitudes in Sect.~\ref{nlamp} 
referred to decays where the emission particle $M_2$ -- the meson that does 
{\em not\/} pick up the 
spectator quark -- is a light meson. We now discuss the two other 
possibilities, $M_2$ a heavy meson (for example, $D$) and $M_2$ an 
onium such as $J/\psi$. 

\boldmath
\subsubsection{$M_2$ a heavy-light meson ($\bar{B}_d\to\pi^0 D^0,\,D^+ D^-$)}
\unboldmath

Suppose that $M_2$ is a $D$ meson and the meson that picks up the 
spectator quark is heavy or light. Examples of this type are the 
decays $\bar{B}_d\to \pi^0 D^0$ and $\bar{B}_d\to D^+ D^-$. It is 
intuitively clear that factorization must be problematic in this case, 
because the heavy $D$ meson has large overlap with the $B\pi$ (or 
$BD$ in case of $\bar{B}_d\to D^+ D^-$) system, which is dominated by 
soft processes.

In more detail, we consider the coupling of a gluon to the two quarks 
that form the emitted $D$ meson, i.e.\ the pairs of diagrams in 
Figs.~\ref{fig6} (a+b), (c+d) and Fig.~\ref{fig8}. Denoting the gluon 
momentum by $k$, the quark momenta by $l_q$ and $l_{\bar{q}}$, and the 
$D$-meson momentum by $q$, we find that the gluon couples to the 
``current''
\begin{equation}
\label{jj}
J_\lambda = 
\frac{\gamma_\lambda(\not\! l_q+\not\! k+m_c)\Gamma}{2 l_q\cdot k+k^2} - 
\frac{\Gamma(\not\! l_{\bar{q}}+\not\! k)\gamma_\lambda}{2 l_{\bar{q}}
\cdot k+k^2},
\end{equation}
where $\Gamma$ is part of the weak decay vertex. 
When $k$ is soft (all components of order $\Lambda_{\rm QCD}$) 
each of the two terms scales as $1/\Lambda_{\rm QCD}$. 
Taking into account 
the complete amplitude as done explicitly in Sect.~\ref{oneloopcancel}, we 
can see that the decoupling of soft gluons requires that the two terms 
in (\ref{jj}) cancel, leaving a remainder of order $1/m_b$. This 
cancellation does indeed occur when $M_2$ is a light meson, since in 
this case $l_q$ and $l_{\bar{q}}$ are dominated by their longitudinal 
components. When $M_2$ is heavy the momenta $l_q$ and $l_{\bar{q}}$ are 
asymmetric, with all components of the light antiquark momentum 
$l_{\bar{q}}$ of order $\Lambda_{\rm QCD}$ in the $B$- or $D$-meson 
rest frame, while the zero-component of $l_q$ is of order $m_c\sim m_b$. 
Hence the current can be approximated by 
\begin{equation}
J_\lambda \approx \frac{\delta_{\lambda 0}\Gamma}{k_0}- 
\frac{\Gamma(\not\! l_{\bar{q}}+\not\! k)\gamma_\lambda}{2 l_{\bar{q}}
\cdot k+k^2} \sim \frac{1}{\Lambda_{\rm QCD}},
\end{equation}
and the soft cancellation does not occur. (The on-shell condition 
for the charm quark has been used to arrive at the previous equation.)

It follows that the emitted $D$ meson does not factorize from the 
rest of the process and that a factorization formula analogous 
to (\ref{fff}) does not apply to decays such as 
$\bar{B}_d\to \pi^0 D^0$ and $\bar{B}_d\to D^+ D^-$. An important 
implication of this is that one should also not expect naive 
factorization to work in this case. In other words, non-factorizable 
corrections such as those shown in Fig.~\ref{fig6} modify the 
(naively) factorized decay amplitude by terms of order 1. 

There are decay modes, such as $B^-\to D^0\pi^-$, in which the 
spectator quark can go to either of the two final-state mesons. The 
factorization 
formula (\ref{fff}) applies to the contribution that arises when 
the spectator quark goes to the $D$ meson, but not when the spectator 
quark goes to the pion. However, even in the latter case we may use 
naive factorization to estimate the power behaviour of the decay 
amplitude. Adapting (\ref{abd}) and (\ref{abpi}) to the 
decay $B^-\to D^0\pi^-$, we find that the non-factorizing (class-II) 
amplitude is suppressed compared to the factorizing (class-I) 
amplitude:
\begin{equation}
\label{onetwo}
\frac{{\cal A}(B^-\to D^0\pi^-)_{\rm class-II}}
{{\cal A}(B^-\to D^0\pi^-)_{\rm class-I}} \sim 
\frac{F^{B\to \pi}(m_D^2) f_D}{F^{B\to D}(0) f_\pi} 
\sim \left(\frac{\Lambda_{\rm QCD}}{m_b}\right)^2.
\end{equation}
Here we use that $F^{B\to \pi}(q^2) \sim 1/m_b^{3/2}$ even for 
$q^2\sim m_b^2$ as long as $q_{max}^2-q^2$ is also of order $m_b^2$. 
(It follows from our definition of heavy final-state mesons that 
these conditions are fulfilled.) As a consequence, factorization 
does hold for $B^-\to D^0\pi^-$ in the sense that the class-II 
contribution is power suppressed. It should be mentioned that 
(\ref{onetwo}) refers to the heavy-quark limit and that the scaling 
behaviour for real $B$ and $D$ mesons is far from the 
estimate (\ref{onetwo}). This will be discussed briefly later in 
this section and in more detail in Sect.~\ref{bdpi}.

\boldmath
\subsubsection{$M_2$ an onium ($\bar{B}_d\to J/\psi K$)}
\unboldmath

The case where $M_2$ is a heavy quarkonium is special, because then  
additional momentum scales are involved. We consider the decay into 
charmonium and suppose that $m_c\sim m_b \to \infty$, bearing in mind 
that this limit is hardly realistic. 

The gluon coupling to the $c\bar{c}$ pair analogous to (\ref{jj}) 
is now given by
\begin{equation}
\label{jjcc}
J_\lambda = 
\frac{\gamma_\lambda(\not\! l_q+\not\! k+m_c)\Gamma}{(l_q+k)^2-m_c^2} - 
\frac{\Gamma(\not\! l_{\bar{q}}+\not\! k-m_c)\gamma_\lambda}{(l_{\bar{q}}
+k)^2-m_c^2}.
\end{equation}
In the heavy-quark limit we may write $l_q=q/2+p$, $l_{\bar{q}}=
q/2-p$, where $p$ is of order $m_c \alpha_s$, the inverse size of 
the charmonium. A second important difference to the case considered 
previously is that the charm-quark lines directed upwards in 
Figs.~\ref{fig4} and 6 must be considered off-shell by an amount 
$\delta\sim (m_c\alpha_s)^2$. When $k$ is soft (all components of 
order $\Lambda_{\rm QCD}$), the denominators in (\ref{jjcc}) are 
dominated by the off-shellness $\delta$, and the current simplifies 
to
\begin{equation}
\label{jjccapp}
J_\lambda \approx 
\frac{1}{\delta}\left(4 p_\lambda\Gamma+
(-\!\not\! l_q+m_c)\gamma_\lambda\Gamma-
\Gamma\gamma_\lambda(-\!\not\! l_{\bar{q}}-m_c)\right)
\approx
\frac{4 p_\lambda\Gamma}{\delta}\sim \frac{1}{m_b\alpha_s}.
\end{equation}
Here we used that $-\!\!\not\! l_q+m_c$ acting to the left (and similarly 
$-\!\!\not\! l_{\bar{q}}-m_c$ acting to the right) gives a contribution of 
order $m_c\alpha_s^2$, and we identified $m_c$ and $m_b$ in our formal 
scaling limit. (Note that the scale 
of $\alpha_s$ that appears here is $p\sim m_c\alpha_s$.) It follows that 
factorization does hold for decay modes like $J/\psi K$, although the 
soft gluon contribution is suppressed only by a factor 
$\Lambda_{\rm QCD}/(m_b\alpha_s)$ rather than $\Lambda_{\rm QCD}/m_b$. 
This reflects the fact that an onium is small in the heavy-quark limit, 
but that its Bohr radius is larger than $1/m_b$. For $J/\psi$ the 
suppression is probably only marginal. On the other hand,
factorization is also recovered in the limit $m_c/m_b\to 0$, i.e.\ when
the $J/\psi$ is treated as a light meson relative to the $B$ meson.

\subsection{Non-leading Fock states}
\label{otherfock}

The discussion of the previous subsections concentrated on contributions 
related to the quark-antiquark components of the meson wave functions. 
We now present qualitative arguments that justify this restriction to 
the valence-quark Fock components. Some of these arguments are standard 
\cite{LB80,EfRa80}.

\begin{figure}[t]
   \vspace{-3.3cm}
   \epsfysize=27cm
   \epsfxsize=18cm
   \centerline{\epsffile{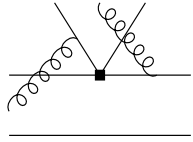}}
   \vspace*{-21.1cm}
\caption[dummy]{\label{fig10}\small Diagram that contributes to the 
hard-scattering kernel involving a quark-antiquark-gluon distribution 
amplitude of the $B$ meson and the emitted light meson.}
\end{figure}

An example of a diagram that would contribute to a hard-scattering 
function involving quark-antiquark-gluon components of the emitted 
meson and the $B$ meson is shown in Fig.~\ref{fig10}. 
For light mesons higher Fock components are related to higher-order 
terms in the collinear expansion, including the effects of intrinsic 
transverse momentum and off-shellness of the partons by gauge 
invariance. The assumption is that the additional partons 
are collinear and carry a finite fraction of the meson's momentum in 
the heavy-quark limit. Under this assumption, it is easy to see that 
adding additional partons to the Fock state increases the number 
of off-shell propagators in a given diagram (compare Fig.~\ref{fig10} to 
Fig.~\ref{fig4}). This implies power suppression in the heavy-quark 
expansion. Additional partons 
in the $B$-meson wave function are always soft, as is the spectator 
quark. Nevertheless, when these partons are connected to the 
hard-scattering amplitudes the virtuality of the additional propagators 
is still of order $m_b\Lambda_{\rm QCD}$, which is sufficient to 
guarantee power suppression.

\begin{figure}[t]
   \vspace{-3.3cm}
   \epsfysize=27cm
   \epsfxsize=18cm
   \centerline{\epsffile{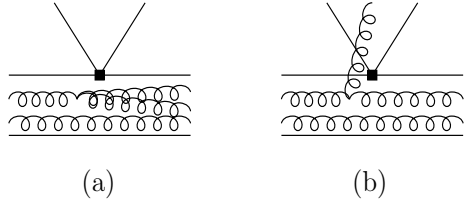}}
   \vspace*{-20.5cm}
\caption[dummy]{\label{fig11}\small (a) Soft overlap contribution 
which is part of the $B\to M_1$ form factor. (b) Soft overlap with 
$M_2$ which would violate factorization, if it were unsuppressed. }
\end{figure}

A more precarious situation may arise when the additional Fock 
components carry only a small fraction of the meson's momentum, 
contrary to the assumption made above. It is usually argued 
\cite{LB80,EfRa80} 
that these configurations are suppressed, because they occupy only 
a small fraction of the available phase space (since $\int d u_i \sim 
\Lambda_{\rm QCD}/m_b$ when the parton that carries momentum 
fraction $u_i$ is soft). 
This argument does not apply when the process involves heavy mesons. 
Consider for example a diagram such as the one in Fig.~\ref{fig11}a 
for $B\to D\pi$. This denotes the overlap of the $B$-meson 
wave function involving additional soft gluons with the wave function 
of the $D$ meson, also containing soft gluons. There is no reason 
to suppose that this overlap is suppressed relative to the soft overlap 
of the valence-quark wave functions. It represents (part of) the 
overlap of the ``soft cloud'' around the $b$ quark with (part of) the ``soft 
cloud'' around the $c$ quark after the weak decay of the $b$ quark. 
The partonic decomposition of this cloud is unrestricted up to 
global quantum numbers. In the case where the $B$ meson decays into two 
light mesons, there is a form-factor suppression 
$\sim(\Lambda_{\rm QCD}/m_b)^{3/2}$ for the overlap of 
the valence-quark wave functions (see Sect.~\ref{formfactor}), but 
once this price is paid there is again no reason for further suppression 
of additional soft gluons in the overlap of the $B$-meson wave function 
and the wave function of the recoiling meson $M_1$.

\begin{figure}[t]
   \vspace{-4.1cm}
   \epsfysize=27cm
   \epsfxsize=18cm
   \centerline{\epsffile{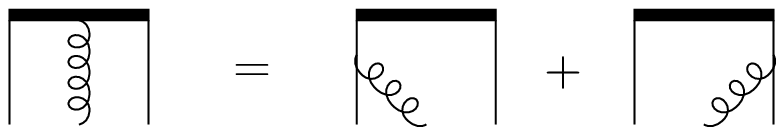}}
   \vspace*{-21.1cm}
\caption[dummy]{\label{fig12}\small Quark-antiquark-gluon distribution 
amplitude in the gluon endpoint region.}
\end{figure}

The previous paragraph essentially repeated our earlier argument 
against the hard-scattering approach, and in favour of using the 
$B\to M_1$ form factor as an input for the factorization formula. 
However, given the presence of additional soft partons in the 
$B\to M_1$ transition, we must now argue that it is unlikely that 
the emitted meson $M_2$ drags with it one of these soft partons, 
for instance a soft gluon that goes into the wave function of $M_2$, 
as shown in Fig.~\ref{fig11}b. Notice that if the $q\bar{q}$ pair is 
produced in a colour-octet state at the weak interaction vertex, at least 
one gluon (or further $q\bar{q}$ pair) must be pulled into the 
emitted meson, if the decay is to result in a two-body final state. 
What suppresses the process shown in Fig.~\ref{fig11}b relative to the one 
shown in Fig.~\ref{fig11}a even if the emitted $q\bar{q}$ pair is 
in a colour-octet state? The dominant configuration has both quarks 
carry a large fraction of the momentum of $M_2$, and only the gluon 
might be soft. In this situation we can apply a non-local 
``operator product expansion'' to determine the coupling of the 
soft gluon to the small $q\bar{q}$ pair. The gluon endpoint 
behaviour of the $q\bar{q}g$ wave function shown in Fig.~\ref{fig12} 
is then determined by the sum of the two diagrams on the 
right-hand side of this figure. The leading term (for small gluon 
momentum) cancels in the sum of the two diagrams, because the meson 
(represented by the black bar) is a colour singlet. This cancellation, 
which is exactly the same cancellation needed to demonstrate that 
``non-factorizable'' vertex corrections (Fig.~\ref{fig6}) are dominated 
by hard gluons, provides one factor of $\Lambda_{\rm QCD}/m_b$ needed 
to show that Fig.~\ref{fig11}b is power suppressed relative to 
Fig.~\ref{fig11}a. An explicit calculation of this soft, 
non-factorizable contribution is presented in Sect.~\ref{snf} for 
the decay $B\to D\pi$, which confirms that it is 
power suppressed in the heavy-quark limit. 
We have thus covered (qualitatively) all possibilities for 
non-valence contributions to the decay amplitude and find that they 
are all suppressed in the heavy-quark limit.

\subsection{Limitations of the factorization approach}
\label{limitations}

Above we argued that the factorization formula (\ref{fff}) holds 
in the heavy-quark limit $m_b\to\infty$. Since $m_b$ is fixed to about 
5 GeV in the real world one may question the accuracy of the heavy-quark 
limit. Indeed, we have seen that corrections to the asymptotic 
limit are of order $\Lambda_{\rm QCD}/m_b$ and, generally speaking, 
do not assume a factorized form. In this subsection we discuss 
several reasons why some power corrections could turn out to be 
numerically larger than suggested by the parametric 
suppression factor $\Lambda_{\rm QCD}/m_b$. 

\subsubsection{Several small parameters}

Large non-factorizable power corrections may arise if the leading-power, 
factorizable term is somehow suppressed. There are several possibilities 
for such a suppression, given a variety of small parameters that may enter 
into the non-leptonic decay amplitudes.

I) The hard, ``non-factorizable'' effects computed using the factorization 
formula occur at order $\alpha_s$. Some interesting effects such as 
final-state interactions appear first at this order. However, for realistic 
$B$ mesons $\alpha_s$ is not large compared to $\Lambda_{\rm QCD}/m_b$. 
Strong-interaction phases are a particularly important example. Since the 
phases due to hard interactions are of order $\alpha_s$ and soft 
phases are of order $\Lambda_{\rm QCD}/m_b$, one should not expect 
that these phases can be calculated with great precision. However, it 
is probably more important to know that the imaginary part is small 
compared to the real part, which is of order $\alpha_s^0$, and hence 
that strong phases should be small. This does not apply if the real part is 
suppressed for other reasons (see below).

II) Colour suppression. Either the leading-order contribution or 
the order-$\alpha_s$ correction to naive factorization may be 
colour suppressed. In the second case, which occurs for 
the class-I decays $\bar{B}_d\to D^+ \pi^-$ and $\bar{B}_d\to\pi^+\pi^-$, 
the first-order correction to naive factorization is small. In this 
case, the two-loop correction may be as important as the first-order 
correction computed later in this paper and in \cite{BBNS99}, 
but both are small. If, on the other hand, the lowest-order term 
is colour suppressed, as occurs for the class-II decays 
$\bar{B}_d\to J/\psi K$ and $\bar{B}_d\to\pi^0\pi^0$, perturbative and 
power corrections can be sizeable. Then even the hard strong-interaction 
phase can be large \cite{BBNS99}. But at the same time soft contributions 
could be potentially important, so that in some cases only an order-of-magnitude 
estimate of the decay rate may be possible.

III) Small Wilson coefficients. The effective Hamiltonian (\ref{effham}) 
contains small coefficients $C_i$ relative to 
$C_1\approx 1$, in particular the coefficients of the penguin 
operators ($i=3,\ldots$). 
The distinction of class-I and class-II decays mentioned 
in II) is in part also a manifestation of this effect. In addition 
there are decays for which the entire leading-power contribution 
is suppressed by small Wilson coefficients, but some power-suppressed 
effects are not. An example of this type is $B^-\to K^- K^0$. 
The decay proceeds through a penguin operator $b\to d s\bar{s}$ 
at leading power. But the annihilation contribution which is power 
suppressed can occur through the current-current operator with large 
Wilson coefficient $C_1$. Our approach does not apply to such 
(presumably) annihilation-dominated decays, unless a systematic treatment 
of annihilation amplitudes can be found.

IV) Small CKM elements. Some amplitudes may be suppressed by a combination 
of small CKM elements. For example, $B\to \pi K$ decays receive large 
penguin contributions despite their small Wilson coefficients, because 
the so-called tree amplitude is CKM suppressed. This is not a problem 
for factorization, since it applies to the penguin and the tree 
amplitudes. 
We are not aware of any case (for ordinary $B$ mesons) in which a 
purely power-suppressed term is CKM enhanced and which
would therefore dominate the decay, as in the example of III) above.
This situation could occur for $B^-_c\to \bar D^0 K^-$,
where the QCD dynamics is similar, if we consider the
charm as a light quark.

\subsubsection{Power corrections enhanced by small quark masses}
\label{chiral}

There is another enhancement of power-suppressed effects for some 
decays into two light mesons connected 
with the curious numerical fact that 
\begin{equation}
\label{mupi}
2\mu_\pi\equiv 
\frac{2 m_\pi^2}{m_u+m_d} = -\frac{4 \langle\bar{q}q\rangle}{f_\pi^2} 
\approx 3\,\mbox{GeV}
\end{equation}
is much larger than its naive scaling estimate $\Lambda_{\rm QCD}$. 
(Here $\langle \bar{q}q\rangle = \langle 0|\bar{u} u|0\rangle= 
\langle 0|\bar{d} d|0\rangle$ is the 
quark condensate.) Consider 
the contribution of the penguin 
operator ${\cal O}_6=(\bar{d}_i b_j)_{V-A} (\bar{u}_j u_i)_{V+A}$ to the 
$\bar{B}_d\to\pi^+\pi^-$ decay amplitude. The leading-order graph 
of Fig.~\ref{fig4} results in the expression
\begin{equation}
\label{lotw3}
\langle \pi^+\pi^-|(\bar{d}_i b_j)_{V-A} (\bar{u}_j u_i)_{V+A}|
\bar{B}_d\rangle =i m_B^2 F_+^{B\to \pi}(0) f_\pi \times \frac{2\mu_\pi}{m_b},
\end{equation}
which is formally a $\Lambda_{\rm QCD}/m_b$ power correction compared 
to (\ref{abpi}) but numerically large due to (\ref{mupi}).
We would not have to worry about such terms if they could all be 
identified and the factorization formula (\ref{fff}) applied 
to them, since in this case higher-order perturbative 
corrections would not contain non-factorizing
infrared logarithms. However, this is 
not the case.

After including radiative corrections, the matrix element on the 
left-hand side of (\ref{lotw3}) is expressed as a non-trivial 
convolution with the pion light-cone 
distribution amplitude. The terms involving $\mu_\pi$ can be 
related to two-particle twist-3 (rather than leading twist-2) 
distribution amplitudes, conventionally called $\Phi_p(u)$ and 
$\Phi_\sigma(u)$. The distribution amplitude $\Phi_p(u)$ does not 
vanish at the endpoint. As a consequence the hard spectator interaction 
(Fig.~\ref{fig8}) contains an endpoint divergence. In other words, 
the ``correction'' relative to (\ref{lotw3}) is of the form 
$\alpha_s\times\,$logarithmic divergence, which we interpret as 
being of the same order as (\ref{lotw3}). The non-factorizing 
character of the ``chirally-enhanced'' \cite{BBNS99} power 
corrections can introduce a substantial uncertainty in some decay 
modes. As in the related situation for the pion form factor 
\cite{GT82} one may argue that the endpoint divergence is suppressed 
by a Sudakov form factor. However, it is likely that when $m_b$ 
is not large enough to suppress these chirally-enhanced terms, then it is 
also not large enough to make Sudakov suppression effective. 
Given the importance of this issue, it deserves further investigation.

Notice that the chirally-enhanced terms do not appear in decays into 
a heavy and a light meson such as $B\to D\pi$, which we 
treat in detail later in this paper, because these decays have 
no penguin contribution and no contribution from the hard spectator 
interaction. Hence the twist-3 light-cone distribution amplitudes 
responsible for chirally-enhanced power corrections do not enter 
in the evaluation of the decay amplitude.

We conclude this subsection with a side remark: when (\ref{lotw3}) 
is applied to the $\pi^0\pi^0$ final state, naive application of 
the equations of motion to the factorized matrix element would result 
in $m_\pi^2/m_u$ rather than $2m_\pi^2/(m_u+m_d)$. This statement 
can sometimes be found in the literature but it is incorrect. The 
distinction of $2 m_u$ and $m_u+m_d$ is an isospin-breaking effect. 
In the presence of isospin breaking the $\pi^0$ has a small iso-singlet 
component, which leads to a non-vanishing vacuum-to-$\pi^0$ 
matrix element of the anomaly term in the divergence of the 
singlet axial-vector current. When this term is taken into account in the 
equation of motion one obtains $m_\pi^2/(m_u+m_d)$ also for 
the factorized matrix element in $\bar{B}_d\to \pi^0\pi^0$ decay. 
Note that as $(m_u+m_d)/(2 m_u)\approx 1.5$, keeping track of the 
light quark masses is important to correctly 
estimate the factorized amplitude.

\subsubsection{Difficulties with charm}
\label{subsec:diffcharm}

For the purposes of power counting we treated the charm quark as 
heavy, taking the heavy-quark limit for fixed $m_c/m_b$. This simplified 
the discussion, since we did not have to introduce $m_c$ as a separate 
scale. However, in reality charm is somewhat intermediate between 
a heavy and a light quark, since $m_c$ is not particularly large 
compared to $\Lambda_{\rm QCD}$.

It is worth noting that the first hard-scattering kernel in 
(\ref{fff}) cannot have $1/m_c$ corrections, since there is  a smooth 
transition to the case of two light mesons. The situation is 
different with the hard spectator interaction term, which we argued 
to be power suppressed for decays into a $D$ meson and a light meson. 
We shall come back to this in Sect.~\ref{bdpi}, where we estimate 
the magnitude of this term for the $D\pi$ final state relaxing 
the assumption that the $D$ meson is heavy.

The power-counting estimates based on $m_c\sim m_b\to \infty$ are 
particularly suspicious in case of the suppression of the class-II 
amplitude in $B^-\to D^0\pi^-$ in (\ref{onetwo}). Since the 
class-I amplitude dominates, we expect
\begin{equation}
R \equiv 
\frac{\mbox{Br}(B^-\to D^0\pi^-)}{\mbox{Br}(\bar{B}_d\to D^+\pi^-)}
=1
\end{equation}
in the heavy-quark limit.
This contradicts existing data which yield $R=1.89\pm 0.35$, 
despite the additional colour suppression of the class-II amplitude. 
One reason for the failure of power counting lies in the departure of 
the decay constants and form factors from naive power counting. The 
following compares the power counting to the actual numbers 
(square brackets):
\begin{equation}\label{scalingviolations}
\frac{f_D}{f_\pi}\sim \left(\frac{\Lambda_{\rm QCD}}{m_c}\right)^{1/2} 
\,\,[\approx 1.5]
\qquad
\frac{F_+^{B\to \pi}(m_D^2)}{F_+^{B\to D}(0)}
\sim \left(\frac{\Lambda_{\rm QCD}}{m_b}\right)^{3/2} 
\,\,[\approx 0.5].
\end{equation}
However, it is unclear whether the failure of power counting can be 
attributed to the form factors and decay constants alone.


\boldmath
\section{$B\to D\pi$: Factorization to one-loop order}
\unboldmath
\label{oneloop}

In this section we begin a more detailed and quantitative treatment 
of exclusive $B$ decays into a heavy meson (a $D$ or $D^*$ meson) and 
a light meson, governed by a $b\to c\bar ud$ transition. 
Following the general discussion of Sect.~\ref{arguments}, we
shall illustrate explicitly how factorization emerges at the one-loop 
order in this specific case, and in the heavy-quark limit, defined as
$m_b$, $m_c\gg\Lambda_{\rm QCD}$ with $m_c/m_b$ fixed.
In particular, we will compute at order $\alpha_s$ the hard-scattering 
kernel $T^I(u)$ in the factorization formula (\ref{fff}) for the decays 
$\bar B_d\to D^{(*)+} L^-$, where $L$ is a light meson. 
For each final state $f$,
we will finally express the decay amplitudes in terms 
of parameters $a_{1}(f)$, defined in analogy with similar 
parameters used in the literature on naive factorization.
The numerical analysis of 
one-loop corrections to factorization and a comparison of 
our results to the existing branching ratio measurements
are postponed to Sect.~\ref{bdpi}.
For notational convenience we shall in this section mostly speak
about $B\to D\pi$ decays, but a similar treatment
applies also to transitions such as $B\to D^*\pi$, $D\rho$, or $D^*\rho$.

\subsection{Generalities}

The effective Hamiltonian relevant for $B\to D\pi$ can be written as
\begin{equation}\label{heff18}
{\cal H}_{\rm eff}=\frac{G_F}{\sqrt{2}}V^*_{ud}V_{cb}
\left( C_0 O_0+C_8 O_8\right) ,
\end{equation}
with the operators
\begin{eqnarray}\label{o18}
O_0 &=& \bar c\gamma^\mu(1-\gamma_5)b\, 
        \bar d\gamma_\mu(1-\gamma_5)u , \\
O_8 &=& \bar c\gamma^\mu(1-\gamma_5)T^A b\, 
        \bar d\gamma_\mu(1-\gamma_5)T^A u .
\end{eqnarray}
Here we have chosen to write the two independent operators
in the singlet-octet basis, which is most convenient for
our purposes, rather than in the more conventional bases
of $Q_1$, $Q_2$ or $Q_+$, $Q_-$ \cite{BBL}. 
The Wilson coefficients $C_0$, $C_8$ describe the exchange of
hard gluons in the weak transition with virtualities between
the high-energy matching scale $M_W$ and a renormalization scale
$\mu$ of order $m_b$ in the low-energy effective theory. 
These coefficients have been calculated at next-to-leading order
in renormalization-group improved perturbation theory
\cite{ACMP,BW} and are given by
\begin{equation}\label{c18}
C_0=\frac{N_c+1}{2N_c}C_++\frac{N_c-1}{2N_c}C_- ,\qquad
C_8=C_+-C_- ,
\end{equation}
where
\begin{equation}\label{cpm}
C_\pm(\mu)=\left(1+\frac{\alpha_s(\mu)}{4\pi}B_\pm\right)\,
\bar C_\pm(\mu) ,
\end{equation}
\begin{equation}\label{cpmb}
\bar C_\pm(\mu)=\left[\frac{\alpha_s(M_W)}{\alpha_s(\mu)}\right]^{d_\pm}
\left[1+\frac{\alpha_s(M_W)-\alpha_s(\mu)}{4\pi}(B_\pm-J_\pm)\right] .
\end{equation}
(The coefficients $C_0$, $C_8$ are related
to the ones of the standard basis by $C_0=C_1+C_2/3$ and 
$C_8=2C_2$.) 
We employ the next-to-leading order expression for the running 
coupling,
\begin{equation}\label{als}
\alpha_s(\mu)=\frac{4\pi}{\beta_0\ln(\mu^2/\Lambda_{\rm QCD}^2)}
 \left[1-\frac{\beta_1}{\beta^2_0}
  \frac{\ln\ln(\mu^2/\Lambda_{\rm QCD}^2)}{\ln(\mu^2/\Lambda_{\rm
      QCD}^2)} \right] ,
\end{equation}
\begin{equation}\label{b0b1}
\beta_0=\frac{11N_c-2f}{3} ,\qquad 
\beta_1=\frac{34}{3}N^2_c-\frac{10}{3}N_c f-2 C_F f ,\qquad
C_F=\frac{N^2_c-1}{2N_c} ,
\end{equation}
where $N_c$ is the number of colours, and $f$ the number of light 
flavours. $\Lambda_{\rm QCD}\equiv\Lambda^{(f)}_{\overline{\rm MS}}$ is 
the QCD scale in the $\overline{\rm MS}$ scheme with $f$ flavours.
Next we have
\begin{equation}\label{dgbpm}
d_\pm=\frac{\gamma^{(0)}_\pm}{2\beta_0} ,\qquad
\gamma^{(0)}_\pm=\pm 12\frac{N_c\mp 1}{2N_c} ,\qquad
B_\pm=\pm\frac{N_c\mp 1}{2N_c}B .
\end{equation}
The general definition of $J_\pm$ may be found in \cite{BBL}.
Numerically, for $N_c=3$ and $f=5$
\begin{equation}\label{dbjpm}
d_\pm=\left\{ 
 \begin{array}{c}
    \phantom{-}\frac{6}{23} , \\[0.2cm]
    -\frac{12}{23} ,
 \end{array} \right.
\qquad
B_\pm-J_\pm=\left\{ 
 \begin{array}{c}
   \phantom{-}\frac{6473}{3174} , \\[0.2cm]
   -\frac{9371}{1587} .
 \end{array} \right.
\end{equation}
The quantities $\beta_0$, $\beta_1$, $d_\pm$, $B_\pm-J_\pm$
are scheme independent. The scheme dependence of the coefficients
at next-to-leading order is parameterized by $B_\pm$ in (\ref{cpm}). In 
the naive dimensional regularization (NDR) and `t~Hooft-Veltman (HV)
schemes, this scheme dependence is expressed in a single number
$B$ with $B_{\rm NDR}=11$ and $B_{\rm HV}=7$.
The dependence of the Wilson coefficients on the renormalization
scheme and scale is cancelled by a corresponding scale and scheme dependence 
of the hadronic matrix elements of the operators $O_0$ and $O_8$.

\begin{figure}
 \vspace{4cm}
\includegraphics{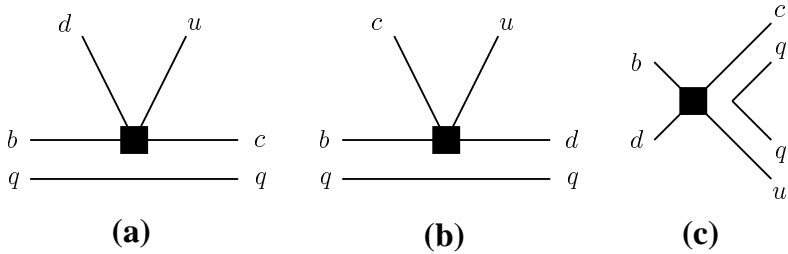}
 \caption{\small Basic quark-level topologies for $B\to D\pi$
   decays ($q=u$, $d$): (a) class-I, (b) class-II,  
   (c) weak annihilation. $\bar B_d\to D^+\pi^-$ receives 
   contributions from (a) and (c), $\bar B_d\to D^0\pi^0$ from
   (b) and (c), and $B^-\to D^0\pi^-$ from (a) and (b).
   Only (a) contributes in the heavy-quark limit.
    \label{fig:bdpi}}
\end{figure}

Before continuing with a discussion 
of these matrix elements, it is useful to consider the
flavour structure for the various contributions to $B\to D\pi$ decays.  
The possible quark-level topologies are depicted in 
Fig.~\ref{fig:bdpi}. 
In the terminology generally adopted for
two-body non-leptonic decays, the decays $\bar B_d\to D^+\pi^-$,
$\bar B_d\to D^0\pi^0$ and $B^-\to D^0\pi^-$ are referred to
as class-I, class-II and class-III, respectively (see e.g.\ \cite{NeSt97}).
In both $\bar B_d\to D^+\pi^-$ and $B^-\to D^0\pi^-$ decays
the pion can be directly created from the weak current.
We may call this a class-I contribution, following the above
terminology. In addition, in the case of $\bar B_d\to D^+\pi^-$
there is a contribution from weak annihilation and a class-II
amplitude contributes to $B^-\to D^0\pi^-$, see Fig.~\ref{fig:bdpi}. 
The important point is that the spectator quark goes into the 
light meson in the case of the class-II amplitude. According 
to Sect.~\ref{hl} this amplitude is therefore suppressed in the 
heavy-quark limit, as is the annihilation amplitude. 
It follows that the amplitude for $\bar B_d\to D^0\pi^0$, receiving
only class-II and annihilation contributions, is subleading compared
with $\bar B_d\to D^+\pi^-$ and $B^-\to D^0\pi^-$, which are
dominated by the class-I topology.
The treatment of this leading class-I mechanism will be the
main subject of the following sections. (With reference to the 
general discussion in Sect.~\ref{arguments}, it should be noted that 
in the case of decays into light-light final states class-I and 
class-II amplitudes are both of leading power in the heavy-quark 
expansion, and the factorization formula applies to both of them.)

We shall use the one-loop analysis for $\bar B_d\to D^+\pi^-$ as a
concrete example on which we will illustrate explicitly
the various steps involved in establishing the factorization
formula. We emphasize that most of the arguments used below are
standard from the theory of hard exclusive processes involving 
light hadrons \cite{LB80}. However, we find it instructive to repeat 
those arguments in the context of $B$ decays.

\subsection{Soft and collinear cancellations at one-loop order}
\label{oneloopcancel}

In order to demonstrate the property of factorization for
$\bar B_d\to D^+\pi^-$, we will now analyze the ``non-factorizable''
one-gluon exchange contributions (Fig.~\ref{fig6}) to the
$b\to c\bar ud$ transition in some detail. 
Recall from Sect.~\ref{arguments} that this is the only type 
of one-loop corrections that we need to consider for heavy-light 
final states. 

We consider the leading, valence Fock state
of the emitted pion. This is justified since higher Fock components
only give power-suppressed contributions to the decay
amplitude in the heavy-quark limit, as 
discussed in Sect.~\ref{otherfock} and demonstrated below.
The valence Fock state of the pion can be written as
\begin{equation}\label{piwf}
|\pi(q)\rangle =\int\frac{du}{\sqrt{u\bar u}}
\frac{d^2 l_\perp}{16\pi^3}\frac{1}{\sqrt{2N_c}}
\left(a^\dagger_\uparrow(l_q)b^\dagger_\downarrow(l_{\bar q})-
      a^\dagger_\downarrow(l_q)b^\dagger_\uparrow(l_{\bar q})\right)
|0\rangle\, \Psi(u,\vec l_\perp) ,
\end{equation}
where $a^\dagger_s$ ($b^\dagger_s$) denotes the creation operator
for a quark (antiquark) in a state with spin $s=\uparrow$ or
$s=\downarrow$, and we have suppressed colour indices. This 
representation of the pion state is adequate for a leading-power 
analysis. The wave function $\Psi(u,\vec l_\perp)$ is defined as the
amplitude for the pion to be composed of two on-shell quarks,
characterized by longitudinal momentum fraction $u$ and
transverse momentum $l_\perp$. The on-shell momenta
($l^2_{q,\bar q}=0$) of the quark ($l_q$) and the antiquark ($l_{\bar q}$)
are given by
\begin{equation}\label{q1u}
l_q=u q + l_\perp+\frac{\vec{l}^{\,2}_\perp}{4 u E}n_- ,
\qquad
l_{\bar q}=\bar u q - l_\perp+\frac{\vec{l}^{\,2}_\perp}{4\bar u E}n_- .
\end{equation}
Here $q=E(1,0,0,1)$ is the pion momentum, $E=p_B\cdot q/m_B$
the pion energy and $n_-=(1,0,0,-1)$. 
Furthermore $l_\perp\cdot q=l_\perp\cdot n_-=0$.
For the purpose of power counting 
$l_\perp\sim\Lambda_{\rm QCD}\ll E\sim m_b$.
Note that the invariant mass of the valence state is
$(l_q+l_{\bar q})^2=\vec{l}^{\,2}_\perp/(u\bar u)$, which is of order
$\Lambda^2_{\rm QCD}$ and hence negligible in the heavy-quark limit,
unless $u$ is in the vicinity of the endpoints ($0$ or $1$).
In this case the invariant mass of the quark-antiquark pair
becomes large and the valence Fock state is no longer
a valid representation of the pion. However, in the heavy-quark
limit the dominant contributions to the decay amplitude come from
configurations where both partons are hard
($u$ and $\bar u$ both of order 1) and the two-particle Fock state
yields a consistent description. The suppression of the
soft regions ($u$ or $\bar u\ll 1$) is related to the endpoint
behaviour of the pion wave function, as discussed in previous
sections. We will provide an explicit consistency
check of this important feature later on.

As a next step we write down the amplitude
\begin{equation}\label{piudb}
\langle\pi(q)|u(0)_\alpha\bar d(y)_\beta|0\rangle =
\int du\frac{d^2l_\perp}{16\pi^3}\frac{1}{\sqrt{2 N_c}}
\Psi^*(u,\vec l_\perp)(\gamma_5\!\not\! q)_{\alpha\beta}\,
 e^{i l_q\cdot y} ,
\end{equation}
which appears as an ingredient of the $B\to D\pi$ matrix element.
The right-hand side of (\ref{piudb}) follows directly from (\ref{piwf}).
Using (\ref{piudb}) it is straightforward to write down the
one-gluon exchange contribution to the $B\to D\pi$ matrix
element of the operator $O_8$ (Fig.~\ref{fig6}). We have
\begin{eqnarray}\label{o8a1a2}
\langle D^+\pi^-|O_8|\bar B_d\rangle_{\rm 1-gluon} &=& 
\\
&&\hspace*{-4cm}
i g_s^2\frac{C_F}{2}\int\frac{d^4k}{(2\pi)^4}
\langle D^+|\bar c A_1(k) b|\bar B_d\rangle\frac{1}{k^2}
\int^1_0 du\frac{d^2l_\perp}{16\pi^3}
\frac{\Psi^*(u,\vec l_\perp)}{\sqrt{2 N_c}}\,
{\rm tr}[\gamma_5\!\not\! q A_2(l_q,l_{\bar q},k)] , \nonumber
\end{eqnarray}
where
\begin{equation}\label{a1bc}
A_1(k)=
\frac{\gamma^\lambda(\not\! p_c-\not\! k+m_c)\Gamma}{2p_c\cdot k-k^2}-
\frac{\Gamma(\not\! p_b+\not\! k+m_b)\gamma^\lambda}{2p_b\cdot k+k^2} ,
\end{equation}
\begin{equation}\label{a2ud}
A_2(l_q,l_{\bar{q}},k)=
\frac{\Gamma(\not\! l_{\bar q}+\not\! k)\gamma_\lambda}{2l_{\bar q}\cdot k+
      k^2}-
\frac{\gamma_\lambda(\not\! l_q+\not\! k)\Gamma}{2l_q\cdot k+k^2} .
\end{equation}
Here $\Gamma=\gamma^\mu(1-\gamma_5)$ and $p_b$, $p_c$ are the
momenta of the $b$ quark and the $c$ quark, respectively.
Note that this expression holds in an arbitrary covariant gauge. 
The gauge-parameter dependent part of the gluon propagator gives no
contribution to (\ref{o8a1a2}), as can be easily seen
from (\ref{a1bc}) and (\ref{a2ud}). There is no correction 
to the matrix element of $O_0$ at order $\alpha_s$, because in this 
case the $(d\bar{u})$ pair is necessarily in a colour-octet 
configuration and cannot form a pion.

In (\ref{o8a1a2}) the pion wave function $\Psi(u,l_\perp)$ 
appears separated
from the $B\to D$ transition. This is merely
a reflection of the fact that we have represented the pion state 
by (\ref{piwf}). It does not, by itself,
imply factorization, since the right-hand side 
of (\ref{o8a1a2}) involves
still nontrivial integrations over $\vec l_\perp$ and gluon
momentum $k$, and long- and short-distance contributions are not
yet disentangled. In order for (\ref{o8a1a2}) to make sense 
we need to show that the integral over $k$ receives only subdominant 
contributions from the region of small $k^2$. This is equivalent 
to showing that the integral over $k$ does not contain 
infrared divergences at leading power in $1/m_b$. 

To demonstrate infrared finiteness of the one-loop integral
\begin{equation}
 \label{dka1a2}
J\equiv
\int d^4k\, \frac{1}{k^2}\, A_1(k)\otimes A_2(l_q,l_{\bar{q}},k)
\end{equation}
at leading power, the heavy-quark limit and the
corresponding large light-cone momentum of the pion
are again essential. First note that when $k$ is of order $m_b$, 
$J\sim 1$ for dimensional reasons. Potential infrared divergences 
could arise when $k$ is soft or when $k$ is collinear to the 
pion momentum $q$. We need to show that the contributions from 
these regions are power suppressed in $m_b$. (Note that we do not 
need to show that $J$ is infrared finite. It is enough that 
logarithmic divergences have coefficients that are 
power suppressed.)

We treat the soft region first. Here all components of $k$ become
small simultaneously, which we describe by scaling $k\sim\lambda$.
Counting powers of $\lambda$
($d^4k\sim \lambda^4$, $1/k^2\sim\lambda^{-2}$,
$1/p\cdot k\sim\lambda^{-1}$) reveals that each of the four diagrams
(corresponding to the four terms in the product in
(\ref{dka1a2})) is logarithmically divergent. However, because 
$k$ is small, the integrand can be simplified. For instance, 
the second term in $A_2$ can be approximated as
\begin{equation}
\label{a2lperp}
\frac{\gamma_\lambda(\not\! l_q+\not\! k)\Gamma}{2 l_q\cdot k+k^2}=
\frac{\gamma_\lambda(\!u\not\! q+\not\! l_\perp+
  \frac{\vec l^2_\perp}{4uE}\not\! n_-+\not\! k)\Gamma}{2u q\cdot k+
   2l_\perp\cdot k+\frac{\vec l^2_\perp}{2uE} n_-\cdot k+k^2}
   \simeq \frac{q_\lambda}{q\cdot k}\,\Gamma ,
\end{equation}
where we used that $\not\!q$ to the extreme left or right of an 
expression gives zero due to the on-shell condition for the external 
quark lines. We get exactly the same expression but with an opposite 
sign from the other term in $A_2$ and hence the soft 
divergence cancels out when adding the two terms in $A_2$. More 
precisely, we find that the integral is infrared finite in the 
soft region when $l_\perp$ is neglected. When $l_\perp$ is not 
neglected, there is a divergence from soft $k$ which is proportional 
to $l^2_\perp/m_b^2\sim \Lambda^2_{\rm QCD}/m_b^2$. In either case, the 
soft contribution to $J$ is of order $\Lambda_{\rm QCD}/m_b$ 
or smaller and 
hence suppressed relative to the hard contribution.
This corresponds to the standard soft cancellation mechanism,
which is a technical manifestation of colour transparency.

Each of the four terms in (\ref{dka1a2}) is also divergent 
when $k$ becomes collinear with the light-cone
momentum $q$. This implies the scaling 
\begin{equation}\label{kcoll}
k^+\sim \lambda^0,\quad k_\perp\sim\lambda,\quad k^-\sim\lambda^2 .
\end{equation}
Then $d^4k\sim dk^+ dk^- d^2k_\perp\sim\lambda^4$ and
$q\cdot k=q^+ k^-\sim\lambda^2$, $k^2=2k^+ k^-+k^2_\perp\sim\lambda^2$.
The divergence is again logarithmic and it is thus sufficient
to consider the leading behaviour in the collinear limit. 
Writing $k=\alpha q+\ldots$ we can now simplify the second term 
of $A_2$ as
\begin{equation}
\label{a2lperp2}
\frac{\gamma_\lambda(\not\! l_q+\not\! k)\Gamma}{2 l_q\cdot k+k^2}
\simeq 
q_\lambda\,\frac{2 (u+\alpha)\Gamma}{2 l_q\cdot k+k^2} .
\end{equation}
No simplification occurs in the denominator (in particular $l_\perp$ 
cannot be neglected), but the important point is that the leading-power 
contribution is proportional to $q_\lambda$. Therefore, 
substituting $k=\alpha q$ into $A_1$ and using $q^2=0$, we obtain 
\begin{equation}\label{a1coll}
q_\lambda A_1\simeq
\frac{\not\! q(\not\! p_c+m_c)\Gamma}{2\alpha p_c\cdot q}-
\frac{\Gamma(\not\! p_b+m_b)\!\not\! q}{2\alpha p_b\cdot q}=0 ,
\end{equation}
employing the equations of motion for the heavy quarks.
Hence the collinearly 
divergent region is seen to cancel out via the standard
collinear Ward identity.
This completes the proof of the absence of infrared divergences 
at leading power in the hard-scattering kernel for 
$\bar B_d\to D^+\pi^-$ to one-loop order. In other words, we have shown 
that the ``non-factorizable'' diagrams of Fig.~\ref{fig6} are 
dominated by hard gluon exchange. 

Since we have now established that the leading contribution to 
$J$ arises from $k$ of order $m_b$ (``hard'' $k$), and 
since $|\vec l_\perp|\ll E$, we may expand $A_2$
in $|\vec l_\perp|/E$. To leading power the expansion simply 
reduces to neglecting $l_\perp$ altogether, which implies
$l_q=uq$ and $l_{\bar q}=\bar uq$ in (\ref{a2ud}). 
As a consequence,
we may perform the $l_\perp$ integration in (\ref{o8a1a2})
over the pion wave function. Defining
\begin{equation}\label{psiphi}
\int\frac{d^2l_\perp}{16\pi^3}
\frac{\Psi^*(u,\vec l_\perp)}{\sqrt{2N_c}}\equiv
\frac{i f_\pi}{4 N_c}\Phi_\pi(u) , 
\end{equation}
the matrix element of $O_8$ in (\ref{o8a1a2}) becomes
\begin{eqnarray}\label{o8phi}
\langle D^+\pi^-|O_8|\bar B_d\rangle_{\rm 1-gluon} &=& \\
&&\hspace*{-4cm}-g_s^2\frac{C_F}{8N_c}\int\frac{d^4k}{(2\pi)^4}
\langle D^+|\bar c A_1(k) b|\bar B_d\rangle\frac{1}{k^2} f_\pi
\int^1_0 du\, \Phi_\pi(u)\,
{\rm tr}[\gamma_5\!\not\! q A_2(uq,\bar uq,k)] . \nonumber
\end{eqnarray}
Putting $y$ on the light-cone in (\ref{piudb}),
$y^+=y_\perp=0$, hence $l_q\cdot y=l^+_q y^-=uq y$, and comparing
with (\ref{distamps}), we see that the $l_\perp$-integrated wave
function $\Phi_\pi(u)$ in (\ref{psiphi}) is precisely the
light-cone distribution amplitude of the pion in the usual
definition (\ref{distamps}). This demonstrates the relevance of the 
light-cone wave function for the factorization formula. Note 
that the collinear approximation for the quark-antiquark momenta 
emerges automatically in the heavy-quark limit. 

After the $k$ integral is performed, the expression (\ref{o8phi}) 
can be cast into the form 
\begin{equation}\label{o8fth}
\langle D^+\pi^-|O_8|\bar B_d\rangle_{\rm 1-gluon} \sim  
F_{B\to D}(0)\, \int^1_0 du\, T_8(u,z) \Phi_\pi(u) , 
\end{equation}
where $z=m_c/m_b$, $T_8(u,z)$ is the hard-scattering kernel, and 
$F_{B\to D}(0)$ the form factor that parameterizes the 
$\langle D^+|\bar c [\ldots] b|\bar B_d\rangle$ matrix element. 
The result for $T_8(u,z)$ is given in Sect.~\ref{menlo} below.

\subsection{Higher Fock states and soft non-factorizable
contributions}
\label{snf}

The discussion of the previous subsection relied on the dominance
of the valence Fock state of the high-energy pion emitted
in $B\to D\pi$.
In the following section we will argue that higher Fock states
yield only subleading contributions in the heavy-quark limit.

\subsubsection{Additional hard-collinear partons}

\begin{figure}
 \vspace{4.3cm}
\includegraphics{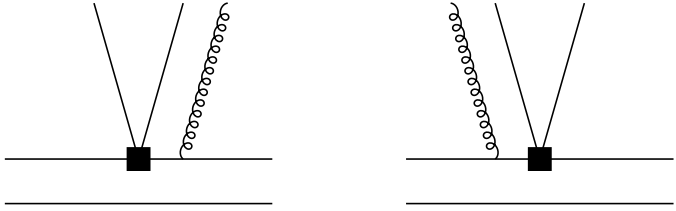}
 \caption{\small The contribution of the $q\bar{q}g$ Fock state to the
  $\bar B_d\to D^+\pi^-$ amplitude.
    \label{fig:qqg}}
\end{figure}

Generally, if additional collinear partons beyond the valence
quarks are present in the pion state, the $B\to D\pi$ amplitude
will contain additional hard propagators that lead to a power
suppression in $\Lambda_{\rm QCD}/m_b$.
We illustrate this property by considering the simplest
nontrivial example, where the pion is composed of three partons,
the quark, the antiquark, and an additional gluon.
The contribution of this 3-particle Fock state to the $B\to D\pi$
decay amplitude is shown in Fig.~\ref{fig:qqg}.
To evaluate this contribution it is convenient to use the
Fock-Schwinger gauge, which allows one to express the gluon field
$A_\lambda$ in terms of the field-strength tensor $G_{\rho\lambda}$ via
\begin{equation}\label{axg}
A_\lambda(x)=\int^1_0 dv\, v x^\rho G_{\rho\lambda}(vx) .
\end{equation}
Up to twist 4 there are three quark-antiquark-gluon matrix elements 
that could potentially contribute to the diagrams shown in 
Fig.~\ref{fig:qqg}. Due to the $V-A$ structure of the weak interaction 
vertex, the only relevant three-particle light-cone wave function has 
twist 4 and is given by \cite{Khod98a,BF90}
\begin{eqnarray}\label{pidgu}
\langle\pi(q)|\bar d(0)\gamma_\mu\gamma_5\,
g_s G_{\alpha\beta}(vx)u(0)|0\rangle &=& 
f_\pi(q_\beta g_{\alpha\mu}-q_\alpha g_{\beta\mu})
   \int{\cal D}u\, \phi_\perp(u_i)e^{ivu_3 q\cdot x} 
\nonumber\\
&&\hspace*{-5cm} +\,
f_\pi\frac{q_\mu}{q\cdot x}(q_\alpha x_\beta-q_\beta x_\alpha)
\int{\cal D}u\, \left(\phi_\perp(u_i)+\phi_\parallel(u_i)\right)
e^{ivu_3 q\cdot x} .
\end{eqnarray}
Here 
$\int{\cal D}u\equiv\int^1_0 du_1\, du_2\, du_3\, \delta(1-u_1-u_2-u_3)$,
with $u_1$, $u_2$ and $u_3$ the fractions of the pion momentum carried by 
the quark, the antiquark and the gluon, respectively. $\phi_\perp$ and 
$\phi_\parallel$ are twist-4, 3-particle light-cone distribution amplitudes. 
Evaluating the diagrams in Fig.~\ref{fig:qqg}, and
neglecting the charm-quark mass for simplicity, we find
\begin{equation}\label{o8qqg}
\langle D^+\pi^-|O_8|\bar B_d\rangle_{q\bar q g}=
if_\pi\langle D^+|\bar c\!\not\! q(1-\gamma_5)b|\bar B_d\rangle
\int{\cal D}u\frac{2\phi_\parallel(u_i)}{u_3\, m^2_b} .
\end{equation}
Since $\phi_\parallel\sim\Lambda^2_{\rm QCD}$, the suppression by two
powers of $\Lambda_{\rm QCD}/m_b$ compared to the leading-order matrix 
element is obvious. We remark that due to G-parity $\phi_\parallel$ is 
antisymmetric in $u_1\leftrightarrow u_2$ for a pion, so that
(\ref{o8qqg}) actually vanishes in this case. (It would be
non-zero if the pion were replaced by a $K$ meson.)

There are higher-twist corrections also
in the two-particle Fock state of the pion itself. They could also
contribute to power corrections. The leading ones could come
from the twist-3, two-particle pion wave functions, which can be
important numerically due to a chiral enhancement factor
$m^2_\pi/(m_u+m_d)$ (despite the suppression by a power of 
$\Lambda_{\rm QCD}/m_b$ as $m_b\to\infty$).
However, their contribution vanishes identically for $B\to D\pi$.
This comes about because the twist-3 wave function is proportional to
an even number of $\gamma$-matrices 
($\gamma_5$, $\sigma_{\mu\nu}\gamma_5$)
and therefore the projection of the light-quark ``current'' 
$A_2$ in (\ref{a2ud}) onto this wave function is zero.

\subsubsection{Additional soft partons}

Finally, we consider the case where the non-factorizable gluon, 
i.e.\ the gluon exchanged between the pion and the $(\bar{B}D)$ 
system, is soft. In this case, the ``$q\bar qg$ Fock state'' cannot be
described by a light-cone wave function as in (\ref{pidgu}),
which requires the partons to be energetic. As we shall see now, such a
contribution still receives a power suppression in the heavy-quark
limit, which arises from the soft-cancellation mechanism discussed 
already in Sect.~\ref{oneloopcancel}. Here 
we will derive an explicit expression for the soft non-factorizable
gluon correction. Note that the soft gluon
can interact with the spectator degrees of freedom in the
$B\to D$ transition; this was not possible for the mechanism of
Fig.~\ref{fig:qqg}, which requires a collinear, energetic gluon,
whereas the spectator cloud is always soft.

We start from (\ref{a2ud}), take $l_q=uq$ and $l_{\bar q}=\bar uq$,
and put the gluon on-shell (since now we are interested in an
external gluon field). The resulting expression describes the interaction 
of a soft gluon with the collinear light-quark pair, since both quarks 
are energetic. Re-introducing colour, the coupling constant $g_s$ and 
the gluon polarization vector 
$\varepsilon^\lambda$, the expression (\ref{a2ud}) projected
onto the pion state becomes
\begin{equation}\label{udpi}
\bar d A_2 u \,\to\,
-\frac{q^\kappa k^\alpha\varepsilon^\lambda
\epsilon_{\kappa\alpha\mu\lambda}}{2 q\cdot k}
\frac{f_\pi\Phi_\pi(u)}{u\bar u} \frac{g_s \mbox{Tr}(T^A T)}{N_c} ,
\end{equation}
where $T$ denotes the colour matrix at the weak vertex ($T=1$ for 
$O_0$, $T=T^B$ for $O_8$).
We also used the $\epsilon$-tensor with $\epsilon^{0123}=-1$ and
\begin{equation}\label{pipro}
\bar d\gamma_\mu(1-\gamma_5)u\,\to\, 
i f_\pi\Phi_\pi(u) q_\mu
\end{equation}
for projecting the current on the pion wave function, see (\ref{curproj}). 
To simplify the result, we have used the symmetry of $\Phi_\pi(u)$ under 
$u\leftrightarrow\bar u$. The dependence on the gluon momentum $k$ in 
(\ref{udpi}) involves the eikonal propagator $i/q\cdot k$, which has
the Fourier-decomposition
\begin{equation}\label{epft}
\frac{i}{q\cdot k+i\epsilon}=\int d^4x\,e^{ik\cdot x}
\int^\infty_0 d\tau\, \delta^{(4)}(x-\tau q) .
\end{equation}
Hence we see that in configuration space the right-hand side 
of (\ref{udpi}) corresponds to the operator expression 
(at space-time point $x=0$)
\begin{equation}\label{gtil}
-\frac{f_\pi\Phi_\pi(u)}{4N_c u\bar u}\int^\infty_0 ds\,g_s\,
\mbox{Tr}[\tilde G_{\mu\nu}(-s n)\,T]\,n^\nu ,
\end{equation}
where we defined $n$ as a dimensionless null-vector
describing the pion flight direction, i.e.\ $q=E n$, and  
\begin{equation}\label{pgg}
\tilde G_{\mu\nu}=\epsilon_{\mu\nu\alpha\beta}\,G^{\alpha\beta},
\qquad G_{\mu\nu}=G^A_{\mu\nu}\,T^A .
\end{equation}
Note that the expression (\ref{gtil}) corresponds to the 
right-hand side of Fig.~\ref{fig12}. 
With this result we can write down the soft non-factorizable (SNF)
contribution to the matrix elements of $O_{0,8}$ from one-gluon 
exchange as
\begin{eqnarray}
\label{o1snf}
\langle D^+\pi^-|O_0|\bar B_d\rangle_{\rm SNF}&=&0 ,
\\
\label{o8snf}
\langle D^+\pi^-|O_8|\bar B_d\rangle_{\rm SNF}&=& 
\nonumber \\
&&\hspace*{-3cm}
-\int^\infty_0 ds\,\langle D^+|\bar c\gamma^\mu(1-\gamma_5)
g_s\tilde G_{\mu\nu}(-s n) n^\nu b|\bar B_d\rangle\,
\int^1_0 du\,\frac{f_\pi\Phi_\pi(u)}{8N_c u\bar u} .
\end{eqnarray}
Because of the colour structure of the one-gluon contribution,
only the matrix element of $O_8$ is non-vanishing.
The result in (\ref{o8snf}) can be compared to the leading
contribution to the matrix element of $O_0$, 
\begin{equation}\label{o1lf}
\langle D^+\pi^-|O_0|\bar B_d\rangle_{\rm lead}=
\langle D^+|\bar c\gamma^\mu(1-\gamma_5)b|\bar B_d\rangle\,
i f_\pi q_\mu \int^1_0 du\,\Phi_\pi(u) .
\end{equation}
In the heavy-quark limit ($m_b, m_c\to\infty$) the
dependence of the matrix elements of the $(\bar cb)$ currents
in (\ref{o8snf}), (\ref{o1lf}) on the heavy-quark masses can be
extracted using heavy-quark effective theory. Up to logarithms,
this dependence arises only from trivial factors related to the 
normalization of the $B$- and $D$-meson states, i.e.\
\begin{equation}
\langle D^+|(\bar cb)_{V-A}|\bar B_d\rangle \sim \sqrt{m_c m_b} . 
\end{equation}
From dimensional counting one then finds for the matrix elements in
(\ref{o8snf}) and (\ref{o1lf})
\begin{equation}\label{o18fml}
\langle O_0\rangle_{\rm lead}\sim f_\pi m^2_b ,\qquad
\langle O_8\rangle_{\rm SNF}\sim f_\pi m_b\Lambda_{\rm QCD} . 
\end{equation}
It follows that the soft non-factorizable interactions of the pion
with the spectator, and soft partons in the $B\to D$ transition in 
general, are suppressed by one power of 
$\Lambda_{\rm QCD}/m_b$. This result 
is general as argued in Sect.~\ref{otherfock} above. 
Note that in the present case the contribution of
$\langle O_8\rangle_{\rm SNF}$ to the decay amplitude is further
suppressed as it occurs only at relative order $1/N^2_c$ in
colour counting.

\subsection{Matrix elements at next-to-leading order}
\label{menlo}

As we have seen above, the $\bar B_d\to D^+\pi^-$ amplitude
factorizes in the heavy-quark limit into a matrix element of the 
form $\langle D^+|\bar{c}[\ldots]b|\bar B_d\rangle$ for the 
$B\to D$ transition and a 
matrix element $\langle\pi^-|\bar{d}(x)[\ldots]u(0)|0\rangle$ 
with $x^2=0$ that gives rise to the pion light-cone distribution 
amplitude. Leaving aside power-suppressed contributions, the essential
requirement for this conclusion was the absence of both soft
and collinear infrared divergences in the gluon exchange between
the $(\bar cb)$ and $(\bar du)$ currents.
This gluon exchange is therefore calculable in QCD perturbation
theory. We now compute these corrections explicitly to order 
$\alpha_s$.

The effective Hamiltonian (\ref{heff18}) can be written as
\begin{eqnarray}\label{bheff}
   {\cal H}_{\rm eff} &=& \frac{G_F}{\sqrt2}\,V^*_{ud} V_{cb}\,
    \Bigg\{ \left[ \frac{N_c+1}{2N_c}\bar C_+(\mu)
    + \frac{N_c-1}{2N_c}\bar C_-(\mu)
    + \frac{\alpha_s(\mu)}{4\pi}\,\frac{C_F}{2N_c}\,B C_8(\mu)
    \right] O_0 \nonumber\\
   &&\qquad \mbox{}+ C_8(\mu)\,O_8 \Bigg\} ,
\end{eqnarray}
where the scheme-dependent terms in the coefficient of the operator 
$O_0$, proportional to the constant $B$ 
defined after (\ref{dbjpm}), have been written explicitly.

Schematically, the matrix elements of $O_0$ and $O_8$ can be
expressed in the form of (\ref{fff}). Because the light-quark
pair has to be in a colour singlet to produce the pion in the leading
Fock state, only $O_0$ gives a contribution to zeroth order in
$\alpha_s$. Similarly, to first order in $\alpha_s$ only $O_8$
can contribute. The result of 
computing the diagrams in Fig.~\ref{fig6} with an insertion of $O_8$ can 
be presented in a form that holds simultaneously for $H=D,D^*$ and 
$L=\pi, \rho$, using only that the $(\bar{u}d)$ pair is a colour 
singlet and that the external quarks can be taken on-shell.
We obtain ($z=m_c/m_b$)
\begin{eqnarray}\label{delo8}
   \langle H(p') L(q)|O_8|\bar{B}_d(p)\rangle
   &=& \frac{\alpha_s}{4\pi}\frac{C_F}{2N_c}\,i f_L \int_0^1 du\,
   \Phi_L(u) \\ 
   &&\hspace{-4cm}\times 
    \left[ - \left( 6\ln\frac{\mu^2}{m_b^2} + B \right)
    (\langle J_V\rangle - \langle J_A\rangle)
    + F(u,z)\,\langle J_V\rangle - F(u,-z)\,\langle J_A\rangle \right] ,
    \nonumber
\end{eqnarray}
where 
\begin{equation}\label{qva}
   \langle J_V\rangle
   = \langle H(p')|\bar c \!\not\!q \,b|\bar{B}_d(p)\rangle , \qquad
   \langle J_A\rangle
   = \langle H(p')|\bar c\!\not\!q\gamma_5 b\,|\bar{B}_d(p)\rangle .
\end{equation}
In obtaining (\ref{delo8}) we have used the equations of motion for the 
quarks to reduce the operator basis to $J_V$ and $J_A$. It is 
worth noting that even after computing the one-loop correction the 
$(\bar{u}d)$ pair retains its $V-A$ structure. This, together 
with (\ref{distamps}), implies that the form of (\ref{delo8}) is identical 
for pions and longitudinally polarized $\rho$ mesons. The production of  
transversely polarized $\rho$ mesons is power suppressed in 
$\Lambda_{\rm QCD}/m_b$, as follows from the 
$[\not\!\varepsilon_\perp^*,\not\!q\,]$ structure in the third line of 
(\ref{distamps}). 

In the case of a distribution amplitude $\Phi_L(u)$ that is
symmetric under $u\leftrightarrow \bar u$, which is relevant
for $L=\pi, \rho$, the function $F(u,z)$ appearing in (\ref{delo8}) 
can be compactly written as
\begin{equation}\label{ffsym}
   F(u,z) = 3 \ln z^2 - 7 + f(u,z) + f(u,1/z) ,
\end{equation}
with
\begin{equation}\label{fxzsym}
 f(u,z) = - \frac{u(1-z^2)[3(1-u (1-z^2))+z]}{[1-u(1-z^2)]^2}
    \ln[u(1-z^2)] - \frac{z}{1-u(1-z^2)} .
\end{equation}
In the general case, where $\Phi_L(u)$ is not necessarily symmetric,
the function $F(u,z)$ is given by
\begin{equation}\label{ff}
   F(u,z) = \left( 3 + 2 \ln\frac{u}{\bar u} \right) \ln z^2 - 7
   + f(u,z) + f(\bar u,1/z) ,
\end{equation}
where 
\begin{eqnarray}\label{fxz}
   &&f(u,z) = - \frac{u(1-z^2)[3(1-u (1-z^2))+z]}{[1-u(1-z^2)]^2}
    \ln[u(1-z^2)] - \frac{z}{1-u(1-z^2)} \nonumber\\
   &&~~\mbox{}+ 2 \Bigg[ \frac{\ln[u(1-z^2)]}{1-u(1-z^2)} - \ln^2[u(1-z^2)] 
    - \mbox{Li}_2[1-u(1-z^2)] - \{ u\to\bar u \} \Bigg] ,
\end{eqnarray}
and
\begin{equation}\label{l2def}
   \mbox{Li}_2(x) = - \int_0^x dt\,\frac{\ln(1-t)}{t}
\end{equation}
is the dilogarithm. The contribution of $f(u,z)$ in (\ref{ff}) comes from 
the first two diagrams in Fig.~\ref{fig6} with the gluon coupling to the $b$ 
quark, whereas $f(\bar u,1/z)$ arises from the last two diagrams with the 
gluon coupling to the charm quark. Note that the terms in the 
large square brackets 
in the definition of the function $f(u,z)$ are odd under the exchange 
$u\leftrightarrow\bar u$ and thus vanish for a symmetric light-cone 
distribution amplitude. These terms can be dropped if the light final-state
meson is a pion or a $\rho$ meson, but they are relevant, e.g., for the 
discussion of Cabibbo-suppressed decays such as $\bar B_d\to D^{(*)+} K^-$ 
and $\bar B_d\to D^{(*)+} K^{*-}$. 
The discontinuity of the amplitude, which is responsible for the occurrence 
of strong rescattering phases, arises from $f(\bar u,1/z)$ and can be 
obtained by recalling that $z^2$ is $z^2-i\epsilon$ with $\epsilon>0$ 
infinitesimal. We then find 
\begin{eqnarray}
   \frac{1}{\pi}\,\mbox{Im}\,F(u,z)
   &=& - \frac{(1-u)(1-z^2)[3(1-u (1-z^2))+z]}{[1-u(1-z^2)]^2} \nonumber\\
   &&\hspace{-2.2cm}
    \mbox{}- 2 \Bigg[ \ln[1-u(1-z^2)] + 2\ln u
    + \frac{z^2}{1-u(1-z^2)} - \{ u\to\bar u \} \Bigg] .
\end{eqnarray}
For $z\to 0$ and the special case of a symmetric wave function these results 
coincide with the results already presented in \cite{BBNS99}.

As mentioned above, (\ref{delo8}) is applicable to all decays
of the type $\bar B_d\to D^{(*)+}L^-$, where $L$ is a light
hadron such as a pion or a (longitudinally polarized) $\rho$ meson.
Only the operator $J_V$ contributes to $\bar B_d\to D^+ L^-$, and
only $J_A$ contributes to $\bar B_d\to D^{*+} L^-$. (Due to helicity 
conservation the vector current $B\to D^*$ matrix element contributes
only in conjunction with a transversely polarized $\rho$ meson and hence 
is power suppressed in the heavy-quark limit.) Our final result 
can therefore be written as 
\begin{equation}\label{bdpi18}
\langle D^+ L^-|O_{0,8}|\bar B_d\rangle=
\langle D^+|\bar c\gamma^\mu(1-\gamma_5)b|\bar B_d\rangle
\cdot i f_L q_\mu\int^1_0 du\,T_{0,8}(u,z)\,\Phi_L(u) ,
\end{equation}
where $L=\pi$, $\rho$, and the hard-scattering kernels are
\begin{eqnarray}
\label{t1uz}
   T_0(u,z)&=& 1+O(\alpha^2_s) , \\
\label{t8uz}
   T_8(u,z)&=&\frac{\alpha_s}{4\pi}\frac{C_F}{2N_c} \left[
    - 6\ln\frac{\mu^2}{m_b^2} - B + F(u,z) \right]
    + O(\alpha^2_s) .
\end{eqnarray}
When the $D$ meson is replaced by a $D^*$ meson, the result is 
identical except that $F(u,z)$ in (\ref{t8uz}) must be replaced 
by $F(u,-z)$. Since no order $\alpha_s$ corrections exist for $O_0$, 
the matrix element retains its leading-order factorized form
\begin{equation}\label{o1me}
\langle D^+L^-|O_0|\bar B_d\rangle= i f_L q_\mu\,
\langle D^+|\bar c\gamma^\mu(1-\gamma_5)b|\bar B_d\rangle
\end{equation}
to this accuracy. 
From (\ref{fxz}) it follows that $T_8(u,z)$ tends to a 
constant as $u$ approaches the endpoints ($u\to 0$, $1$). 
(This is strictly true for the part of $T_8(u,z)$ that is 
symmetric in $u\leftrightarrow\bar u$; the asymmetric part diverges 
logarithmically ($\propto \ln u$) as $u\to 0$, which however
does not affect the power behaviour and the convergence properties
in the endpoint region.)  Therefore the contribution
to (\ref{bdpi18}) from the endpoint region is suppressed,
both by phase space and by the endpoint suppression
intrinsic to $\Phi_L(u)$. Consequently the emitted light meson is indeed
dominated by energetic constituents, as required for the self-consistency
of the factorization formula (\ref{bdpi18}).  

Combining (\ref{bheff}), (\ref{bdpi18}), (\ref{t1uz}) and (\ref{t8uz}), we 
obtain our final result for the class-I, non-leptonic $\bar B_d\to D^{(*)+} 
L^-$ decay amplitudes in the heavy-quark limit, and at next-to-leading order 
in $\alpha_s$. The results can be compactly expressed in terms of the matrix
elements of a ``transition operator''
\begin{equation}\label{heffa1}
{\cal T}=\frac{G_F}{\sqrt{2}} V^*_{ud}V_{cb}
\left[ a_1(D L)\, Q_V - a_1(D^* L)\, Q_A\right] ,
\end{equation}
where
\begin{equation}\label{qva2}
Q_V=\bar c\gamma^\mu b\,\otimes\, \bar d\gamma_\mu(1-\gamma_5)u ,
\qquad
Q_A=\bar c\gamma^\mu\gamma_5 b\,\otimes\, \bar d\gamma_\mu(1-\gamma_5)u ,
\end{equation}
and hadronic matrix elements of $Q_{V,A}$ are understood to be evaluated
in factorized form, i.e.\ 
\begin{equation}
   \langle D L|j_1\otimes j_2|\bar B\rangle
   \equiv \langle D|j_1|\bar B\rangle\,\langle L|j_2|0\rangle .
\end{equation} 
Eq.~(\ref{heffa1}) defines the quantities $a_1(D^{(*)} L)$, which include 
the leading ``non-factorizable'' corrections, in a renormalization-scale and 
-scheme independent way. To leading power in $\Lambda_{\rm QCD}/m_b$ these 
quantities should not be interpreted as phenomenological parameters (as is 
usually done), because they are dominated by hard gluon exchange and thus 
calculable in QCD. At next-to-leading order we get
\begin{eqnarray}\label{a1dpi}
   a_1(D L) &=& \frac{N_c+1}{2N_c}\bar C_+(\mu)
    + \frac{N_c-1}{2N_c}\bar C_-(\mu) \nonumber\\
   &&\mbox{}+ \frac{\alpha_s}{4\pi}\frac{C_F}{2N_c}\,C_8(\mu) \left[
    - 6\ln\frac{\mu^2}{m_b^2} + \int^1_0 du\,F(u,z)\,\Phi_L(u)
    \right] , \\
\label{a1dspi}
   a_1(D^* L) &=& \frac{N_c+1}{2N_c}\bar C_+(\mu)
    + \frac{N_c-1}{2N_c}\bar C_-(\mu) \nonumber\\
   &&\mbox{}+ \frac{\alpha_s}{4\pi}\frac{C_F}{2N_c}\,C_8(\mu) \left[
    - 6\ln\frac{\mu^2}{m_b^2} + \int^1_0 du\,F(u,-z)\,\Phi_L(u) \right] .
\end{eqnarray}
These expressions generalize the well-known leading-order formula
\begin{equation}\label{a1lo}
a^{\rm LO}_1=\frac{N_c+1}{2N_c}C^{\rm LO}_+(\mu)
+\frac{N_c-1}{2N_c}C^{\rm LO}_-(\mu) .
\end{equation}
We observe that the scheme dependence, parameterized by $B$, is cancelled
between the coefficient of $O_0$ in (\ref{bheff}) and the matrix
element of $O_8$ in (\ref{bdpi18}). Likewise, the $\mu$ dependence of the 
terms in brackets
in (\ref{a1dpi}) and (\ref{a1dspi}) cancels against the scale dependence of 
the coefficients $\bar C_\pm(\mu)$, ensuring a consistent physical result 
at next-to-leading order in QCD. 

The coefficients $a_1(D L)$ and $a_1(D^* L)$ are seen to be non-universal, 
i.e.\ they are explicitly dependent on the nature of the final-state mesons. 
This dependence enters via the light-cone distribution amplitude $\Phi_L(u)$ 
of the light emission meson and via the analytic form of the hard-scattering 
kernel ($F(u,z)$ vs.\ $F(u,-z)$). However, the non-universality enters only 
at next-to-leading order.

Politzer and Wise have computed the ``non-factorizable'' vertex corrections 
to the decay rate ratio of the $D\pi$ and $D^*\pi$ final states \cite{PW91}. 
This requires only the symmetric part (with respect to 
$u\leftrightarrow\bar u$) of the difference $F(u,z)-F(u,-z)$. Explicitly,
\begin{equation}\label{dpdsp}
   \frac{\Gamma(\bar B_d\to D^+\pi^-)}{\Gamma(\bar B_d\to D^{*+}\pi^-)}
   = \left|
   \frac{\langle D^+|\bar c\!\not\! q(1-\gamma_5)b|\bar B_d\rangle}
        {\langle D^{*+}|\bar c\!\not\! q(1-\gamma_5)b|\bar B_d\rangle}
   \right|^2 \left| \frac{a_1(D\pi)}{a_1(D^*\pi)} \right|^2 ,
\end{equation}
where for simplicity we neglect the light meson masses as well as the
mass difference between $D$ and $D^*$ in the phase-space for the two decays.
At next-to-leading order 
\begin{equation}
   \left| \frac{a_1(D\pi)}{a_1(D^*\pi)} \right|^2
   = 1 + \frac{\alpha_s}{4\pi}\frac{C_F}{N_c}\frac{C_8}{C_0}\,
   {\rm Re}\int^1_0 du \left[ F(u,z)-F(u,-z) \right]\,\Phi_\pi(u) .
\end{equation}
Our result for the symmetric part of $F(u,z)-F(u,-z)$ coincides with that
of Politzer and Wise. Haas and Youssefmir have considered the decay 
rate ratio of the $D\rho$ and $D\pi$ final states \cite{HY}, which requires 
the calculation of the symmetric part of $F(u,z)$. Neglecting again the
light meson masses we obtain
\begin{equation}\label{drdp}
   \frac{\Gamma(\bar B_d\to D^+\rho^-)}{\Gamma(\bar B_d\to D^+\pi^-)}
   = \frac{f^2_\rho}{f^2_\pi} \left| \frac{a_1(D\rho)}{a_1(D\pi)} \right|^2 ,
\end{equation}
where
\begin{equation}\label{drdp2}
   \left| \frac{a_1(D\rho)}{a_1(D\pi)} \right|^2
   = 1 + \frac{\alpha_s}{4\pi}\frac{C_F}{N_c}\frac{C_8}{C_0}\,
   {\rm Re}\int^1_0 du\,F(u,z) \left[ \Phi_\rho(u)-\Phi_\pi(u) \right] .
\end{equation}
Our result for $F(u,z)$ does not agree with that obtained in \cite{HY}. 
To compare the two results, the kernel $F(u,z)$ has to be symmetrized in 
$u\leftrightarrow\bar u$ and $u$-independent constants can be dropped,
because they do not contribute to (\ref{drdp2}). The result of Haas and 
Youssefmir would agree with ours if the term $a+\ln|a|/(1-a)$ in the 
function $I_{1,1}(a)$ defined in \cite{HY} were substituted by 
$1+\ln|a|/(1-a)$, and the function $J(a)$ were multiplied by 4.


\boldmath
\section{$B\to D\pi$: Factorization in higher orders}
\unboldmath
\label{allorders}

The one-loop expression for $T_8$ in (\ref{t8uz}) has no
infrared singularities, as required for the validity of the factorization
formula. In order for the factorization formula for $\bar{B}_d\to
D^+\pi^-$ decays (here and below we suppress a factor $i f_\pi m_B^2$),
\begin{equation}
\label{x1}
\langle\pi^- D^+|O_{0,8}|\bar{B}_d\rangle=F^{B\to D}(0)
\int_0^1 du\,T_{0,8}(u)\,\Phi_\pi(u),
\label{eq:fact}\end{equation}
to be valid, the amplitude $T_{0,8}(u)$ must be free of infrared 
singularities to all orders in perturbation
theory. (In this section we use again the decay $\bar{B}_d\to
D^+\pi^-$ as a representative for all decays into a
heavy and light meson for which the spectator quark in the $B$ meson
goes to the heavy meson in the final state.)
This requires demonstrating that the long-distance contributions 
to the $\bar{B}_d\to D^+\pi^-$ amplitude match those contained in the 
form factor and the pion light-cone distribution amplitude on 
the right-hand side of (\ref{x1}), i.e.\ that the 
sum of all the infrared singularities in Feynman diagrams for the
$\bar{B}_d\to D^+\pi^-$ amplitude must be precisely that present in 
the Feynman diagrams for the form factor
$F^{B\to D}(0)$ and for the light-cone 
distribution amplitude $\Phi_\pi(u)$ of the pion.
In this section we analyze in detail the infrared singularities
for $\bar{B}_d\to D^+ \pi^-$ decays at two-loop order and demonstrate 
that this is indeed the case. Some of the arguments we use have 
straightforward extensions to all orders, but the following does 
not accomplish an all-order ``proof''. We hope that the arguments used  
to prove infrared finiteness at two-loop order are sufficiently 
convincing to make infrared finiteness at all orders plausible. 

The content of this section is rather technical. The 
phenomenologically oriented reader may want to proceed directly with 
Sect.~\ref{bdpi}, where we discuss practical applications of the 
factorization formula and comparisons of our results with 
experimental data.

\subsection{Structure of the factorization proof at two-loop order}

To state more precisely what needs to be demonstrated, we
write the factorization formula schematically as
\begin{equation}
A(B\to D\pi)= F_{B\to D}(0)\cdot T*\Phi_\pi,
\label{eq:factform}
\end{equation}
where the $*$ represents the convolution, $A(B\to D\pi)$ represents 
the matrix element on the left-hand side of (\ref{x1}), and the
subscript `0,8' is omitted. In order to extract $T$,
one computes $A$, $F_{B\to D}$ and $\Phi_\pi$ in perturbation theory and
uses (\ref{eq:factform}) to determine $T$. We therefore
rewrite (\ref{eq:factform}) in perturbation theory, 
\begin{eqnarray}
A^{(0)} + A^{(1)} +A^{(2)}+\,\cdots
&=&\left(F_{B\to D}^{(0)}+F_{B\to D}^{(1)}+F_{B\to D}^{(2)}+\cdots\right)
\cdot\nonumber\\ 
&&\hspace{-2.5cm}\left(T^{(0)}+T^{(1)}+T^{(2)}+\cdots\right)
*\left(\Phi_\pi^{(0)}+\Phi_\pi^{(1)}+\Phi_\pi^{(2)}+\cdots\right),
\label{eq:ffpert}
\end{eqnarray}
where the superscripts in parentheses indicate the order of perturbation
theory, and then compare terms of the same order. Thus up to 
two-loop order
\begin{eqnarray}
F_{B\to D}^{(0)}\cdot T^{(0)}*\Phi_\pi^{(0)}& = & 
A^{(0)},
\label{eq:t20}\\
F_{B\to D}^{(0)}\cdot T^{(1)}*\Phi_\pi^{(0)}& = &
A^{(1)} 
-F_{B\to D}^{(1)}\cdot T^{(0)}*\Phi_\pi^{(0)}
-F_{B\to D}^{(0)}\cdot T^{(0)}*\Phi_\pi^{(1)},
\label{eq:t21}\\
F_{B\to D}^{(0)}\cdot T^{(2)}*\Phi_\pi^{(0)}& = &
A^{(2)}
-F_{B\to D}^{(0)}\cdot T^{(1)}*\Phi_\pi^{(1)}
-F_{B\to D}^{(1)}\cdot T^{(1)}*\Phi_\pi^{(0)}
\nonumber\\ & & \hspace{-0.8in}
-F_{B\to D}^{(2)}\cdot T^{(0)}*\Phi_\pi^{(0)}
-F_{B\to D}^{(0)}\cdot T^{(0)}*\Phi_\pi^{(2)}
-F_{B\to D}^{(1)}\cdot T^{(0)}*\Phi_\pi^{(1)}.
\label{eq:t22}
\end{eqnarray}
By perturbative expansion of the $B\to D$ form factor, we mean the
perturbative expansion of the matrix element of $\bar{c}\Gamma b$,
evaluated between on-shell $b$- and $c$-quark states. By perturbative 
expansion of the pion light-cone distribution amplitude, we mean 
the perturbative expansion of the light-cone matrix element 
in (\ref{distamps}), but with the pion state replaced by an on-shell 
quark with momentum $u q$ and an on-shell antiquark with momentum 
$\bar{u} q$.

The zeroth order term in (\ref{eq:t20}) is trivial and states that 
$T^{(0)}$ is given by the diagram of Fig.~\ref{fig4}. The two terms 
that need to be subtracted from $A^{(1)}$ at first order exactly cancel 
the ``factorizable'' contributions to $A^{(1)}$ shown in 
Fig.~\ref{fig5}. The first order term in (\ref{eq:t21}) therefore 
states that $T^{(1)}$ is given by the ``non-factorizable'' diagrams 
of Fig.~\ref{fig6}. If $T^{(1)}$ is to be infrared finite, the 
sum of these diagrams must be infrared finite, which is indeed the 
case as we have already seen.

In this section we will be concerned with the second order term 
(\ref{eq:t22}). The last three terms on the right-hand side exactly cancel 
the ``factorizable'' corrections to the two-loop amplitude 
$A^{(2)}$. The remaining two terms that need to be subtracted from 
$A^{(2)}$ are non-trivial. We must show that the infrared divergences 
in the sum of ``non-factorizable'' contributions to $A^{(2)}$ have 
precisely the right structure to match the infrared divergences 
in $F_{B\to D}^{(1)}$ and $\Phi_\pi^{(1)}$, such that
\begin{equation}
\label{cc1}
A^{(2)}_{\rm non-fact.}
-F_{B\to D}^{(0)}\cdot T^{(1)}*\Phi_\pi^{(1)}
-F_{B\to D}^{(1)}\cdot T^{(1)}*\Phi_\pi^{(0)} = 
\mbox{ infrared finite.}
\end{equation}

The discussion above implies that in order to demonstrate the
validity of the factorization formula we need to identify the regions
of phase space which can give rise to infrared singularities. In
general these arise when massless lines become soft or 
collinear with the direction of $q$, the momentum of the pion. This 
requires that one or both of the loop momenta in a two-loop diagram 
become soft or collinear. Rather than computing the two-loop diagrams,
we will analyze the Feynman integrands corresponding to 
these diagrams in those momentum configurations that can 
give rise to singularities, as we did at one-loop order, and show that 
(\ref{cc1}) is valid. The potentially singular regions are: 
one momentum soft or collinear, the other hard; both momenta soft 
or collinear; one momentum soft, the other collinear. In the
following subsections we consider each of the five regions in 
turn. 

\begin{figure}[t]
\vspace*{-0.3cm}
\hbox to\hsize{
\hss
\epsfxsize=1.23\textwidth\epsfbox[0 480 600 750]{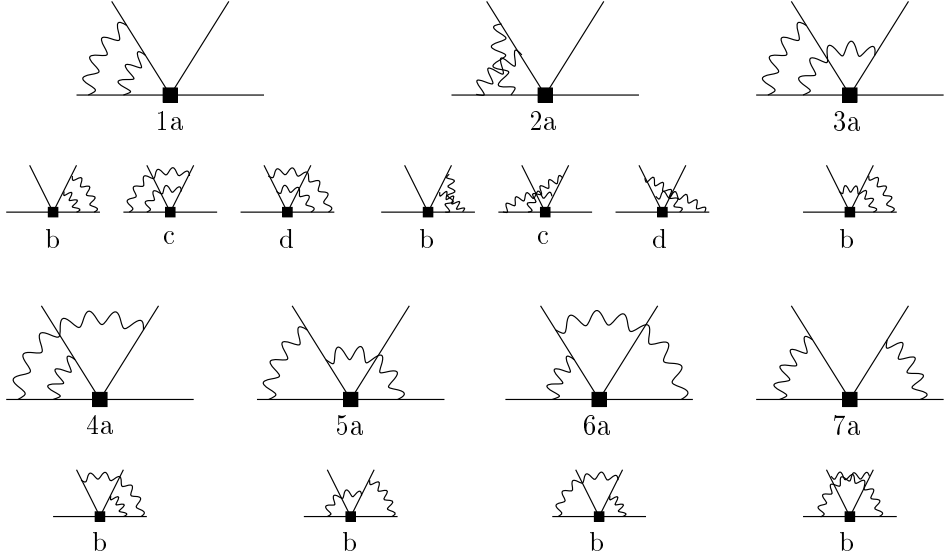}
\hss}
\caption{Two-loop diagrams 1a-7b, contributing to the amplitudes
for $B\to D\pi$ decays.}
\label{fig:diagsone}\end{figure}

\begin{figure}[p]
\hbox to\hsize{
\hss
\epsfxsize=1.23\textwidth\epsfbox[0 120 600 750]{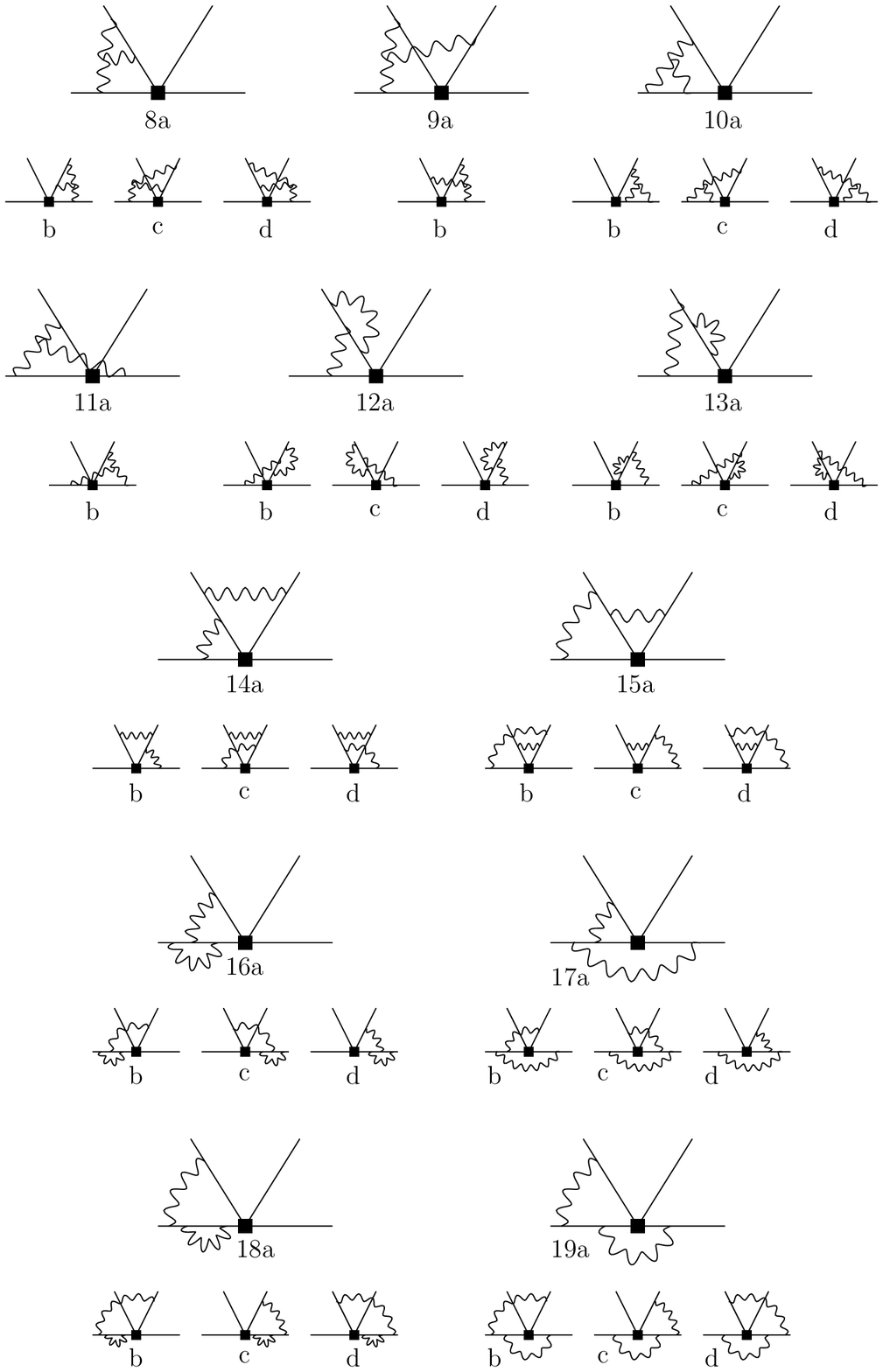}
\hss}
\caption{Two-loop diagrams 8a-19d, contributing to the amplitudes
for $B\to D\pi$ decays.}
\label{fig:diagstwo}
\end{figure}

At two-loop order there are 
62 ``non-factorizable'' diagrams that contribute to $A^{(2)}$ 
which we label from 1a to 19d and exhibit
in Figs.~\ref{fig:diagsone} and
\ref{fig:diagstwo}. (We do not exhibit diagrams with vacuum
polarization insertions in gluon propagators, external self-energy
insertions in quark propagators, or one-loop counterterm insertions. The
demonstration of factorization for these sets of diagrams is a simple
extension of that for one-loop diagrams.)
In these diagrams the
inserted four-quark current-current operators can have either a
singlet-singlet ($O_0$) or octet-octet ($O_8$) colour structure.  The
corresponding colour factors for each of the 62 diagrams are tabulated
in Table~\ref{tab:colour} and these 
will be used extensively in the discussion
below. In evaluating the entries we have taken a
factor $T^A_{ij}$ at each quark-gluon vertex ($A=1\ldots 8$ labels the
colour of the gluon and $i=1,2,3$ ($j=1,2,3$) that of the
outgoing (incoming) quark) and a factor of $-i f^{ABC}$ at each
three-gluon vertex where $A,B,C$ are the colour labels of the three
gluons, with $A\to B\to C$ taken in a clockwise direction.

The diagrams are taken with on-shell external quark lines. We never
write down the corresponding on-shell spinors, but we use extensively 
the equation of motion for the on-shell spinors to simplify the 
Feynman amplitude. We also use Feynman gauge for gluon propagators.

\begin{table}[t]
\caption{Colour factors for the two-loop diagrams.
Here the normalization is such that the tree-level
diagram with $O_0$ insertion has a colour factor $N_c$.}
\label{tab:colour}
\vspace*{0.2cm}
\begin{center}
\begin{tabular}{|ccc|}\hline
\rule[-0.25cm]{0cm}{0.7cm} Diagrams & \ $O_0$\  &\  $O_8$\ \\ 
\hline
\rule[-0.3cm]{0cm}{0.9cm}
1a,b\ \ 2c,d\ \  4a,b\ \  6a,b\ \ 7a & $\frac{C_F}{2}$
& $\frac{C_F}{2}\,\frac{(N_c^2-2)}{2N_c}$\\ \hline   
\rule[-0.3cm]{0cm}{0.9cm}
1c,d\ \ 2a,b\ \ 3a,b\ \ 5a,b\ \ 7b & $\frac{C_F}{2}$
& $-\frac{C_F}{2}\,\frac{1}{N_c}$\\ 
\hline
\rule[-0.3cm]{0cm}{0.9cm}
8a,b\ \ 10a-d & 0 & $\frac{C_F}{2}\,\frac{N_c}{2}$\\ 
\hline
\rule[-0.3cm]{0cm}{0.9cm}
8c,d\ \ 9a,b\ \ 11a,b & 0 & $-\frac{C_F}{2}\,\frac{N_c}{2}$\\ 
\hline
\rule[-0.3cm]{0cm}{0.9cm}
12a-d\ \ 15a-d\ \ 16a-d\ \ 19a-d & 0 & 
$-\frac{C_F}{2}\,\frac{1}{2N_c}$\\ 
\hline
\rule[-0.3cm]{0cm}{0.9cm}
13a-d\ \ 14a-d\ \ 17a-d\ \ 18a-d & 0 & 
$\frac{C_F}{2}\,C_F$\\ \hline
\end{tabular}
\end{center}
\end{table}

\subsection{The soft-soft region}
\label{subsec:softsoft}

We start by considering the region of phase space in which both loop 
momenta are soft, i.e.\ they have components of momentum
(in the rest-frame of the $b$ quark) which are much smaller than $m_b$.
There are different ways of routing the large external momenta 
through a two-loop diagram. It is easy to see that one obtains 
an infrared divergence only if the large momentum is routed through 
the quark lines and all the gluons in each diagram are soft. 
The generic power counting in this region of phase-space, taking all
components of the gluons' momenta to be of order $\lambda$, gives a factor
of $\lambda^8$ from the two-loop phase space, factors of $\lambda^{-1}$
and $\lambda^{-2}$ for each quark and gluon propagator and $\lambda^0$
and $\lambda^1$ for each quark-gluon and triple-gluon vertex,
respectively.  Thus, for example, the counting for diagrams 1a-d would
give $\lambda^8$ from phase space, $\lambda^{-4}$ from the two gluon
propagators, $\lambda^{-4}$ from the four quark propagators and
$\lambda^0$ for the four quark-gluon vertices, giving a total factor of
$\lambda^0$ corresponding to a logarithmic divergence. Similarly for
diagrams 8a-d we have factors of $\lambda^8$ (phase-space),
$\lambda^{-3}$ (quark propagators), $\lambda^{-6}$ (gluon propagators),
$\lambda^0$ (quark-gluon vertices) and $\lambda^1$ (triple-gluon vertex)
giving $\lambda^0$ and a logarithmic divergence again. The analysis 
of all diagrams shows that the divergence is at most logarithmic. In this
subsection we demonstrate the cancellation of these logarithmic
divergences.

For diagrams in which a single gluon is attached to a constituent of the
pion (diagrams 10a-11b, 16a-19d) the cancellation can be
demonstrated in an exactly analogous way
to the one-loop case, i.e.\ the two contributions in which the gluon
is attached to the quark and antiquark in the pion cancel
(e.g.\ the contribution of diagram 10a cancels that of 10c). We need
not discuss such diagrams further in this subsection.

When one of the gluons is attached at both ends to the constituents
of the pion (diagrams 12a-15d), each diagram is logarithmically 
divergent according to the generic power-counting rules above, but 
the divergence is in fact absent for each diagram separately. This 
follows because we can use the eikonal approximation for the 
quark propagators and the on-shell condition for the external quark 
lines to show that the 
logarithmic divergence is proportional to $q^\rho q_\rho=q^2=0$. 
(Of course, these diagrams have divergences in other momentum 
configurations, see the following subsections).

\begin{figure}[t]
   \vspace{-3.7cm}
   \hspace*{-1.5cm}
   \epsfysize=24cm
   \epsfxsize=16cm
   \centerline{\epsffile{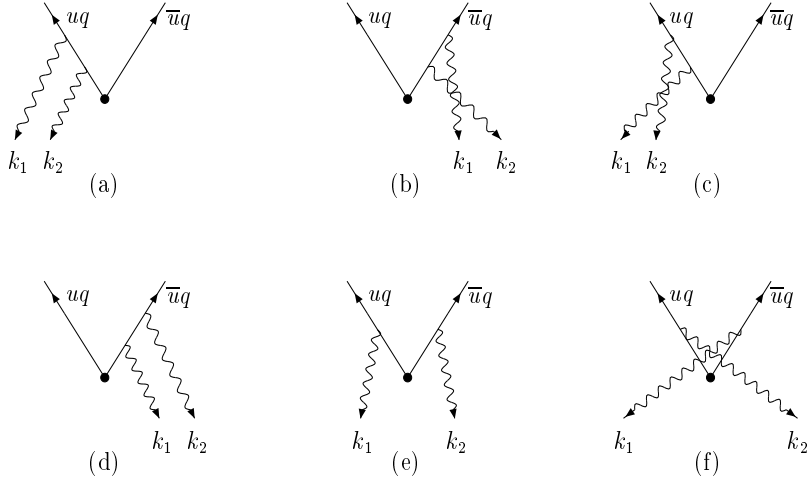}}
   \vspace*{-13.5cm}
\caption[dummy]{\label{fig:ss1}\small 
The six diagrams representing the possible attachments of two
gluons to the quark and antiquark constituents of the pion.}
\end{figure}

Consider now the 18 diagrams 1a-7b which contain two gluons, each
attached to a heavy and a light quark, and take both gluons to be soft.
The cancellation of the corresponding infrared divergences can readily
be demonstrated using standard eikonal combinatorics. In
Fig.~\ref{fig:ss1} we draw the six possible ways of attaching two gluons
to the constituents of the pion. Thus for each of the three distinct ways
of attaching the two gluons to the heavy quarks there are six diagrams,
\{1a, 1c, 2a, 2c, 3a, 4a\}, \{5a, 5b, 6a, 6b, 7a, 7b\} and 
\{1b, 1d, 2b, 2d, 3b, 4b\}.
In this way the 18 diagrams 1a-7b get divided into three sets of six
diagrams. By studying the structure of the diagrams in
Fig.~\ref{fig:ss1} we now demonstrate the cancellation of the
divergences from the soft-soft region of phase space within each of the
sets of six diagrams.

In the six diagrams of Fig.~\ref{fig:ss1} let the gluon with momentum
$k_1$~($k_2$) have Lorentz and colour indices $\mu_1$ and $A_1$ ($\mu_2$
and $A_2$) and let the Dirac matrix of the current be denoted
by $\Gamma$ and its colour matrix by $T$. The diagram in
Fig.~\ref{fig:ss1}a has a factor:
\begin{equation}
\frac{\gamma_{\mu_1}(u\qslash+\kslash_1)\gamma_{\mu_2}(u\qslash+\kslash_1
+\kslash_2)\Gamma}{(uq+k_1)^2\,(uq+k_1+k_2)^2}\, C_1
\label{eq:fact1}\end{equation}
where $C_1\equiv\textrm{Tr}\{T^{A_1}T^{A_2}T\}$. The trace arises 
because the two external light quark lines are projected on a colour 
singlet combination. Since we are neglecting the masses
of the light quarks we can use the equation of motion $\qslash=0$ when
the $\qslash$ is the first or last factor in the product of
gamma-matrices associated with the light quark and antiquark. Then, 
recalling that the components
of $k_1$ and $k_2$ are small, and that we need to keep only the 
leading term because the divergence is logarithmic, 
the term in~(\ref{eq:fact1}) reduces to
\begin{equation}
\frac{q_{\mu_1}q_{\mu_2}}{(q\cdot k_1)\, (q\cdot(k_1+k_2))}\,\Gamma\,C_1.
\label{eq:fact2}\end{equation}
This is the standard eikonal approximation. 
The corresponding factors for all six of the diagrams in
Fig.~\ref{fig:ss1} are:
\begin{center}
\vspace{0.2cm}
\begin{tabular}{lcl}
\rule[-10pt]{0pt}{25pt}
Fig.~\ref{fig:ss1}a: ${\displaystyle \frac{q_{\mu_1}\,q_{\mu_2}}
{(q\cdot k_1)\, (q\cdot(k_1+k_2))}\,\Gamma\,C_1}$ &
\hspace{1.5in}&
Fig.~\ref{fig:ss1}b: ${\displaystyle \frac{q_{\mu_1}\,q_{\mu_2}}
{(q\cdot k_1)\, (q\cdot(k_1+k_2))}\,\Gamma\,C_2}$ 
\\[0.5cm] 
\rule[-10pt]{0pt}{25pt}
Fig.~\ref{fig:ss1}c: ${\displaystyle \frac{q_{\mu_1}\,q_{\mu_2}}
{(q\cdot k_2)\, (q\cdot(k_1+k_2))}\,\Gamma\,C_2}$ &
\hspace{1.5in}&
Fig.~\ref{fig:ss1}d: ${\displaystyle \frac{q_{\mu_1}\,q_{\mu_2}}
{(q\cdot k_2)\, (q\cdot(k_1+k_2))}\,\Gamma\,C_1}$ 
\\[0.5cm] 
\rule[-10pt]{0pt}{25pt}
Fig.~\ref{fig:ss1}e: ${\displaystyle -\frac{q_{\mu_1}\,q_{\mu_2}}
{(q\cdot k_1)\, (q\cdot k_2)}\,\Gamma\,C_2}$ &
\hspace{1.5in}&
Fig.~\ref{fig:ss1}f: ${\displaystyle -\frac{q_{\mu_1}\,q_{\mu_2}}
{(q\cdot k_1)\, (q\cdot k_2)}\,\Gamma\,C_1}$,  
\end{tabular}
\end{center}
where $C_2\equiv\textrm{Tr}\{T^{A_2}T^{A_1}T\}$. We see that the
coefficients of both $C_1$ and $C_2$ vanish when the six contributions
are summed, and hence the divergences of the 18 diagrams 1a-7b cancel.

An analogous cancellation occurs when the two gluons in the diagrams of
Fig.~\ref{fig:ss1} end at a triple gluon vertex. Thus, choosing 
momentum assignments that correspond to (a), (d), (e) in
Fig.~\ref{fig:ss1}, but with the labelling of $k_1$ and $k_2$ in (e) 
interchanged, and accounting for the signs of the colour factors
given in Table~\ref{tab:colour}, the 
diagrams \{8a, 8c, 9a\} have relative contributions
\begin{equation}
\label{3g11}
\frac{1}{(q\cdot k_1) (q\cdot(k_1+k_2))}\ \ :\  \
\frac{1}{(q\cdot k_2) (q\cdot(k_1+k_2))}\ \ :
\ \ -\frac{1}{(q\cdot k_1) (q\cdot k_2)},
\end{equation}
and hence cancel. A similar cancellation occurs between the soft-soft
contributions in diagrams \{8b, 8d, 9b\}. (If one chooses the
momentum labelling of (e) as given in Fig.~\ref{fig:ss1}, one obtains 
the third term in (\ref{3g11}) with an opposite sign. The cancellation
still occurs because the entire expression is multiplied by an odd
expression in $k_1\leftrightarrow k_2$ and then integrated over $k_1$
and $k_2$.)

This completes our proof of the cancellation of potentially
non-factorizing long-distance contributions from the soft-soft region of
phase space at two-loop order. It is easy to see that the 
combinatorics of eikonal propagators is such that the cancellation 
generalizes to all orders, assuming that all loop momenta are 
soft. Furthermore, we did not use that the $c$ quark is heavy. Hence 
the same argument also shows the cancellation of purely soft 
divergences for the form-factor term in the factorization formula 
for decays into two light mesons.

\subsection{The hard-soft region}
\label{subsec:hardsoft}

We now turn our attention to the region of phase space in which one
loop momentum is hard (i.e.\ all its components are of order $m_b$) 
and the other is soft (i.e.\ all its components are much smaller 
than $m_b$). In order
to have a potentially divergent subgraph, we require one of the 
gluons (rather than quarks) to be soft.

\begin{figure}[t]
   \vspace{-4.1cm}
\hspace*{-2cm}
   \epsfysize=24cm
   \epsfxsize=16cm
   \centerline{\epsffile{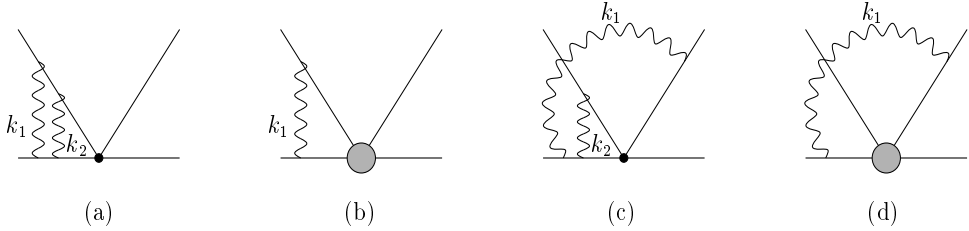}}
   \vspace*{-16.8cm}
\caption[dummy]{\label{fig:hs1}\small 
(a) Diagram 1a, (b) representation of diagram 1a in the
hard-soft region, (c) diagram 4a, (d) representation of diagram 4a in
the hard-soft region.}
\end{figure}

\begin{figure}[t]
   \vspace{-4.1cm}
   \epsfysize=24cm
   \epsfxsize=16cm
   \centerline{\epsffile{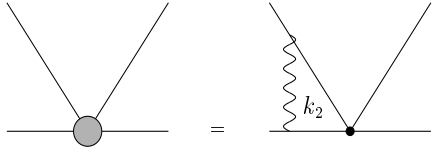}}
   \vspace*{-17.2cm}
\caption[dummy]{\label{fig:hs2}\small 
Representation of the hard inner loop.}
\end{figure}

\paragraph{Diagrams 1a-7b:}
We start by considering diagrams 1a-7b. As an example consider diagram
1a, which we redraw in Fig.~\ref{fig:hs1}a. This has a potentially
non-factorizing contribution from the hard-soft region in which the
inner loop is hard (i.e.\ $k_2$ is hard) and the external loop is soft
(i.e.\ $k_1$ is soft). The key point is that we can neglect $k_1$ in the
inner loop, allowing us, in this region of phase space, to represent
diagram 1a as in Fig.~\ref{fig:hs1}b. The grey circle represents the
inner loop (which contains the integral over $k_2$, see
Fig.~\ref{fig:hs2}), but is independent of $k_1$. Similarly for diagram
4a, which we redraw in Fig.~\ref{fig:hs1}c, the contribution from the
region in which $k_1$ is soft but $k_2$ is hard can be represented as in
Fig.~\ref{fig:hs1}d, where the grey circle represents the same factor
as in Fig.~\ref{fig:hs1}b (see Fig.~\ref{fig:hs2}). The cancellation
of the hard-soft contributions from diagrams 1a and 4a now follows
in the same way as for the infra-red divergences in one-loop graphs in
Sect.~\ref{oneloopcancel}.

Similarly the pairs of diagrams \{1b, 4b\}, \{1c, 3a\}, \{1d, 3b\},
\{5a, 7b\}, \{5b, 7b\}, \{6a, 7a\} and \{6b, 7a\} each cancel by the same
argument. Diagrams 7a and 7b appear twice since they each contain two
divergent hard-soft contributions (each of the two gluons may be soft).
Diagrams 2a-d have no divergent hard-soft contributions.

\paragraph{Diagrams 9a,b, 11a,b and 14a-d:} By inspection we see that
diagrams 9a,b, 11a,b and 14a-d have no divergent contributions in the
hard-soft region.

\paragraph{Diagrams 10a-d, 15a-d, 16a-d, 18a-d and 19a-d:} The
singular terms from the hard-soft contributions from each of the pairs
of diagrams \{10a, 10c\}, \{10b, 10d\}, \{15a, 15b\}, \{15c, 15d\},
\{16a, 16b\}, \{16c, 16d\}, \{18a, 18b\}, \{18c, 18d\}, \{19a, 19b\} and
\{19c, 19d\} cancel by the one-loop mechanism as described above. In
each case the singular contribution comes from the region of phase space
in which the gluon attached to one of the constituents of the pion is
soft (in diagrams 15a-d it is the gluon which is attached to both a
light and a heavy quark).

\paragraph{Diagrams 17a-d:} We now turn to diagrams 17a-d. In this
case the contribution from the region in which the gluon attached to a
light quark is soft and the other one is hard does not give a singular
contribution. On the other hand, the region in which the gluon attached
at both ends to heavy quarks is soft and the other one is hard does
lead to singular contributions, which do not cancel, but which are
necessary for the validity of the factorization formula. In other words,
this contribution does not lead to a singular term in the hard-scattering 
kernel $T$, but is absorbed into the form-factor. 

\begin{figure}[t]
   \vspace{-3.8cm}
   \hspace*{-0.5cm}
   \epsfysize=27cm
   \epsfxsize=18cm
   \centerline{\epsffile{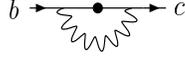}}
   \vspace*{-22cm}
\caption[dummy]{\label{fig:fbd1}\small 
One-loop contribution to the $b\to c$ transition form-factor
(and hence to $F_{B\to D}$)}
\end{figure}

To see this, observe that the momentum of the soft gluon can be 
neglected in the hard subgraph. Hence the four diagrams 17a-d contain
the diagrams in Fig.~\ref{fig6} as hard subgraph. Since, according 
to (\ref{eq:t21}), these give $T^{(1)}$, the contribution of diagrams 
17a-d to the two-loop $B\to D\pi$ amplitude 
$A^{(2)}$ in the hard-soft region is of the form
\begin{equation}
A^{(2)}_{\rm 17a-d, hard-soft} = f_{B\to D}^{(1)}\cdot 
T^{(1)}*\Phi_\pi^{(0)},
\end{equation}
where $f_{B\to D}^{(1)}$ stands for the soft contribution to the 
diagram shown in Fig.~\ref{fig:fbd1}. The factor $f_{B\to D}^{(1)}$ 
contains non-cancelling infrared divergences. 
But this factor is identical to the 
soft contribution to the $B\to D$ form factor $F_{B\to D}^{(1)}$ 
and hence 
\begin{equation}
A^{(2)}_{\rm 17a-d, hard-soft} - F_{B\to D}^{(1)}\cdot 
T^{(1)}*\Phi_\pi^{(0)} = \mbox{ infrared finite}.
\end{equation}
Hence we have recovered (and ``used up'') one of two subtraction 
terms that appear in (\ref{cc1}). Any other non-cancelling divergence
that we may (and must) still find must therefore be cancelled by 
$F_{B\to D}^{(0)}\cdot T^{(1)}*\Phi_\pi^{(1)}$.

\begin{figure}[t]
   \vspace{-3cm}
   \hspace*{-1cm}
   \epsfysize=24cm
   \epsfxsize=16cm
   \centerline{\epsffile{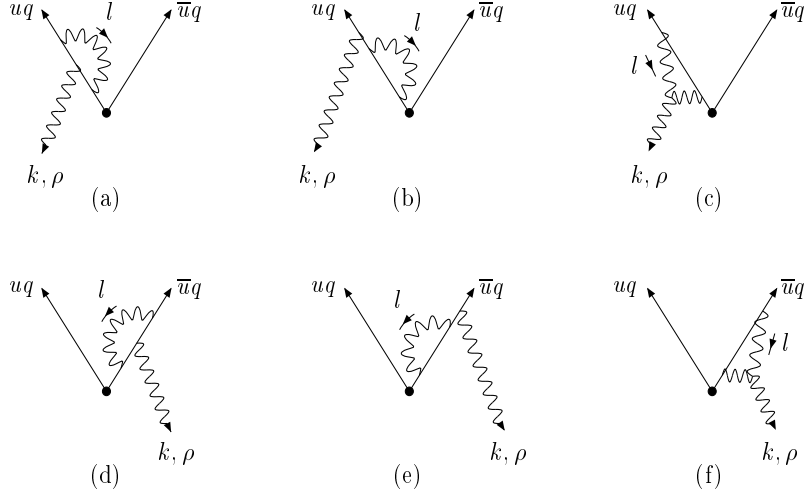}}
   \vspace*{-13.5cm}
\caption[dummy]{\label{fig:inserts}\small 
The six vertex and self-energy insertions on the 
constituents of the pion.}
\end{figure}

\paragraph{Diagrams 8a-d, 12a-d, 13a-d:}
We are now left with diagrams 8a-d, 12a-d and 13a-d representing
vertex and self-energy insertions on the valence quarks in the pion. 
A potentially non-factorizing (infrared divergent) contribution 
arises when the gluon that attaches to the heavy quark line is soft. In
order to facilitate the discussion we draw and label these insertions as
in Fig.~\ref{fig:inserts}. The insertion of Fig.~\ref{fig:inserts}a 
contains a factor
\begin{equation}
\ref{fig:inserts}\mbox{a} = \frac{\gamma^\mu (u\qslash+\lslash)\gamma^\rho
(u\qslash+\kslash+\lslash)\gamma_\mu(u\qslash+\kslash)\Gamma}
{(uq+l)^2\,(uq+k+l)^2\,(uq+k)^2\,l^2}
\simeq\frac{\gamma^\mu (u\qslash+\lslash)\gamma^\rho
(u\qslash+\lslash)\gamma_\mu}     
{\left[(uq+l)^2\right]^2\,l^2}\,\,\frac{\qslash\,\Gamma}{2q\cdot k},
\label{eq:inserta}\end{equation}
where we have used the fact that the components of $k$ are small with
respect to the remaining momenta. This is multiplied by the colour factor
$c_a=-1/2N_c$. In the same region of phase-space the
corresponding factor from the diagram in Fig.~\ref{fig:inserts}(c) 
is $c_c=N_c/2$ times
\begin{eqnarray}
\ref{fig:inserts}\mbox{c}&\simeq& (l_\mu g_{\nu\rho}+l_\nu 
g_{\mu\rho} -2l_\rho g_{\mu\nu})\,
\frac{\gamma^\mu\,(u\qslash+\lslash)\gamma^\nu}{(uq+l)^2\,\left[l^2\right]^2}
\,\frac{\qslash\,\Gamma}{2q\cdot k}\nonumber\\
&=& \frac{2q^\rho\,\Gamma}{\left[l^2\right]^2\,q\cdot k}
-2l_\rho\frac{\gamma^\mu (u\qslash+\lslash)\gamma_\mu}{(uq+l)^2\,
\left[l^2\right]^2}\,\,\frac{\qslash\,\Gamma}{2q\cdot k}.
\label{eq:insertc}\end{eqnarray}
The first term on the right-hand side of (\ref{eq:insertc}) cancels
with the corresponding contribution from diagram~\ref{fig:inserts}f 
and hence we do
not consider it further here.

Finally we turn to the self-energy insertion in Fig.~\ref{fig:inserts}b,
\begin{equation}
\ref{fig:inserts}\mbox{b}=\frac{\gamma^\rho (u\qslash+\kslash)
\left[(u\qslash+\kslash)
\Sigma\right](u\qslash+\kslash)\Gamma}
{\left[(uq+k)\right]^2}
=\frac{\gamma^\rho\Sigma\ (u\qslash+\kslash)\Gamma}{(uq+k)^2},
\label{eq:insertb}\end{equation} 
where the self-energy $\Sigma$ is given by
\begin{equation}
\int_l\frac{\gamma^\mu\,(u\qslash+\kslash+\lslash)\gamma_\mu}
{(uq+k+l)^2\,l^2}=(u\qslash+\kslash)\Sigma.
\label{eq:Sigmadef}\end{equation}
In (\ref{eq:Sigmadef}) $\int_l$ represents the integration over loop 
momentum $l$.
The terms above are to be multiplied by the colour factor $c_b=(N_c^2-1)
/2N_c$. Differentiating both sides of (\ref{eq:Sigmadef}) with
respect to $(uq+k)_\rho$ we obtain:
\begin{equation}
-\int_l\frac{\gamma^\mu\,(u\qslash+\kslash+\lslash)\,\gamma^\rho
\,(u\qslash+\kslash+\lslash)\,\gamma_\mu}
{\left[(uq+k+l)^2\right]^2\,l^2}
=\gamma^\rho\Sigma + 2 (uq+k)^\rho (u\qslash+\kslash)\frac{d\Sigma}
{d(uq+k)^2}.
\label{eq:Sigma}\end{equation}
Now in this paragraph we are only considering the contribution from the
integral over the region of hard $l$. In this case $\Sigma$ is 
analytic in $(u q+k)^2$ and 
does not have a logarithmic singularity in $(uq+k)^2$ at 
small $(uq+k)^2$, so the second term
on the right-hand side of (\ref{eq:Sigma}) is of higher order in
$(uq+k)^2 \ll m_b^2$ 
and hence can be neglected. Inserting the expression (\ref{eq:Sigma}) 
for $\gamma^\rho\Sigma$ into (\ref{eq:insertb}) and 
keeping only the terms which give a singularity
from the infrared region of the integration over $k$ we obtain
\begin{equation}
\ref{fig:inserts}\mbox{b}=-\frac{\gamma^\mu (u\qslash+\lslash)\gamma^\rho
(u\qslash+\lslash)\gamma_\mu}     
{\left[(uq+l)^2\right]^2\,l^2}\,\,\frac{\qslash\,\Gamma}{2q\cdot k},
\label{eq:insertb2}\end{equation}
which apart from the colour factor is minus the contribution from the diagram
of Fig.~\ref{fig:inserts}a as expected from the Ward identity.

Thus we have two different integrals over $l$ to consider. The first is the
one which appears in diagrams \ref{fig:inserts}a and \ref{fig:inserts}b:
\begin{equation}
J_1^\rho\equiv\int_l\frac{\gamma^\mu(u\qslash+\lslash)
\gamma^\rho(u\qslash+\lslash)\gamma_\mu}
{\left[(uq+l)^2\right]^2\,l^2}=A_1\,\gamma^\rho+A_2\,\qslash\,q^\rho,
\label{eq:j1def}\end{equation}
and the second comes from diagram \ref{fig:inserts}c:
\begin{equation}
J_2^\rho\equiv\int_l\frac{-2l_\rho\,\gamma^\mu (u\qslash+\lslash)\gamma_\mu}
{(uq+l)^2\,\left[l^2\right]^2}=B_1\,\gamma^\rho+B_2\,\qslash\,q^\rho.
\label{eq:j2def}\end{equation}
Since the $\qslash\,q^\rho$ gives zero when multiplied by
$\qslash\,\Gamma/(2 q\cdot k)$, the total contribution from 
21a-c (up to the term in (\ref{eq:insertc}), which cancels with 21f) is
\begin{equation}
21\mbox{a}+21\mbox{b}+21\mbox{c(part)}
=\left[(c_a-c_b) A_1+c_c B_1\right]\,\frac{\qslash\,\Gamma}{2q\cdot k}.
\end{equation}
But multiplying $J_i^\rho$ by $\gamma_\rho$ and using the on-shell 
condition and $q^2=0$ gives
\begin{equation}
A_1=B_1=\int_l\frac{1}{l^2\,(uq+l)^2},
\end{equation}
and thus the total contribution is proportional to the combination of colour
factors $c_a-c_b+c_c=0$. 

Similarly the soft-hard singularity from
figures \ref{fig:inserts}d, \ref{fig:inserts}e and the remaining
contribution from \ref{fig:inserts}f cancel. Thus we have demonstrated
the cancellation of the singularities from the soft-hard region of phase
space in each of the sets of six diagrams \{8a, 8c, 12a, 12b, 13a, 13c\}
and \{8b, 8d, 12c, 12d, 13b, 13d\}. This completes our proof that the
contributions from this region of phase-space satisfy the factorization
formula.

\subsection{The collinear-collinear contribution}
\label{subsec:colcol}

We now consider the (logarithmic) singularities from the region of phase
space in which the momenta in both loops are collinear with the pion's
momentum. Let $k_i$ be the momentum of a gluon collinear with the pion,
and write it in terms of Sudakov (light-cone) variables,
\begin{equation}
k_i=\alpha_i q + \beta_i \bar q + k_{\perp,i}, 
\label{eq:ksudakov}\end{equation}
with $\bar q$ a second light-like vector such that $q\cdot\bar q\sim 
m_b^2$ (e.g.\ in a frame in which $q=E\,(1,0,0,1)$, 
where $E$ is the energy of the pion in the B-meson rest-frame,
it is convenient to define $\bar q=E\,(1,0,0,-1))$. 
$k_{\perp,i}$ contains only components
which are perpendicular to both $q$ and $\bar q$. By ``collinear" we
mean that $\alpha_i\sim 1,\,\beta_i\sim \lambda^2, \,k_{\perp,i}
\sim\lambda m_b$, with $\lambda\sim\Lambda_{\rm QCD }/m_b\ll 1$, 
and hence $k^2\sim \lambda^2 m_b^2$. The loop integration measure 
scales as $d^4k_i\sim \lambda^4 m_b^4$ for collinear $k_i$. 

\begin{figure}[t]
   \vspace{-4.1cm}
\hspace*{-1.2cm}
   \epsfysize=24cm
   \epsfxsize=16cm
   \centerline{\epsffile{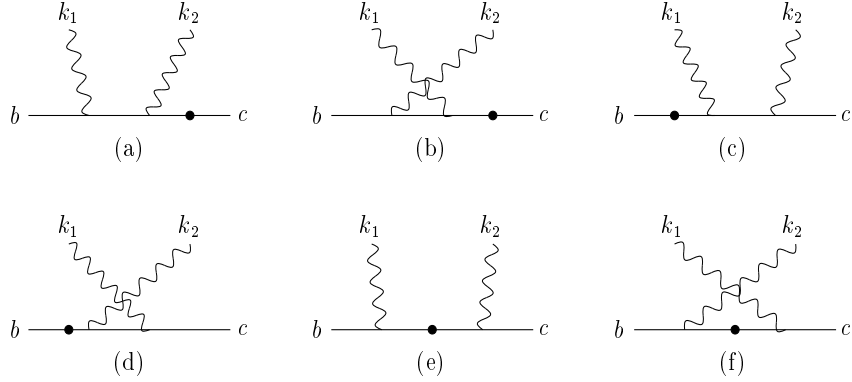}}
   \vspace*{-15cm}
\caption[dummy]{\label{fig:colcol}\small 
The six insertions of two gluons onto the heavy-quark
propagators. The momenta $k_i$ are parallel to the pion's momentum $q$
and so we write $k_i\approx \alpha_i q$. The black circle represents the
heavy-quark weak current.}
\end{figure}

\paragraph{Diagrams 1a-7b:}
We start with diagrams 1a-7b, and consider the six ways of attaching
two gluons onto the heavy-quark propagators as illustrated in 
Fig.~\ref{fig:colcol}.
The heavy quarks are off-shell in this region of phase space 
and hence their propagators are not singular, which
significantly simplifies the discussion. Furthermore, the singularity 
is at most logarithmic and hence we may approximate the collinear 
momenta by $k_i\approx\alpha_i\,q$ ($i=1,2$). The diagrams in 
Fig.~\ref{fig:colcol} have two Lorentz indices $\mu_1$ and $\mu_2$ for the 
external gluon lines. After approximating $k_i\approx\alpha_i\,q$, 
the remainder of a diagram, i.e.\ the light quark lines with their
couplings to the two gluons, depends only on the vector $q$, hence 
the open indices in the diagrams of Fig.~\ref{fig:colcol} must be 
contracted with $q^{\mu_1} q^{\mu_2}$. Including this factor, we
obtain by repeated use of the on-shell condition for the $b$ quark:
\begin{eqnarray}
\textrm{Fig.}~\ref{fig:colcol}\mbox{a} &=& \frac{1}{\alpha_1\alpha_2}\,\Gamma\,
\frac{1}{\pslash+\alpha_1\qslash+\alpha_2\qslash-m_b}\,\alpha_2\qslash\,
\frac{1}{\pslash+\alpha_1\qslash-m_b}\,\alpha_1\qslash
\nonumber\\
&=& \frac{1}{\alpha_1\alpha_2}\,\Gamma\,
\frac{1}{\pslash+\alpha_1\qslash+\alpha_2\qslash-m_b}\,\alpha_2\qslash
\,\,= \,\,\frac{1}{\alpha_1(\alpha_1+\alpha_2)}\,\Gamma
\end{eqnarray}
Similar simplifications occur for the other five terms shown in 
Fig.~\ref{fig:colcol}. The relative factors of the six insertions in 
Fig.~\ref{fig:colcol} are:
\begin{eqnarray}
\textrm{Fig.}~\ref{fig:colcol}\mbox{a} & = & 
\frac{1}{\alpha_1\,(\alpha_1+\alpha_2)}\, C_1
\qquad
\textrm{Fig.}~\ref{fig:colcol}\mbox{b} \,\,= \,\,
\frac{1}{\alpha_2\,(\alpha_1+\alpha_2)}\, C_2
\label{eq:cca}\\
\textrm{Fig.}~\ref{fig:colcol}\mbox{c} & = & 
\frac{1}{\alpha_2\,(\alpha_1+\alpha_2)}\, C_1
\qquad
\textrm{Fig.}~\ref{fig:colcol}\mbox{d} \,\, = \,\, 
\frac{1}{\alpha_1\,(\alpha_1+\alpha_2)}\, C_2
\label{eq:ccc}\\ 
\textrm{Fig.}~\ref{fig:colcol}\mbox{e} & = & 
-\frac{1}{\alpha_1\,\alpha_2}\, C_2
\qquad\qquad\,\,
\textrm{Fig.}~\ref{fig:colcol}\mbox{f} \,\, = \,\, 
-\frac{1}{\alpha_1\,\alpha_2}\, C_1.
\label{eq:ccf}\end{eqnarray}
$C_1$ and $C_2$ are the two
distinct colour factors, $C_1=\textrm{Tr}(T\,T_2\,T_1)$
($C_2=\textrm{Tr}(T\,T_1\,T_2$)) where $T,\,T_1$ and $T_2$
are the colour matrices of the heavy-quark weak current and the gluons
with momenta $k_1$ and $k_2$, respectively. Adding up the contributions
in (\ref{eq:cca})-(\ref{eq:ccf}) we see that the coefficients of
both $C_1$ and $C_2$ vanish. In this way the singularities from the
collinear-collinear region of phase space in the 18 diagrams 1a-7b
cancel in each of the following sets of six diagrams \{1a, 2a, 2d,
1d, 5a, 6a\}, \{2c, 1c, 1b, 2b, 5b, 6b\} and \{3a, 4a, 3b, 4b, 7a,
7b\}.

\paragraph{Diagrams 10a-11b:} A similar cancellation occurs for the the
six diagrams 10a-11b, which are also individually logarithmically
divergent. Indeed the cancellation occurs in each of the two sets of
three diagrams \{10a, 10d, 11a\} and \{10c, 10b, 11b\}.

\paragraph{Diagrams 8a-9b and 12a-15d:} The diagrams 8a-9b and
12a-15d require special attention as they each contain five light
propagators (either light quark or gluon) which are collinear with the
pion and these diagrams are hence potentially quadratically divergent.
In each case, however, the numerators are suppressed in 
the collinear-collinear
region, so that the leading divergence is logarithmic. The cancellation
of this singularity occurs between pairs of diagrams (such as 8a and 8d
or 14a and 14d) in the same way as in the one-loop graphs.

As an example consider the numerator that comes from the light quark 
lines in diagram 12a,
\begin{equation}
N_{12}\equiv \gamma^\mu (u\qslash+\!\not\!k_2)\gamma^\rho 
(u\qslash+\!\not\!k_1+\!\not\!k_2)\gamma_\mu (u\qslash+\!\not\!k_1) 
\,\Gamma \sim m_b^3.
\end{equation}
The scaling estimate is the naive scaling estimate, 
according to which diagram 12a is 
quadratically divergent, not taking into 
account cancellations.
Inserting the Sudakov decomposition (\ref{eq:ksudakov}) and using 
only the on-shell condition and $q^2=0$, this transforms into
\begin{eqnarray}
N_{12} &=& 
(u+\alpha_2) \,\gamma^\mu \!\qslash \,\gamma^\rho 
(\not\!k_{\perp,1}+\!\not\!k_{\perp,2})\gamma_\mu \!\not\!k_{\perp,1}
\,\Gamma
+ 
(u+\alpha_1+\alpha_2) \,\gamma^\mu \!\not\!k_{\perp,2}\gamma^\rho 
\qslash \,\gamma_\mu \!\not\!k_{\perp,1}
\,\Gamma
\nonumber\\
&&\,+ \,(u+\alpha_1) \,\gamma^\mu \!\not\!k_{\perp,2}\gamma^\rho 
(\not\!k_{\perp,1}+\!\not\!k_{\perp,2})\gamma_\mu  
\qslash\,\Gamma \sim   \lambda^2 m^3_b
\end{eqnarray}
The two leading powers have cancelled, so that the divergence is at 
most logarithmic. We can now use that $q\cdot k_{\perp,1}=
q\cdot k_{\perp,2}=0$ to commute $\qslash$ to the left. The 
result is that 
\begin{equation}
N_{12}\sim q^\rho.
\end{equation}
But the logarithmically divergent terms proportional to $q^\rho$ 
(where $\rho$ is the index that couples the gluon in diagram 12a to 
the heavy-quark line) cancel pairwise according to the one-loop 
collinear cancellation mechanism discussed in
Sect.~\ref{oneloopcancel}. Similar manipulations apply to all 
other diagrams 8a-9b and 12b-15d.

\paragraph{Diagrams 16a-19d:} The diagrams 16a-19d are not divergent
in the collinear-collinear region and we do not consider them further in
this subsection. They do however have a collinear divergence, when the
gluon attached to one of the constituents of the pion is collinear with
the pion's momentum, while the other momentum is hard. 
These divergences will be considered when
considering the collinear-hard region of phase-space in
Sect.~\ref{subsec:hardcol} below.

This concludes our demonstration that the singularities in diagrams
from the collinear-collinear region of phase space cancel and 
hence do not invalidate the factorization formula.

\subsection{The soft-collinear contributions}
\label{subsec:softcol}

We now present, in some detail, a demonstration that the singularities
in the soft-collinear region of phase space do not invalidate the
factorization formula, i.e.\ that soft-collinear singularities cancel 
in the sum of all diagrams.

\subsubsection{Introduction}

We start by explaining more carefully what we mean by both ``soft" and
``collinear". Let $l$ be the momentum of a gluon collinear with the pion;
we write it in terms of Sudakov variables as in (\ref{eq:ksudakov}), 
$l=\alpha q+\beta\bar{q}+l_\perp$. The 
scaling of the components of $l$ is as described at the beginning of 
Sect.~\ref{subsec:colcol}. 

Consider now a soft gluon with momentum $k$, by which we mean that all
components of $k$ are much smaller than $m_b$. It is now important to 
consider what we mean by ``much smaller". To illustrate this point
imagine that we have a diagram with a propagator $1/(uq+l)^2$, where $l$
is collinear to the pion's momentum $q$ as described above. By the power
counting introduced above we find that this propagator is of order 
$1/(\lambda^2\, m_b^2)$. Now if in addition we have the radiation
of a soft gluon, there may be a propagator such as $1/(uq+l+k)^2$.
If the components of $k$ are of order $\lambda m_b$ then
$1/(uq+l+k)^2\sim 1/(\lambda\,m^2_b)$ and not of order 
$1/(\lambda^2\, m_b^2)$. In this case $k+l$ is not collinear 
(as defined above), neither is it soft. Indeed in many
of the diagrams there are propagators whose denominators are linear
combinations of $q\cdot k,\,q\cdot l$ and $l^2$. Since 
$q\cdot l, l^2\sim \lambda^2 m_b^2$, it is therefore also
necessary to consider the region of phase space in which $q\cdot
k\sim \lambda^2m_b^2$, which implies that $k\sim \lambda^2 m_b$. 
We therefore distinguish the regions of
phase-space in which the components of $k$ are of order $\lambda m_b$
(which we call ``soft") and where they are of order $\lambda^2 m_b$ 
(which we call ``supersoft"). When $l$ is collinear and $k$ is 
supersoft, then $k+l$ satisfies the scaling conditions for a 
collinear momentum. This means that when a loop momentum is 
collinear, one can classify the set of collinear lines of a 
graph in terms of one-particle-irreducible subgraphs, just as 
in the case of hard subgraphs. As far as we know, the only previous 
discussion of the distinction of soft and supersoft in 
the context of QCD factorization ``proofs'' is in \cite{CFS97}. If 
we think of $\lambda$ as being of order $\Lambda_{\rm QCD}/m_b$, 
one may consider the supersoft region as unphysical, since 
non-perturbative modifications of the quark and gluon propagators 
would prevent the momentum from becoming supersoft. On the other 
hand, the structure of denominators clearly indicates that 
the infrared singularities of Feynman integrals in 
dimensional regularization originate from supersoft momentum.

Table~\ref{tab:pc} contains a summary of the rules for determining the
order of the divergence from the soft-collinear and supersoft-collinear
regions of phase space. We imagine that the gluons have momenta $k$ and
$l$ (and $k\pm l$ if the diagram contains a third gluon) and each
propagator scales like $1/\lambda^\delta$ in these regions. The table
contains the powers $\delta$ for the collinear gluon with momentum $l$,
the soft (S) or supersoft (SS) gluon with momentum $k$, the gluon with
momentum $k\pm l$ and the light and heavy quarks which radiate these
gluons. Thus, for example, if there is a light quark propagator $1/(uq + k +
l)^2$ in a diagram, then it scales like $1/\lambda$ for $k$ soft and
$1/\lambda^2$ for $k$ supersoft. For the heavy quarks
$1/[(p_{b,c}+k+l)^2-m^2_{b,c}]$ 
scales like $\lambda^0$ (where $p_{b,c}$ are the
momenta of the $b$ and $c$ quarks respectively).

\begin{table}
\caption{\label{tab:pc} The power $\delta$ corresponding
to the $1/\lambda^\delta$ scaling behaviour of each quark and gluon propagator
in the soft-collinear (S) and supersoft-collinear (SS) regions of phase space.
$l$ is the momentum collinear to the pion's momentum $q$ and $k$ is soft
or supersoft.}
\vspace*{0.2cm}
\begin{center}
\begin{tabular}{|cc|ccc|}\hline
&&&&\\[-0.4cm]
& & Gluon & Light quark & Heavy quark \\
&&&&\\[-0.4cm]
\hline
&&&&\\[-0.4cm]
& S &2&1&1\\
\raisebox{8pt}[0cm][0cm]{$k$}& SS & 4 & 2 & 2 \\
&&&&\\[-0.4cm]
\hline
&&&&\\[-0.4cm]
$l$ & & 2 & 2 & 0\\
&&&&\\[-0.4cm]
\hline
&&&&\\[-0.4cm]
&S& 1 & 1& 0 \\
\raisebox{8pt}[0cm][0cm]{$k\pm l$}& SS & 2 & 2 & 0\\[0.1cm]
\hline 
\end{tabular}
\end{center}
\end{table}

To illustrate the use of the entries in Table~\ref{tab:pc} consider 
diagram 1a in Fig.~\ref{fig:diagsone} and start with the region in which
one gluon is supersoft and the other is collinear. The phase space for
this region is of $O(\lambda^{12})$. One possibility is that the
outer gluon is supersoft and the inner one is collinear in which
case we obtain factors from the propagators proportional to:
\begin{equation}
\begin{array}{cccccc}
\displaystyle 
\frac{1}{\lambda^4} & \displaystyle \frac{1}{\lambda^2}&
\displaystyle \frac{1}{\lambda^2}&\displaystyle 
\frac{1}{\lambda^0}&\displaystyle \frac{1}{\lambda^2}&
\displaystyle \frac{1}{\lambda^2},\\[0.4cm]
\textrm{supersoft}&\textrm{collinear} &\textrm{outer}
&\textrm{inner}&\textrm{outer}&\textrm{inner}\\
\textrm{gluon}&\textrm{gluon}&\textrm{heavy quark}&
\textrm{heavy quark}& \textrm{light quark}
&\textrm{light quark}
\end{array}
\label{eq:koutlin}\end{equation}
which combine to give $1/\lambda^{12}$. Thus we have a logarithmic
divergence from this region. If instead the outer gluon is collinear
and the inner one is supersoft, then the outer heavy-quark
propagator scales like $1/\lambda^0$ so that the combined scaling factor
for all six propagators is $1/\lambda^{10}$ and there is no divergence.

Now consider the region in which $k$ is soft rather than supersoft, and
again we illustrate the power counting in diagram 1a of
Fig.~\ref{fig:diagsone}. In this case the phase space is of
$O(\lambda^8)$ and if the outer gluon is the soft one then the
propagators scale as:
\begin{equation}
\begin{array}{cccccc}
\displaystyle \frac{1}{\lambda^2} & \displaystyle \frac{1}{\lambda^2}&
\displaystyle \frac{1}{\lambda}&\displaystyle \frac{1}{\lambda^0}
&\displaystyle \frac{1}{\lambda}&
\displaystyle \frac{1}{\lambda},\\[0.4cm]
\textrm{soft}&\textrm{collinear}&\textrm{outer} 
&\textrm{inner}&\textrm{outer}&\textrm{inner}\\
\textrm{gluon}&\textrm{gluon}&\textrm{heavy quark}&
\textrm{heavy quark}&\textrm{light quark}
&\textrm{light quark}
\end{array}
\label{eq:koutlin2}\end{equation}
which gives a combined factor of $O(1/\lambda^7)$. We therefore have no
divergence from this region of phase-space (nor from the region in which
the inner gluon is the soft one). This is not the case for all the
diagrams, however, as we shall demonstrate below.

We now consider the supersoft-collinear and soft-collinear regions in turn.

\subsubsection{The supersoft-collinear region}
\label{subsubsec:sscol}

\begin{figure}[t]
   \vspace{-3.6cm}
   \epsfysize=24cm
   \epsfxsize=16cm
   \centerline{\epsffile{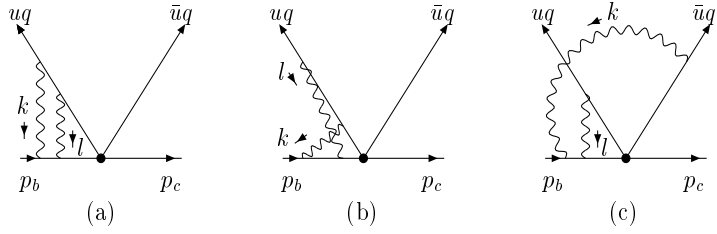}}
   \vspace*{-17.2cm}
\caption[dummy]{\label{fig:colss1}\small 
Supersoft-collinear contributions to diagrams 1a [(a)],
2a [(b)] and 4a [(c)]. The gluon labelled by $l$ is collinear
with the pion's momentum $q$ and the one labelled by $k$ is supersoft.}
\end{figure}

\paragraph{Diagrams 1a-7b} We start by considering the 18 diagrams
1a-7b. We place them into 4 groups \{1a, 5a; 2a, 6a; 4a, 7b\},
\{1b, 5b; 2b, 6b; 4b, 7b\}, \{1c, 6b; 2c, 5b; 3a, 7a\} and 
\{1d, 6a; 2d, 5a; 3b, 7a\}.
We label these groups I-IV, and only consider explicitly the
cancellation within group I (which will require contributions from
additional graphs). The cancellations within groups II-IV proceeds
analogously to that within group I.  Diagrams 5a-7b have two
contributions (where the two gluons are supersoft-collinear or
collinear-supersoft respectively) and hence appear twice in the lists.
Within each group, we identify three pairs (separated by the
semicolons). Consider the first pair \{1a, 5a\} and the region of phase
space for which the gluon with the outer vertex on the light quark
propagator is the supersoft one (see e.g.\ Fig.~\ref{fig:colss1}a). For
the insertion of the singlet-singlet operator $O_0$ the singularities
from these diagrams cancel analogously to the cancellation of collinear
divergences at one-loop level. A similar cancellation occurs between the
singularities in all the pairs of diagrams for the insertion of the 
operator $O_0$.

For the insertion of the octet-octet operator $O_8$ the cancellation
of the singularities is not complete within the set of diagrams
1a-7b. For illustration consider the first group of diagrams
\{1a, 5a; 2a, 6a; 4a, 7b\}. Since, apart from the colour factors, the
contributions of the two diagrams in each of the three pairs are equal
and opposite, and using the colour factors tabulated in
Table~\ref{tab:colour} the total contribution from this group of diagrams
can be written in the form
\begin{equation}
\frac{C_F}{2}\,\frac{N_c}{2}\,\left(\textrm{Diag1a}
-\textrm{Diag2a}+\textrm{Diag4a}\right).
\label{eq:group11}\end{equation}
We therefore need to look at diagrams 1a, 2a and 4a in some more detail
(see Fig.~\ref{fig:colss1}). In this region of phase-space it is straightforward to
establish that the integrands are proportional to
\begin{eqnarray}
\textrm{Diag1a}& \equiv & \frac{2(u+\alpha)}{\alpha}\ \frac{p_b\cdot q}
{(p_b\cdot k)\,(q\cdot k)}\ \frac{1}{k^2\,l^2\,(uq+k+l)^2}\label{eq:css1a}\\ 
\textrm{Diag2a}& \equiv & \frac{4(u+\alpha)^2}{\alpha}\ \frac{p_b\cdot q}
{p_b\cdot k}\ \frac{1}{k^2\,l^2\,(uq+l)^2\,(uq+k+l)^2}\label{eq:css2a}\\ 
\textrm{Diag4a}& \equiv &
-\frac{2(u+\alpha)}{\alpha}\ \frac{p_b\cdot q}
{(p_b\cdot k)\,(q\cdot k)}\ \frac{1}{k^2\,l^2\,(uq+l)^2}.\label{eq:css4a}
\end{eqnarray}
Now using
\begin{equation}
(uq+k+l)^2-(uq+l)^2\simeq 2(u+\alpha)q\cdot k
\end{equation}
we see that $\mbox{Diag2a} = - (\mbox{Diag1a}+\mbox{Diag4a})$
and that the non-cancelling contribution is proportional to the integral over
\begin{equation}
A_1=\frac{N_c}{2}\ \frac{4(u+\alpha)}{\alpha}\ \frac{p_b\cdot q}{(p_b\cdot k)
(q\cdot k)k^2l^2}\left(\frac{1}{(uq+k+l)^2}-\frac{1}{(uq+l)^2}\right),
\end{equation}
where we have included the colour 
factor. Note that here and in the following we do
not include explicitly the colour factor $C_F/2$ present in one-loop 
diagrams.

\paragraph{Diagrams 16a-19d:}
We now consider diagrams 16a-19d in the supersoft-collinear region. By
inspection (power counting) we readily deduce that diagrams 16a-d and
19a-d do not give a singular contribution in this region of phase
space, regardless of which of the  two gluons is supersoft and which is
collinear.

Diagram 17a does have a singular contribution from the region in which
the gluon which is attached (at one end) to a light quark is collinear
and the one which is attached at both ends to heavy quarks is supersoft.
This singularity is cancelled by the corresponding one in diagram 17c by
the same mechanism by which collinear divergences cancel at one-loop
order (diagrams 17a and 17c have the same colour factor). Similarly
the singular contributions from diagrams 17b and 17d cancel.

Diagram 18a has a singular contribution from the region in which the
gluon which is attached at one end to a light quark is supersoft and
the one which is attached at both ends to the $b$ quark is collinear.
This singularity cancels against the corresponding one in diagram 18b,
by the same mechanism by which soft divergences cancel at one-loop
order. Similarly the singularities in diagrams 18c and 18d cancel.

Thus there is no residual singular contribution from diagrams 16a-19d.

\begin{figure}[t]
   \vspace{-4.4cm}
\hspace*{-1.3cm}
   \epsfysize=24cm
   \epsfxsize=16cm
   \centerline{\epsffile{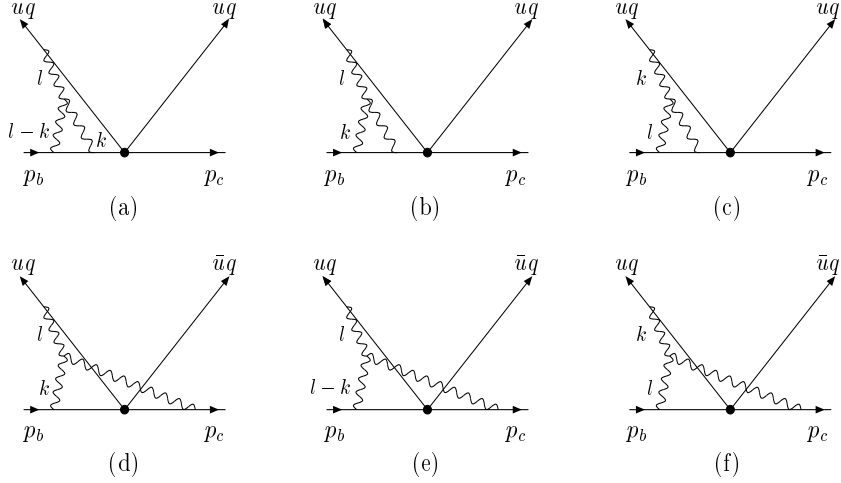}}
   \vspace*{-13.1cm}
\caption[dummy]{\label{fig:tena}\small 
The three possible momentum routings in the supersoft-collinear
region for diagrams 10a [(a)-(c)] and 11a [(d)-(f)]. $l$ represents
the collinear momentum and $k$ the supersoft one.}
\end{figure}

\paragraph{Diagrams 10a-11b:} We now turn to the six diagrams 10a-11b.
Consider diagram 10a and the three collinear-supersoft contributions as
labelled in Fig.~\ref{fig:tena}. The propagators in the region of
Fig.~\ref{fig:tena}a combine to give a factor of $\lambda^{-10}$ and
hence no singularity. In the regions of Fig.~\ref{fig:tena}b and 
Fig.~\ref{fig:tena}c, however, 
they give factors of $\lambda^{-12}$ and hence do yield
singular contributions. The singular contribution from the region in
Fig.~\ref{fig:tena}c can readily be seen to cancel the corresponding
contribution from diagram 10c by the standard mechanism of one-loop soft
cancellations. This leaves us the contribution of Fig.~\ref{fig:tena}b,
which is straightforward to evaluate giving a
term proportional to the integral over
\begin{equation}
\textrm{Diag10a}\equiv -2(u+\alpha)\ \frac{p_b\cdot q}{p_b\cdot k}\ 
\frac{1}{k^2l^2(k+l)^2(uq+k+l)^2},
\label{eq:colss10a}\end{equation}
where we have used a shift of integration variable $l\to l+k$.
The corresponding contributions from diagrams 10b, 10c and 10d
contribute to groups II, III and IV respectively.

Consider now diagram 11a and consider the three regions of phase-space
as labelled in Fig.~\ref{fig:tena}d-f. The region in
Fig.~\ref{fig:tena}f does not give a singular contribution, whereas the
other two regions do. We consider the contribution from
Fig.~\ref{fig:tena}d here, that from Fig.~\ref{fig:tena}e contributes
to group IV. Similarly the two singular contributions from diagram 11b
contribute to groups II and III. The contribution from the region in
Fig.~\ref{fig:tena}d can readily be evaluated and is proportional to
the integral over
\begin{equation}
\textrm{Diag11a(part)}\equiv 2(u+\alpha)\ \frac{p_b\cdot q}{p_b\cdot k}\
\frac{1}{k^2l^2(k+l)^2(uq+k+l)^2},
\label{eq:colss11a}\end{equation}
where we repeat that it is only the part of diagram 11a which
contributes to group I which is being evaluated.

Putting in the colour factor of $N_c/2$ for diagram 10a ($-N_c/2$ for 
diagram 11a) and summing the contributions from (\ref{eq:colss10a}) and
(\ref{eq:colss11a}), we obtain a singular contribution to group I of
\begin{equation}
A_2=-\frac{N_c}{2}\ 4(u+\alpha)\ \frac{p_b\cdot q}{p_b\cdot k}\
\frac{1}{k^2l^2(k+l)^2(uq+k+l)^2}.
\label{eq:colss10a11a}\end{equation}
Next we note that $(k+l)^2\approx l^2+2\alpha q\cdot k$ so that
\begin{equation}
1=\frac{(k+l)^2}{2\alpha q\cdot k} - \frac{l^2}{2\alpha q\cdot k}  
\label{eq:starstar}\end{equation}
and
\begin{eqnarray}
A_2&=&\frac{N_c}{2}\,(-2)\,\frac{(u+\alpha)}{\alpha}\ \frac{p_b\cdot q}
{(p_b\cdot k)(q\cdot k)k^2l^2}\ \left(\frac{1}{(uq+k+l)^2}-
\frac{1}{(uq+l)^2}\right) \label{eq:a2}\\
&=&-\frac12 \,A_1
\label{eq:a1}\end{eqnarray}
where we have made the change of variables $l\to l-k$ in the second term
of (\ref{eq:a2}). (Since $l-k$ remains collinear for supersoft $k$, 
this change of variables is permitted for supersoft $k$, 
but would not be permitted for soft $k$.)

\paragraph{Diagrams 8a-9b and 12a-15d:} Finally we consider the
remaining diagrams in which there is a single gluon attached
(at one end) to one of the heavy quarks. 
Power counting (including the momentum factor in the numerator) shows
that there is no singularity when this gluon is collinear.
Thus we can restrict our attention to
the region in which the gluon which is attached to the heavy quark
is supersoft.
Note that in this case the superficial degree of divergence is
quadratic. However, taking the numerator factors into account,
the actual divergence is again only
logarithmic as can be seen from the expressions below.

Consider the contributions to diagrams 8a, 12a and 13a from this region
of phase space. This contribution can be written in the form:
\begin{equation}
A_3=\frac{N_c}{2}\,\textrm{Diag8a}
+ \left(\frac{-1}{2N_c}\right)\,\textrm{Diag12a}
+\left(\frac{N_c}{2}-\frac{1}{2N_c}\right)\,\textrm{Diag13a},
\label{eq:a3}\end{equation}
where we have explicitly exhibited the colour factors. Straightforward
evaluation of the diagrams shows that they are respectively proportional to
the integrals over
\begin{eqnarray}
\textrm{Diag8a}&\equiv&\frac{4\alpha q\cdot(k+l)+
2(u+\alpha)q\cdot k + 2(uq+l)^2}{(uq+k+l)^2k^2l^2(k+l)^2}\
\frac{p_b\cdot q}{(p_b\cdot k)(q\cdot k)}\label{eq:colss8a}\\
\textrm{Diag12a}&\equiv&\frac{2l_\perp^2}{(uq+l)^2(uq+k+l)^2
k^2l^2}\ \frac{p_b\cdot q}{(p_b\cdot k)(q\cdot k)}
\label{eq:colss12a}\\
\textrm{Diag13a}&\equiv&\frac{-2q\cdot (l+k)}{(uq+l+k)^2k^2l^2}
\ \frac{p_b\cdot q}{(p_b\cdot k)(q\cdot k)^2}\label{eq:colss13a},
\end{eqnarray}
where in (\ref{eq:colss12a}) we use a metric such that $l_\perp^2$
is negative.

Using $l_\perp^2=(uq+l)^2-2(u+\alpha)q\cdot l$, $2(u+\alpha)q\cdot k
=(uq+k+l)^2-(uq+l)^2$, and dropping a term that vanishes after
integration over $k$ because it is antisymmetric in $k$, one 
readily finds that $\mbox{Diag12a}=-\mbox{Diag13a}$, so that
(\ref{eq:a3}) becomes
\begin{equation}
A_3=\frac{N_c}{2}(\textrm{Diag8a}-\textrm{Diag12a}).
\label{eq:a3p}\end{equation}
Similarly using (\ref{eq:starstar}) we find that the first term in the
numerator of (\ref{eq:colss8a}) for diagram 8a gives the same contribution
as diagram 12a, so that $A_3$ is the sum of the contributions of the
second and third terms in the numerator of (\ref{eq:colss8a}):
\begin{eqnarray}
A_3&=&\frac{N_c}{2}\ \frac{2(u+\alpha)q\cdot k + 2(uq+l)^2}
{(uq+k+l)^2k^2l^2(k+l)^2}\ \frac{p_b\cdot q}{(p_b\cdot k)
(q\cdot k)}\\
& = & \frac{N_c}{2}\ \frac{2(uq+k+l)^2-2(u+\alpha)q\cdot k}
{(uq+k+l)^2k^2l^2(k+l)^2}\ \frac{p_b\cdot q}{(p_b\cdot k)
(q\cdot k)}\\
&=&\frac{N_c}{2}\ \left\{\frac{2}
{k^2l^2(k+l)^2}
-\frac{2(u+\alpha)q\cdot k}{(uq+k+l)^2k^2l^2(k+l)^2}\right\}
\ \frac{p_b\cdot q}{(p_b\cdot k)
(q\cdot k)}.\label{eq:a3pp}
\end{eqnarray}
The first term in the braces in (\ref{eq:a3pp}) is independent
of $u$. It cancels against the corresponding term in diagram 8c by the
standard one-loop soft cancellation mechanism and so we drop it from 
the definition of $A_3$. We are left with the
second term which (by direct comparison with the expression in
(\ref{eq:colss10a11a})\,) gives
\begin{equation}
A_3=\frac12 \,A_2.
\label{eq:a3f}\end{equation}

We do not exhibit all the corresponding formulae here for diagrams 9a,
14a and 15a but the steps are very similar. There is one subtlety in
that diagram 9a has two contributions, one to group I and the other to
group III. Keeping only the part which has the right structure to be of
group I we find
\begin{eqnarray}
A_4 & = & \frac{N_c}{2}\textrm{Diag9a(part)} 
+ \left(\frac{N_c}{2}-\frac{1}{2 N_c}\right)\textrm{Diag14a} 
+ \left(-\frac{1}{2 N_c}\right)\textrm{Diag15a}
\nonumber\\ 
&=&\frac12 \,A_2 .
\label{eq:a4}\end{eqnarray}

The evaluation of the contributions to the remaining groups of diagrams
proceeds in a similar way. Group II has contributions from diagrams
\{8b, 9b(part), 12d, 13b, 14b, 15c\}, group III from
\{8c, 9a(part), 12b, 13c, 14c, 15b\} and group IV from
\{8d, 9b(part), 12c, 13d, 14d, 15d\}.

\paragraph{Total:}
The total contribution in group I 
is $A_1+A_2+A_3+A_4$, where $A_3=A_4=A_2/2$
(see (\ref{eq:a3f}) and (\ref{eq:a4})) and $A_1=-2A_2$
(see (\ref{eq:a1})). Thus the total contribution in group I is zero. 
The cancellation of all the contributions to groups II, III and IV
occurs in a similar way. Hence we have demonstrated that there are 
no infrared singularities in the supersoft-collinear region of 
phase-space. 

\subsubsection{The soft-collinear region}
\label{subsubsec:scol}

We also discuss briefly how the cancellation of singular contributions
occurs in the soft-collinear case. This case is simpler than 
the supersoft-collinear case, since some terms 
that contribute to singularities for supersoft $k$ give convergent 
integrals for soft $k$, while the converse never occurs. As a
consequence we shall see that the soft-collinear case is covered 
by the line of argument that applied to the supersoft-collinear case. 

\paragraph{Diagrams 1a-7b:}
By power counting we see that the diagram 1a has no singular
contribution in the soft-collinear region, whereas diagrams 2a and 4a
do. Thus
\begin{equation}
A_{1\ \textrm{\scriptsize{soft}}}=\frac{N_c}{2}\,(-\textrm{Diag2a+Diag4a}).
\end{equation}
Using the relation $(uq+k+l)^2\simeq 2(u+\alpha)q\cdot k$ which is valid
in the soft-collinear region, we find that Diag2a = $-$Diag4a and
\begin{equation}
A_{1\ \textrm{\scriptsize{soft}}}=-4\frac{N_c}{2}\frac{u+\alpha}{\alpha}\
\frac{p_b\cdot q}{(p_b\cdot k)(q\cdot k)} \ \frac{1}{k^2l^2(uq+l)^2}.
\label{eq:a1soft}\end{equation}

\paragraph{Diagrams 16a-19d:} Diagrams 16a and 19a are convergent in this
region of phase space just as they were in the supersoft-collinear one.
Diagram 18a, which is singular in the supersoft-collinear region is now
convergent.

Diagram 17a does have a singular contribution again, from the region in
which the gluon which is attached (at one end) to a light quark is
collinear and the one which is attached at both ends to heavy quarks is
soft. As above, this singularity is cancelled by the corresponding one
in diagram 17c by the same mechanism by which collinear divergences
cancel at one-loop order (diagrams 17a and 17c have the same colour
factor). Similarly the singular contributions from diagrams 17b and 17d
cancel. Thus there is no residual singular contribution from diagrams
16a-19d.

\paragraph{Diagrams 10a-11b:}
Consider diagram 10a and the momentum routings exhibited in
Fig.~\ref{fig:tena}. Power counting now indicates that only the routing
of Fig.~\ref{fig:tena}b gives a singular contribution and this can
readily be evaluated:
\begin{equation}
\textrm{Diag10a}=-2\frac{N_c}{2}(u+\alpha)\
\frac{p_b\cdot q}{p_b\cdot k}
\ \frac{1}{k^2l^2(k-l)^2(uq+l)^2}.
\label{eq:tenasoft}\end{equation}

There is a singular contribution from the diagram 11a from the routing
in Fig.~\ref{fig:tena}d (the routing in Fig.~\ref{fig:tena}e is also
singular but contributes to group IV). Explicit evaluation of this
contribution shows that it is equal to that in (\ref{eq:tenasoft}),
so that the combined contribution from diagrams 10a and the part of
diagram 11a which contributes to group I is
\begin{equation}
A_{2\ \textrm{\scriptsize{soft}}}=-4\frac{N_c}{2}(u+\alpha)\
\frac{p_b\cdot q}{p_b\cdot k} \ \frac{1}{k^2l^2(k-l)^2(uq+l)^2}.
\label{eq:a2soft}\end{equation}
In the soft collinear region $(k-l)^2\simeq -2\alpha q\cdot k$, so that
\begin{equation}
A_{2\ \textrm{\scriptsize{soft}}}=-\frac12\,
A_{1\ \textrm{\scriptsize{soft}}}
\label{eq:a2a1soft}\end{equation}
as in the supersoft-collinear case.

\begin{figure}[t]
   \vspace{-3.8cm}
   \epsfysize=24cm
   \epsfxsize=16cm
   \centerline{\epsffile{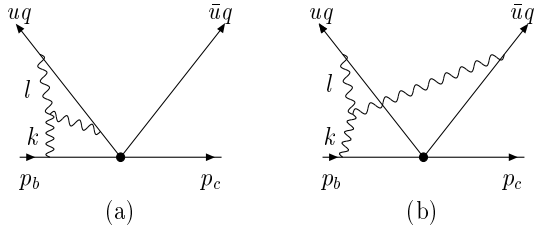}}
   \vspace*{-17.3cm}
\caption[dummy]{\label{fig:eighta}\small 
Momentum routings for diagrams 8a and 9a which are considered
in the text for $l$ collinear to the pion and $k$ soft.}
\end{figure}

\paragraph{Diagrams 8a-9b and 12a-15d:} We start by considering
diagrams 8a, 12a and 13a in the soft collinear region. In general,
and for 13a in particular,
when the
gluon which is attached to a heavy quark is collinear, the
cancellation of the corresponding singularity proceeds in an analogous
way to the cancellation of one-loop collinear divergences. 
In fact for some diagrams (such as 8a or 12a) the potentially
divergent contribution vanishes by itself.
This leaves
us to consider the contributions when this gluon is soft. Although by
power counting it may appear that diagrams 12a and 13a are
logarithmically or linearly divergent, the numerators of these
diagrams vanish sufficiently quickly to render them finite. This
leaves us with the singularity in diagram 8a from the region sketched
in Fig.~\ref{fig:eighta}a which has to be evaluated and is found to be
\begin{equation}
\textrm{Diag8a}=-2\frac{N_c}{2}(u+\alpha)\
\frac{p_b\cdot q}{p_b\cdot k}
\ \frac{1}{k^2l^2(k-l)^2(uq+l)^2}.
\label{eq:eightasoft}\end{equation}
Therefore we have
\begin{equation}
A_{3\ \textrm{\scriptsize{soft}}}=\frac12\,
A_{2\ \textrm{\scriptsize{soft}}}
\label{eq:a3a2soft}\end{equation}
as before.

Finally we have to consider diagrams 14a, 15a and the part of diagram
9a which contributes to group I. The contribution to diagram 14a from
the region in which the gluon which is attached to a heavy quark is
soft, is superficially quadratically divergent, but the numerator
reduces the divergence to a logarithmic one. This singularity is
cancelled by the analogous one in diagram 14c by the standard
mechanism for the cancellation of soft singularities at one-loop
order. The remaining terms in diagram 14a are finite. Diagram 15a is
superficially linearly divergent, but the numerators render it
finite. This leaves the contribution from the diagram 9a to group I,
as sketched in Fig.~\ref{fig:eighta}b. This can be evaluated and is
found to be
\begin{equation}
\textrm{Diag6a}=-4\frac{N_c}{2}(u+\alpha)(\bar u-\alpha)\
\frac{(p_b\cdot q)(q\cdot k)}{p_b\cdot k}
\ \frac{1}{k^2l^2(k-l)^2(uq+l)^2(\bar uq -l+k)^2}.
\label{eq:sixasoft}\end{equation}
But in this region of phase space $(\bar uq-l+k)^2\simeq2(\bar
u-\alpha)\,q\cdot k$, so that again
\begin{equation}
A_{4\ \textrm{\scriptsize{soft}}}=\frac12\,
A_{2\ \textrm{\scriptsize{soft}}}.
\label{eq:a4a2soft}\end{equation}

Thus collecting up (\ref{eq:a2a1soft}), (\ref{eq:a3a2soft}) and
(\ref{eq:a4a2soft}) as before we find that the total contribution is zero:
\begin{equation}
A_{1\ \textrm{\scriptsize{soft}}}+A_{2\ \textrm{\scriptsize{soft}}}+
A_{3\ \textrm{\scriptsize{soft}}}+A_{4\ \textrm{\scriptsize{soft}}}=0.
\end{equation}

\subsection{The hard-collinear contribution}
\label{subsec:hardcol}

In this subsection we consider the region in which one loop momentum is
collinear with the pion's momentum and the other is hard.
We choose loop momenta such that a gluon has collinear momentum $l$ 
and write $l=\alpha q+\beta \bar{q}+l_\perp$. $k$ denotes a 
hard momentum and we choose it such that a gluon line has momentum 
$k$. As in the hard-soft case we can consider 
one-particle-irreducible hard subgraphs (since adding a collinear
momentum to a hard momentum gives a hard momentum). But note that 
since $\alpha q\sim m_b$, the hard subgraph depends on $\alpha$, 
and a convolution in $\alpha$ with the remainder of the graph
remains. 

To anticipate the result of this subsection recall that we need to 
demonstrate that (\ref{cc1}) is valid and 
that we have already identified the second of the two subtraction terms 
with a non-cancelling divergence in the hard-soft region. In order 
to verify that (\ref{cc1}) is valid, it remains to show that there 
is a non-cancelling divergence in the hard-collinear region that 
has the structure of the first subtraction term, $F^{(0)}_{B\to D} 
\cdot T^{(1)}*\Phi_\pi^{(1)}$. It is plausible that we should 
find this term in the hard-collinear region, since $T^{(1)}$ is 
the hard one-loop amplitude and $\Phi_\pi^{(1)}$ has only collinear 
divergences.

\begin{figure}[t]
   \vspace{-3.8cm}
   \epsfysize=24cm
   \epsfxsize=16cm
   \centerline{\epsffile{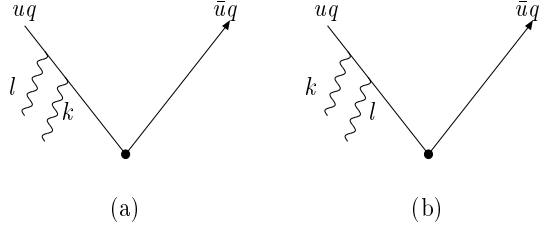}}
   \vspace*{-17.6cm}
\caption[dummy]{\label{fig:hardcol}\small 
Components of two-loop Feynman diagrams. $l$ is 
collinear with the pion's momentum and $k$ is a hard momentum. Routing
(a) leads to potentially singular contributions whereas routing (b)
does not.}
\end{figure}

\paragraph{Diagrams 1a-7b:} It is straightforward to establish that in
order to get a divergence the gluon with the collinear momentum must
be attached to the end of a light-quark or antiquark line and not
internally (see Fig.~\ref{fig:hardcol}).  Since the divergences are
logarithmic we can set $l=\alpha q$, and because the collinear gluon
is attached to the end of a light-quark or antiquark line 
we always get (using the on-shell condition) a factor proportional to
$q_\rho$, where $\rho$ is the Lorentz index of the collinear gluon.

\begin{figure}[t]
   \vspace{-3.5cm}
   \epsfysize=24cm
   \epsfxsize=16cm
   \centerline{\epsffile{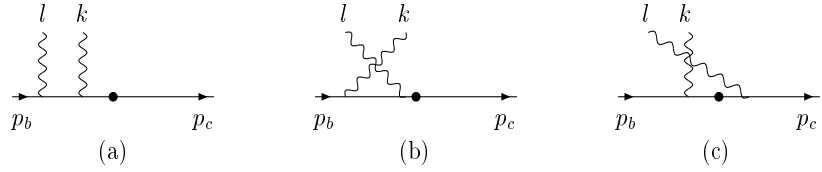}}
   \vspace*{-17.9cm}
\caption[dummy]{\label{fig:klheavy}\small 
Attachments of the collinear gluon (labelled by $l$) and hard gluon
(with momentum $k$) to the heavy quarks (see text).}
\end{figure}

Consider now the diagrams in which both gluons are attached to the
$b$ quark as indicated in Fig.~\ref{fig:klheavy}a. These are
diagrams \{1a, 1c, 3a(part), 4a(part)\}. These diagrams all contain the
factor
\begin{equation}
\label{eq:factor1}
\Gamma\frac{1}{(\pslash_b+\alpha\qslash+\kslash-m_b)}\gamma^\mu
\,\frac{1}{(\pslash_b+\alpha\qslash-m_b)}\,\alpha\qslash,
\end{equation}
which can be simplified to
\begin{equation}
\Gamma\frac{1}{(\pslash_b+\alpha\qslash+\kslash-m_b)}\gamma^\mu.
\label{eq:c1}\end{equation}
(Strictly speaking these diagrams contain the factor
(\ref{eq:factor1}) without the factor $\alpha$. We have included this 
factor for convenience and correct for it later in (\ref{eq:l17}).)

Similarly consider the diagrams in which the two gluons are attached to
the $b$ quark as in Fig.~\ref{fig:klheavy}b, i.e.\ diagrams
\{2a, 2c, 4a(part), 3a(part)\}. These all have a factor
\begin{equation}
\Gamma\frac{1}{(\pslash_b+\alpha\qslash+\kslash-m_b)}\,\alpha\qslash
\,\frac{1}{(\pslash_b+\kslash-m_b)}\,\gamma^\mu,
\end{equation}
which simplifies to
\begin{equation}
\Gamma\frac{1}{(\pslash_b+\kslash-m_b)}\,\gamma^\mu\
-\Gamma\frac{1}{(\pslash_b+\alpha\qslash+\kslash-m_b)}\,\gamma^\mu.
\label{eq:c2}\end{equation}
Apart from the factors in (\ref{eq:c1}) and (\ref{eq:c2}) and the
colour factors, the remaining terms are common in the two sets of
diagrams in the order we have written them (i.e.\ the remaining factors
in diagram 1a are equal to those in diagram 2a etc.). 

Next consider the diagrams which have the hard gluon attached to the
$b$ quark and the collinear one to the $c$-quark as in
Fig.~\ref{fig:klheavy}c, i.e.\ diagrams
\{6a, 5b, 7b(part), 7a(part)\}. These all have a common factor
\begin{equation}
\alpha\qslash\,\frac{1}{(\pslash_c-\alpha\qslash-m_c)}\,\Gamma\,
\frac{1}{(\pslash_b+\kslash-m_b)}\,\gamma^\mu,
\end{equation}
which can be simplified to
\begin{equation}
-\Gamma\frac{1}{(\pslash_b+\kslash-m_b)}\gamma^\mu.
\end{equation}
Again the remaining factors are equal in each of the four diagrams as in
the corresponding diagrams of the previous two sets.

Summing up these contributions we find they are of the form
\begin{equation}
(C_1-C_2)\Gamma\frac{1}{(\pslash_b+\alpha\qslash+\kslash-m_b)}\,\gamma^\mu
+(C_2-C_3)\Gamma\frac{1}{(\pslash_b+\kslash-m_b)}\,\gamma^\mu,
\label{eq:bhard}\end{equation}
where $C_1,C_2$ and $C_3$ are the remaining terms in each of the three
sets of diagrams (they are equal except for the colour factors).

Similarly when the hard gluon couples to the charm quark we obtain
\begin{equation}
(C^\prime_1-C^\prime_2)\gamma^\mu\frac{1}{\pslash_c-\kslash-m_c}\Gamma- 
(C^\prime_3-C^\prime_2)
\gamma^\mu\frac{1}{\pslash_c-\alpha\qslash-\kslash-m_c}\Gamma.
\label{eq:chard}\end{equation}

When we insert the singlet-singlet operator $O_0$, all the colour
factors are equal and hence $C_1-C_2=C_3-C_2=0$ (and similarly for the
$C_i^\prime$) and we get an immediate cancellation.  We therefore
only need to consider the insertion of the octet-octet operator $O_8$.
In this case $C^\prime_1-C_2^\prime=-(C_1-C_2)$ and
$C^\prime_3-C_2^\prime=-(C_3-C_2)$ so that these diagrams give a contribution
which is the integral over
\begin{eqnarray}
&&\hspace*{-0.5cm}
\frac{N_c}{2}\Bigg[\,\Gamma\left(\frac{1}{\pslash_b+\alpha\qslash+\kslash-m_b}
-\frac{1}{\pslash_b+\kslash-m_b}\right)\,\gamma^\mu-
\nonumber\\
&&\hspace*{1.5cm}
\gamma^\mu\left(\frac{1}{\pslash_c-\kslash-m_c}-
\frac{1}{\pslash_c-\alpha\qslash-\kslash-m_c}\right)
\Gamma\,\Bigg] \, \times {\cal L}
\label{eq:bc17}
\end{eqnarray}
where ${\cal L}$ represents the contributions from the remaining parts
of the diagrams, i.e.\ the gluon propagators and the contribution from 
the light quark lines, but with a factor $\alpha q^\rho$, that has
already been included, extracted. Hence ${\cal L}$ is given by 
\begin{eqnarray}
{\cal L}&=&\frac{1}{k^2l^2}\cdot
\nonumber\\
&&\hspace{0.00in}
\left[
\frac{2(u+\alpha)}{\alpha}\,\frac{\gamma^\mu((u+\alpha)\qslash+\kslash)
\Gamma}{(uq+l)^2((u+\alpha)q+k)^2}- \frac{2(\bar u+\alpha)}{\alpha}\,
\frac{\Gamma((\bar u+\alpha)\qslash+\kslash) \gamma^\mu}{(\bar
uq+l)^2((\bar u+\alpha)q+k)^2}\right.\nonumber\\
&&\hspace{0.3cm}\left.  +\,\frac{2(u+\alpha)}{\alpha}\,
\frac{\Gamma(\bar u\qslash+\kslash) \gamma^\mu}{(uq+l)^2(\bar u
q+k)^2}- \frac{2(\bar u+\alpha)}{\alpha}\,
\frac{\gamma^\mu(u\qslash+\kslash) \Gamma}{(\bar uq+l)^2\,(uq+k)^2}
\right].\label{eq:l17}\end{eqnarray} 
The four terms in square brackets in (\ref{eq:l17}) correspond to
the groupings we introduced above, e.g., the diagrams
\{1a, 1c, 3a(part), 4a(part)\} or in the case of the hard gluon being
attached to the charm quark they may be \{5a, 6b, 7a(part), 7b(part)\}.
The common factor of $1/\alpha$ in (\ref{eq:l17}) is present
because we have already included a factor of $\alpha q^\rho$ in the
evaluation of the factors from the heavy quarks. This completes the
consideration of the contribution from diagrams 1a-7b.

\paragraph{Diagrams 8a-9b:}
We now consider diagrams 8a-9b, which have a similar structure
for the light quarks as diagrams 1a-7b. Evaluation of these diagrams
yields the result
\begin{eqnarray}
&&\hspace{-0.4in}\frac{N_c}{2}\,\alpha q^\rho
\left((l-k)_\nu g_{\mu\rho}+(2k+l)_\rho g_{\mu\nu}-(2l+k)_\mu g_{\nu\rho}
\right)\cdot\nonumber\\ 
&&\hspace{0.3in}
\frac{-1}{(k+l)^2}\left(\Gamma\frac{1}{\pslash_b+\kslash
+\lslash-m_b}\gamma^\nu + \gamma^\nu\frac{1}{\pslash_c-\kslash
-\lslash-m_c}\Gamma\right)\, \times {\cal L}.
\label{eq:colhard89}\end{eqnarray}
In the factor coming from the triple gluon vertex in 
(\ref{eq:colhard89}) we can approximate $l$ by $\alpha q$, so that
\begin{eqnarray}
&&q^\rho \left((l-k)_\nu g_{\mu\rho}+(2k+l)_\rho g_{\mu\nu}-(2l+k)_\mu
g_{\nu\rho} \right)
\nonumber\\
&&\hspace*{3cm}=(\alpha q-k)_\nu q_\mu +
2q\cdot kg_{\mu\nu}-(2\alpha q+k)_\mu q_\nu.
\end{eqnarray}
Next we make use of the fact that $q_\mu$ gives zero when contracted
with ${\cal L}$ so that we can drop it and that 
$(k+l)^2=k^2+2\alpha q\cdot k$, to write the contribution from these
diagrams as
\begin{eqnarray}
&&\hspace{-0.4in}
\frac{N_c}{2}\left\{ (k+l)^2g_{\mu\nu}-k^2g_{\mu\nu}
-k_\mu\alpha q_\nu\right\}\cdot\nonumber\\ 
&&\hspace{0.2in}
\frac{-1}{(k+l)^2}\left(\Gamma\frac{1}{\pslash_b+\kslash
+\lslash-m_b}\gamma^\nu + \gamma^\nu\frac{1}{\pslash_c-\kslash
-\lslash-m_c}\Gamma\right)\, \times {\cal L}.
\label{eq:babajaga}\end{eqnarray}
We now consider the three terms in \{ \} in (\ref{eq:babajaga})
in turn. The first term cancels two of the four terms in (\ref{eq:bc17})
from diagrams 1a-7b. The remaining two terms in (\ref{eq:bc17}) 
can be combined with the second term in \{ \} in (\ref{eq:babajaga})
to give
\begin{eqnarray}
X_1&\equiv&
\frac{N_c}{2}\left(\,\frac{2(u+\alpha)}{\alpha}
\frac{1}{(uq+l)^2l^2}[T(u)-T(u+\alpha)]\right.
+\nonumber\\ 
&&\hspace{0.8in}\left.\frac{2(\bar u+\alpha)}{\alpha}
\frac{1}{(\bar uq+l)^2l^2}[T(u)-T(u-\alpha)]\right),
\label{eq:onetoninea}\end{eqnarray}
where 
\begin{equation}
T(u)\equiv\frac{1}{k^2}\left(
\gamma^\mu\frac{1}{u\qslash+\kslash}\Gamma - \Gamma
\frac{1}{\bar u\qslash+\kslash}\gamma^\mu\right)\times
\left(\Gamma\frac{1}{\pslash_b+\kslash-m_b}\gamma^\mu
+\gamma^\mu\frac{1}{\pslash_c-\kslash-m_c}\Gamma\right).
\label{eq:onetonineb}\end{equation}
Note that $T(u)$ is exactly the integrand for the hard amplitude at 
one-loop order, i.e.\ using the notation of (\ref{eq:ffpert})
\begin{equation}
\label{eq:t1def}
\int_k T(u) = T^{(1)},
\end{equation}
where $\int_k$ denotes integration over the loop momentum $k$.

Finally we have to evaluate the third term in \{ \} in
(\ref{eq:babajaga}), which gives a contribution of
\begin{eqnarray}
X_2&\equiv&\frac{N_c}{2}\frac{4}{k^2l^2(k+l)^2}
\left[\frac{u+\alpha}{\alpha}\frac{1}{(uq+l)^2}\Gamma-
\frac{\bar u+\alpha}{\alpha}\frac{1}{(\bar uq+l)^2}\,\Gamma\right]\times
\nonumber\\
&&\hspace{-0.3in}
\left(\Gamma\,\frac{1}{\pslash_b+\alpha\qslash+\kslash-m_b}
\,\alpha\qslash\,+\,\alpha\qslash\,
\frac{1}{\pslash_c-\alpha\qslash-\kslash-m_c}\,\Gamma\,\right)
.
\label{eq:x2def} \end{eqnarray}
$X_1+X_2$ is the total contribution from diagrams 1a-9b.

\paragraph{Diagrams 12 and 15:} We now consider diagrams 12a-12d and
15a-15d. In these diagrams there are singular contributions when either
of the two gluons is collinear (and the other one is hard). However, in
the case where the gluon which is attached at one end to one of the
heavy quarks is collinear, the standard mechanism for the cancellation
of collinear divergences at one-loop level applies (the singularities of
diagrams 15a and 15d cancel as do those in each pair of diagrams
\{15b, 15c\}; \{12a, 12c\} and \{12b, 12d\}).

\begin{figure}[t]
   \vspace{-3.8cm}
\hspace*{-1.5cm}
   \epsfysize=24cm
   \epsfxsize=16cm
   \centerline{\epsffile{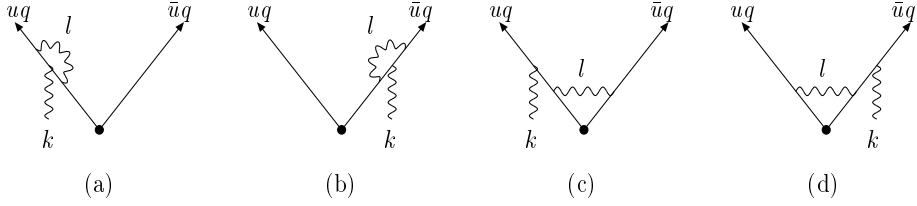}}
   \vspace*{-17.4cm}
\caption[dummy]{\label{fig:kltwelve}\small 
Attachments of the collinear gluon (labelled by $l$) and hard gluon
(with momentum $k$) to the light quarks, as required for the evaluation
of diagrams 12 and 15 (see text).}
\end{figure}

We therefore need only to consider the case where the collinear gluon
is the one attached at both ends to the light quarks. Straightforward
evaluation of the factor coming from the light quarks (see
Fig.~\ref{fig:kltwelve}) shows that this factor is equal to 
\begin{eqnarray}
&&\hspace{-0.3in}
\frac{2(u+\alpha)}{\alpha}\frac{1}{(uq+l)^2}
\bigg(\gamma^\mu\frac{1}{u\qslash+\kslash}\Gamma
-\gamma^\mu\frac{1}{(u+\alpha)\qslash+\kslash}\Gamma
\nonumber\\
&&
-\Gamma\frac{1}{\bar u\qslash+\kslash}\gamma^\mu
+\Gamma\frac{1}{(\bar u-\alpha)\qslash+\kslash}\gamma^\mu\bigg)
\nonumber\\
&&\hspace{-0.3in}
+\frac{2(\bar u+\alpha)}{\alpha}\frac{1}{(\bar uq+l)^2}
\bigg(\gamma^\mu\frac{1}{u\qslash+\kslash}\Gamma
-\gamma^\mu\frac{1}{(u-\alpha)\qslash+\kslash}\Gamma
\nonumber\\
&&
-\Gamma\frac{1}{\bar u\qslash+\kslash}\gamma^\mu
+\Gamma\frac{1}{(\bar u+\alpha)\qslash+\kslash}\gamma^\mu\bigg).
\end{eqnarray}
But this is exactly the same factor as appears in the expression $X_1$
in (\ref{eq:onetoninea}). Indeed, the contribution from diagrams
12 and 15 is identical to (\ref{eq:onetoninea}) 
except that the colour factor is $-1/2N_c$,
instead of $N_c/2$. Combining the contributions from all the
diagrams we have evaluated so far, i.e.\ diagrams 1a-9b, 12a-12d and
15a-15d, we find a total contribution of $2C_F/N_c\,X_1+X_2$.

\begin{figure}[t]
   \vspace{-3.8cm}
\hspace*{-1.5cm}
   \epsfysize=24cm
   \epsfxsize=16cm
   \centerline{\epsffile{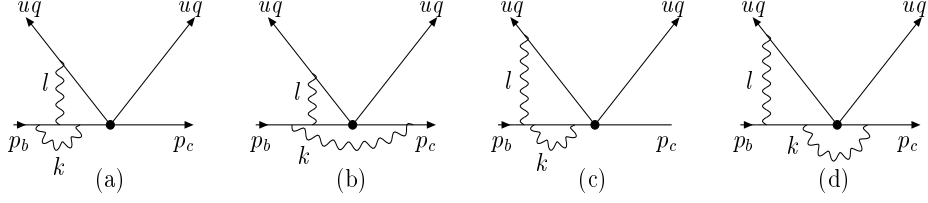}}
   \vspace*{-17.3cm}
\caption[dummy]{\label{fig:klsixteen}\small 
Momentum routings in (a) diagram 16a, (b) diagram 17a
(c) diagram 18a and (d) diagram 19a which are considered in the text.
$l$ is a momentum collinear with that of the pion and $k$ is a hard
momentum.}
\end{figure}

\paragraph{Diagrams 16a-19d:} In diagrams 16a-19d, the singular
contribution comes from the region in which the collinear gluon is the
one attached at one end to the light-quarks. As an example consider
the diagram 16a (see Fig.~\ref{fig:klsixteen}a). The factor coming
from the heavy-quark lines (again multiplying it by $\alpha q^\rho$) 
is of the form:
\begin{eqnarray}
\textrm{Diag16a}&\equiv&\Gamma\,\frac{1}{\pslash_b+\alpha\qslash-m_b}\,
\gamma^\mu
\,\frac{1}{\pslash_b+\alpha\qslash+\kslash-m_b}\,\alpha\qslash\,
\frac{1}{\pslash_b+\kslash-m_b}\,\gamma_\mu\nonumber\\
&&\hspace*{-2cm}=\Gamma\,\frac{1}{\pslash_b+\alpha\qslash-m_b}\,\gamma^\mu
\,\frac{1}{\pslash_b+\kslash-m_b}\,\gamma_\mu-
\Gamma\,\frac{1}{\pslash_b+\alpha\qslash-m_b}\,\gamma^\mu
\,\frac{1}{\pslash_b+\alpha\qslash+\kslash-m_b}\,\gamma_\mu\nonumber\\
&&\hspace*{-2cm}\equiv a_1-a_2.
\label{eq:sixteena}
\end{eqnarray}  
The analogous contribution from diagram 17a is
\begin{equation}
\textrm{Diag17a}\equiv\gamma^\mu\frac{1}{\pslash_c+\kslash-m_c}\,\Gamma\,
\frac{1}{\pslash_b+\kslash-m_b}\,\gamma_\mu-a_3,
\label{eq:seventeena}\end{equation}
where
\begin{equation}
a_3\equiv\gamma^\mu\,\frac{1}{\pslash_c+\kslash-m_c}\,\Gamma\,
\frac{1}{\pslash_b+\alpha\qslash+\kslash-m_b}\,\gamma_\mu.
\label{eq:a3def}\end{equation}
The first term in (\ref{eq:seventeena}) cancels against
the corresponding contribution from diagram~17c and we do not include
it further in the discussion. Finally the corresponding
terms in diagrams~18a and 19a are
\begin{eqnarray}
\textrm{Diag18a}&=&a_2\label{eq:eighteena}\\ 
\textrm{Diag19a}&=&a_3\label{eq:nineteena}.
\end{eqnarray}
All the remaining factors in these diagrams are the same, apart from the
colour factors, so including these we find that the sum of these four
diagrams gives a contribution which is proportional to
\begin{equation}
\textrm{Diag16a+Diag17a+Diag18a+Diag19a} = 
-\frac{1}{2N_c}a_1+\frac{N_c}{2}(a_2-a_3).
\end{equation}

\begin{figure}[t]
   \vspace{-3.8cm}
   \epsfysize=24cm
   \epsfxsize=16cm
   \centerline{\epsffile{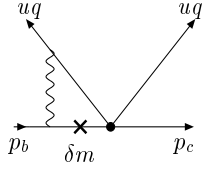}}
   \vspace*{-18cm}
\caption[dummy]{\label{fig:deltam}\small 
Insertion of a mass-renormalization counterterm into a one-loop
diagram.}
\end{figure}

There is a further subtlety, however, and this concerns the mass
renormalization of the $b$ quark (see Fig.~\ref{fig:deltam}). The
singular contribution does not cancel (in contrast to the
wave-function renormalization), and is readily found to be
$-C_F\,a_1$.  Thus the contribution of the diagrams
\{16a, 17a, 18a, 19a\} together with that from the diagram with the mass
counterterm gives a contribution which is proportional to
\begin{equation}
-\frac{N_c}{2}\,(a_1-a_2+a_3).
\label{eq:total1619}\end{equation}
The remaining diagrams in the set 16a-19d yield analogous results, with
different factors from the light quarks naturally.

\paragraph{Diagrams 10a-11b:} Again the only singular contribution
comes from the region in which the collinear gluon is the one which is
attached at one end to a light quark (see
Fig.~\ref{fig:tena}a). Explicit evaluation of diagram 10a gives
the following contribution:
\begin{equation}
\textrm{Diag10a} =  \frac{N_c}{2}(a_1-a_2)\, +\,
\frac{N_c}{2}\,\frac{1}{(k+l)^2}\left[
2\Gamma-\Gamma\,\frac{1}{\pslash_b+\alpha\qslash+\kslash-m_b}\,
\alpha\qslash\,\right], 
\label{eq:tena}\end{equation}
where all remaining factors are as in the previous paragraph. Thus we
see that the first term cancels part of the contribution from diagrams
16a-19d in (\ref{eq:total1619}). The term $2\Gamma$ in the square
brackets cancels against the corresponding contribution in diagram\,10d
and we do not consider it further. This leaves us with the second term
in square brackets, which cancels half of the corresponding contribution
in $X_2$ (by \textit{corresponding term} we mean the product of the
first term in the square brackets in (\ref{eq:x2def}) with the
first term in the brackets on the second line). The remaining terms
in $X_2$ are cancelled by the corresponding contributions from 
the diagrams\,10b-d and 11a-b.

Explicit evaluation of diagram\,11a shows that it also contains the term
$N_c/2\,a_3$ which cancels the remaining term in
(\ref{eq:total1619}). Thus $X_2$, together with diagrams 16a-19d
and 10a-11b cancel. The total contribution from diagrams 1a-12d and
15a-19d is $2C_F/N_c\,X_1$, where $X_1$ is given explicitly in
(\ref{eq:onetoninea}). Finally we need to evaluate
diagrams\,13a-14d.

\paragraph{Diagrams\,13a-14d:} The singular contributions from diagrams
13a-d come from the region in which the collinear gluon is the one
attached at one end to a heavy quark. These singularities in
diagrams 13a-d from the collinear-hard region of phase space can be
seen to cancel by the mechanism which ensures the cancellation of
collinear singularities at one-loop order. Thus the singularity from
diagram 13a cancels that from 13d and similarly the singularities from
diagrams 13b and 13c cancel.

Finally we consider diagrams 14a-d and the singular contributions come
from the region in which the collinear gluon is the one attached at one
end to a light quark and at the other to the light antiquark. Explicit
evaluation of these four diagrams yields the result:
\begin{equation}
\textrm{Diags14a-d}=C_F\,\frac{2l_\perp^2}{(uq+l)^2(\bar uq-l)^2
l^2}\,T(u+\alpha),
\label{eq:fourteen}\end{equation}
In deriving (\ref{eq:fourteen}) we have used that the pion projection 
(\ref{distamps}) inserts a factor $\qslash\gamma^5$ in the trace over 
the light quark lines. 

\subsubsection{The total contribution}

We now combine the result in (\ref{eq:fourteen}) with the remaining
term, $2C_F/N_c\,X_1$, to write the total singular contribution ($S$)
from the hard-collinear region of phase space as
\begin{eqnarray}
S&=&C_F\left(\,\frac{2(u+\alpha)}{\alpha}\,\frac{1}{(uq+l)^2l^2}
-\,\frac{2(\bar u-\alpha)}{\alpha}\,\frac{1}{(\bar uq-l)^2l^2}
\,\right)(\,T(u)-T(u+\alpha)\,)\nonumber\\
&&\hspace{0.0in}-\,C_F\frac{l_\perp^2}{q\cdot l}
\left(\frac{1}{(uq+l)^2l^2} -
\frac{1}{(\bar uq-l)^2l^2}\,\right)\,T(u+\alpha).
\label{eq:tot}\end{eqnarray}
In arriving at (\ref{eq:tot}) we have made the change of integration
variable $l\to -l$ in the second line of (\ref{eq:onetoninea}) and have
used partial fractions to obtain the second line in (\ref{eq:tot}) from
(\ref{eq:fourteen}). From now on we will assume that $T(u)$ contains 
the integration over the hard momentum $k$, so that it equals 
$T^{(1)}$ according to (\ref{eq:t1def}). We now need to convert $S$ 
into an expression of the form $F^{(0)}_{B\to D}\cdot T^{(1)}*
\Phi_\pi^{(1)}$.

We would like to write $\int_l S$ as a convolution in $\alpha$ 
and hence need to do the
integrations over $\beta$ and $l_\perp$.
The $\beta$ integration can be done using Cauchy's theorem, and
vanishes if $\alpha>\bar u$ and $\alpha <-u$, since in that case
all the poles in $\beta$ are on the same side of the contour. 
Including the factors of $i$ and $g_s$ from the Feynman rules
for the propagators and vertices (which we have been neglecting until
now), we find that the total contribution is
\begin{eqnarray}
&&\hspace{-0.2in}
-C_F\frac{\alpha_s}{2\pi}\int\frac{d\,l_\perp^2}{l_\perp^2}\,
d\alpha\left\{\theta(-\alpha)\theta(u+\alpha)
\,\left(-\frac{u+\alpha}{u}\,T(u+\alpha)
+\frac{u+\alpha}{u\alpha}\,[T(u+\alpha)-T(u)]\right)
\right.\nonumber\\
&&\hspace{0.2in}\left.
-\theta(\alpha)\theta(\bar u-\alpha)
\,\left(\frac{\bar u-\alpha}{\bar u}\,T(u+\alpha)
+\frac{\bar u-\alpha}{\bar u\alpha}\,[T(u+\alpha)-T(u)]\right)
\right\}.
\label{eq:final0}\end{eqnarray}
In (\ref{eq:final0}) we have switched metric so that $l_\perp^2$ is
positive. The integral over transverse momentum is both ultraviolet
and infrared divergent. The ultraviolet divergence is an artefact of 
the collinear approximation and is absent in the exact expression 
for the amplitude. 
We write the integral over transverse momentum as
\begin{equation}
\int\,\frac{dl_\perp^2}{l_\perp^2}=2\ln\frac{\mu_{UV}}{\mu_{IR}}, 
\end{equation}
introducing an ultraviolet and an infrared scale,
$\mu_{UV}$ and $\mu_{IR}$, respectively. Then we 
change the longitudinal integration variable from $\alpha$ to $w=u+\alpha$
to write (\ref{eq:final0}) as
\begin{eqnarray}
&&\hspace{-0.3in}
C_F\,\frac{\alpha_s}{\pi}\ln\frac{\mu_{UV}}{\mu_{IR}}
\,\int dw \left\{\theta(w)\theta(u-w)\,\left(
\frac{w}{u}\,T(w)+\frac{w}{u(u-w)}\,[T(w)-T(u)]\,\right)
\right.\nonumber\\
&&+\left.\theta(w-u)\theta(\bar w)\,\left(
\frac{\bar w}{\bar u}\,T(w)+\frac{\bar w}{\bar u(\bar u-\bar w)}
\,[T(w)-T(u)]\,\right)\right\}.
\end{eqnarray}
We rewrite this as 
\begin{equation}
\label{eq:final1}
C_F\,\frac{\alpha_s}{\pi}\,\ln\frac{\mu_{UV}}{\mu_{IR}}
\int_0^1\,dw\,T(w)\,V(w,u),
\end{equation}
where
\begin{equation}
V(w,u)=\theta(u-w)\frac{w}{u}+\left[\theta(u-w)\frac{w}{u(u-w)}\right]_+\,
\,+\,\theta(w-u)\frac{\bar w}{\bar u}+\left[\theta(w-u)\frac{\bar w}
{\bar u(\bar u-\bar w)}\right]_+\,
\end{equation}
with $[f]_+$ defined by
\begin{equation}
[f(w,u)]_+\equiv f(w,u)-\delta(w-u)\,\int^1_0\, dv\, f(v,u).
\label{eq:plusdef}
\end{equation}
This is almost the correct expression for the total hard-collinear
contribution,
except for $\delta$-function terms at $u=w$. These come from the 
diagrams with wave-function renormalization on external 
light-quark lines, which we have not considered so far. They 
modify the expression for $V$ to
\begin{equation}
\label{eq:kernel2}
V(w,u)=\left[\theta(u-w)\frac{w}{u}\left( 1+\frac{1}{u-w}\,\right)
\,+\,\theta(w-u)\frac{\bar w}{\bar u}\left(1+\frac{1}
{\bar u-\bar w}\,\right)\,\right]_+,
\end{equation}
and hence ensure that $\int dw V(w,u)=0$ as required by current conservation. 

Finally, including the convolution with the pion distribution amplitude 
and writing ex\-pli\-cit\-ly also the $B\to D$ form factor
(omitted so far for simplicity)
the total contribution $S$ takes the form
\begin{equation}\label{stvphi}
S=F_{B\to D}^{(0)}\cdot
C_F\frac{\alpha_s}{\pi}\ln\frac{\mu_{UV}}{\mu_{IR}}\,
\int^1_0\, dw\, du\, T(w)\, V(w,u)\, \Phi^{(0)}_\pi(u),
\end{equation}
with $V(w,u)$ as defined in (\ref{eq:kernel2}).
We next recall that the infrared singular contribution to the
pion distribution amplitude is determined by \cite{LB80}
\begin{equation}
\label{eq:erbl1}
\Phi^{(1)}_\pi(w) = 
C_F\,\frac{\alpha_s}{\pi}\,\ln\frac{\mu_{UV}}{\mu_{IR}}\,
\int_0^1\,du\,V(w,u)\,\Phi^{(0)}_\pi(u).
\end{equation}
This expression is equivalent to the familiar evolution
equation for the pion distribution amplitude   
$\Phi_\pi(w)=\Phi^{(0)}_\pi(w)+\Phi^{(1)}_\pi(w)$
to order $\alpha_s$
\begin{equation}\label{phievl}
\frac{d}{d\ln\mu_{UV}}\,\Phi_\pi(w)
=C_F\,\frac{\alpha_s}{\pi}\,\int_0^1\,du\,V(w,u)\,\Phi_\pi(u).
\end{equation}
Using (\ref{eq:erbl1}) and $T(w)\equiv T^{(1)}(w)$, Eq.~(\ref{stvphi}) 
reduces to
\begin{equation}\label{stph}
S=F_{B\to D}^{(0)}\cdot \int^1_0\, dw\, T^{(1)}(w)\, \Phi^{(1)}_\pi(w).
\end{equation}
We see that the collinearly divergent contribution $S$ corresponds 
precisely to the part of
$A^{(2)}_{\rm non-fact.}$ in (\ref{cc1}) that is subtracted
by the second term
$F_{B\to D}^{(0)}\cdot T^{(1)}*\Phi_\pi^{(1)}$.
This completes our demonstration of the validity of the 
factorization formula for
$B\to D\pi$ decays at two-loop order in perturbation theory.

\subsection{Summary}

The result of Sects.~\ref{subsec:softsoft}-\ref{subsec:hardcol} 
is that the non-cancelling infrared divergences are precisely those
which are necessary to cancel the infrared singularities in the 
perturbative expansion of the semi-leptonic form factor and 
pion light-cone distribution amplitude. In this way we have explicitly
verified the factorization formula to two-loop order.

The natural question to ask now is whether our arguments can be
generalized to higher orders of perturbation theory. Although for some
of the singular regions such a generalization is
straightforward (e.g.\ the eikonal combinatorics for soft gluons
discussed in Sect.~\ref{subsec:softsoft}, or the combinatorics for 
collinear gluons discussed in Sect.~\ref{subsec:colcol}) we have not
yet tried to carry out such a
program for all singular regions. This remains an interesting
challenge.

Another important extension concerns $B$ decays into two light 
mesons, for example $\bar{B}_d\to \pi^+\pi^-$. Most of the arguments 
we used apply directly to this case as well. In addition, however, 
one has to consider singularities that arise in momentum 
configurations collinear to the recoil pion. The same method that was  
applied to $D^+\pi^-$ final states should prove that these
singularities cancel in the sum over all diagrams or can be absorbed 
into the $B\to \pi$ form factor. A further complication for final 
states of two light mesons comes from the hard spectator interaction 
in the factorization formula (\ref{fff}). Because the characteristic 
hard gluon virtuality in hard spectator interactions is order 
$m_b\Lambda_{\rm QCD}$, this term requires an analysis of momentum 
configurations different from those considered for heavy-light 
final states. We plan to return to the extension of the factorization 
proof to light-light final states in a future publication.


\boldmath
\section{Phenomenology of $B\to D^{(*)} L$ decays}
\unboldmath
\label{bdpi}

The matrix elements we have computed in Sect.~\ref{menlo} provide
the theoretical basis for a model-independent calculation of the
class-I non-leptonic decay amplitudes for decays of the type
$B\to D^{(*)} L$, where $L$ is a light meson, to leading power
in $\Lambda_{\rm QCD}/m_b$ and at next-to-leading order in 
renormalization-group improved perturbation theory. In this section 
we will discuss phenomenological applications of this formalism and 
confront our numerical results with experiment. We will also provide
some simple estimates of power-suppressed corrections to the 
factorization formula.

\subsection{Basic input}

The results for the class-I decay amplitudes for $B\to D^{(*)} L$  
are obtained by evaluating the (factorized) hadronic matrix elements 
of the transition operator ${\cal T}$ defined in (\ref{heffa1}). They 
are written in terms of products of CKM matrix elements, light-meson 
decay constants, $B\to D^{(*)}$ transition form factors, and the 
QCD parameters $a_1(D^{(*)} L)$, whose explicit expressions at 
next-to-leading order have been given in (\ref{a1dpi}) and (\ref{a1dspi}).

The decay constants of light pseudoscalar and vector mesons are defined 
as
\begin{eqnarray}
   \langle P^-(q)|\bar d\gamma_\mu\gamma_5 u|0\rangle
   &=& -i f_P q_\mu , \label{fpidef}\\
   \langle V^-(q)|\bar d\gamma_\mu u|0\rangle
   &=& -i f_V m_V \eta_\mu^* , \label{frhodef}
\end{eqnarray}
where $\eta_\mu$ is the polarization vector of the vector meson. The 
decay constants can be determined experimentally using data on the 
weak leptonic decays $P^-\to l^-\bar\nu_l(\gamma)$, hadronic 
$\tau^-\to M^-\nu_\tau$ decays, and the electromagnetic decays
$V^0\to e^+ e^-$. Following~\cite{NeSt97}, we use $f_\pi=131$\,MeV, 
$f_K=160$\,MeV, $f_\rho=210$\,MeV, $f_{K^*}=214$\,MeV, and 
$f_{a_1}=229$\,MeV. Here $a_1$ is the pseudovector meson with mass
$m_{a_1}\simeq 1230$\,MeV. Its decay constant is defined in analogy 
with (\ref{frhodef}), but in terms of a matrix element of an axial 
vector current.

The $B\to D^{(*)}$ transition form factors of vector and axial 
vector currents are defined as ($q\equiv p-p'$)
\begin{eqnarray}
   \langle D(p')|\bar c\gamma^\mu b|\bar B(p)\rangle
   = F_+(q^2)\!\left[ (p+p')^\mu-\frac{m^2_B-m^2_D}{q^2}\,q^\mu
    \right] + F_0(q^2) \,\frac{m^2_B-m^2_D}{q^2}\,q^\mu ,
    \nonumber\\[-0.3cm]
   && \label{ffdef}
\end{eqnarray}
\begin{equation}
   \langle D^*(p',\varepsilon)|\bar c\gamma^\mu b|\bar B(p)\rangle
    = \frac{2 i V(q^2)}{m_B+m_{D^*}}\epsilon^{\mu\nu\rho\sigma}
    p'_\nu p_\rho \varepsilon_\sigma^* , 
\label{vdef}
\end{equation}
\begin{eqnarray}
   \langle D^*(p',\varepsilon)|\bar c\gamma^\mu\gamma_5 b
   |\bar B(p)\rangle
   &=& (m_B+m_{D^*}) A_1(q^2) \,\varepsilon^{*\mu} - 
       A_2(q^2)\,\frac{\varepsilon^*\cdot q}{m_B+m_{D^*}} \,(p+p')^\mu
   \nonumber\\
   &&\hspace*{-4cm}-\,\frac{\varepsilon^*\cdot q}{q^2}\,q^\mu 
   \left[(m_B+m_{D^*}) A_1(q^2)-(m_B-m_{D^*}) A_2(q^2)-
   2 m_{D^*}A_0(q^2)\right].
    \label{adef}
\end{eqnarray}
Here the sign conventions are chosen such that all form factors
are positive (in particular $\epsilon^{0123}=-1$). 

\subsection{Non-leptonic amplitudes and decay rates}

Using these definitions, the non-leptonic $\bar B_d\to D^{(*)+} L^-$
decay amplitudes for $L=\pi$, $\rho$ can be expressed as
\begin{eqnarray}
   {\cal A}(\bar B_d\to D^+\pi^-)
   &=& i\frac{G_F}{\sqrt{2}}\,V^*_{ud}V_{cb}\,a_1(D\pi)\,
    f_\pi\,F_0(m^2_\pi)\,(m^2_B-m^2_D) , \label{abdpi}\\
   {\cal A}(\bar B_d\to D^{*+}\pi^-)
   &=& - i\frac{G_F}{\sqrt{2}}\,V^*_{ud}V_{cb}\,a_1(D^*\pi)\,
    f_\pi A_0(m^2_\pi)\,2m_{D^*}\,\varepsilon^*\!\cdot p ,
    \label{abdspi}\\
   {\cal A}(\bar B_d\to D^{+}\rho^-)
   &=& - i\frac{G_F}{\sqrt{2}}\,V^*_{ud}V_{cb}\,a_1(D\rho)\,
    f_\rho\,F_+(m^2_\rho)\,2m_\rho\,\eta^*\!\cdot p . \label{abdrho}
\end{eqnarray}
The decay mode $\bar B_d\to D^{*+}\rho^-$ has a richer structure than 
the decays with at least one pseudoscalar in the final state and 
deserves a more detailed discussion. The most general Lorentz invariant 
decomposition of the corresponding decay amplitude can be written as
\begin{equation}\label{abdsrho}
   {\cal A}(\bar B_d\to D^{*+}\rho^-) = i\frac{G_F}{\sqrt{2}}\,
   V^*_{ud} V_{cb}\,\varepsilon^{*\mu}\eta^{*\nu} \bigg(
   S_1\,g_{\mu\nu} - S_2\,q_\mu p'_\nu
   + iS_3\,\epsilon_{\mu\nu\alpha\beta}\,p'^\alpha q^\beta \bigg) .
\end{equation}
It is convenient to introduce helicity amplitudes corresponding to
the polarization of the vector mesons in the $B$ rest frame. We find
\begin{eqnarray}\label{hpmdef}
   H_0 &=& \frac{1}{2m_{D^*} m_\rho} \left[ (m_B^2-m_{D^*}^2-m_\rho^2)\,
    S_1 - 2m_B^2 |\vec q\,|^2 S_2 \right] , \\
   H_\pm &=& S_1 \mp m_B |\vec q\,|\,S_3 ,
\end{eqnarray}
where
\begin{equation}\label{qvec}
   |\vec q\,| = \frac{1}{2m_B} \sqrt{(m^2_B-m^2_1-m^2_2)^2-4 m^2_1 m^2_2}
\end{equation}
is the momentum of the two final-state mesons in the parent rest frame
(with $m_1=m_{D^*}$, $m_2=m_\rho$ in the present case).
The subscript on the helicity amplitudes refers to the polarization of 
the $D^*$ meson. (Our convention for the helicity amplitudes differs from 
that usually employed in the analysis of semi-leptonic 
$\bar B_d\to D^* l\,\nu$ decays by an overall factor of 
$m_\rho f_\rho\,a_1(D^*\rho)$.) 
To leading power in $\Lambda_{\rm QCD}/m_b$, we obtain
\begin{eqnarray}
   S_1 &=& a_1(D^*\rho)\,m_\rho f_\rho\,(m_B+m_{D^*}) A_1(m_\rho^2)
    , \\
   S_2 &=& a_1(D^*\rho)\,m_\rho f_\rho\,
    \frac{2 A_2(m_\rho^2)}{m_B+m_{D^*}} .
\end{eqnarray}
The contribution proportional to $S_3$ in (\ref{abdsrho}) is associated
with transversely polarized $\rho$ mesons and thus leads to 
power-suppressed effects, which we do not consider here. For the helicity 
amplitudes, it follows that
\begin{eqnarray}
   H_0 &=& \frac{a_1(D^*\rho)\,f_\rho}{2m_{D^*}} \bigg[
    (m_B^2-m_{D^*}^2-m_\rho^2) (m_B+m_{D^*}) A_1(m_\rho^2)
    - \frac{4m_B^2 |\vec q\,|^2}{m_B+m_{D^*}}\,A_2(m_\rho^2) \bigg] ,
    \nonumber\\
\end{eqnarray}
\begin{equation}
   \frac{H_\pm}{H_0} = O(m_\rho/m_B) .
\end{equation}

The decay rates for the non-leptonic decays $\bar B_d\to D^{(*)+} L^-$ are 
given by 
\begin{eqnarray}
   \Gamma(\bar B_d\to D^+\pi^-)
   &=& \frac{G_F^2 (m_B^2-m_D^2)^2 |\vec q\,|}{16\pi m_B^2}\,
    |V_{ud}^* V_{cb}|^2\,|a_1(D\pi)|^2\,f_\pi^2\,F_0^2(m_\pi^2) ,
    \label{gbdpi}\\
   \Gamma(\bar B_d\to D^{*+}\pi^-)
   &=& \frac{G_F^2 |\vec q\,|^3}{4\pi}\,
    |V_{ud}^* V_{cb}|^2\,|a_1(D^*\pi)|^2\,f_\pi^2\,A_0^2(m_\pi^2) ,
    \label{gbdspi}\\
   \Gamma(\bar B_d\to D^{+}\rho^-)
   &=& \frac{G_F^2 |\vec q\,|^3}{4\pi}\,
    |V_{ud}^* V_{cb}|^2\,|a_1(D\rho)|^2\,f_\rho^2\,F_+^2(m^2_\rho) ,
    \label{gbdrho}\\
   \Gamma(\bar B_d\to D^{*+}\rho^-)
   &=& \frac{G_F^2 |\vec q\,|}{16\pi m_B^2}\,
    |V_{ud}^* V_{cb}|^2 \left( |H_0|^2+|H_+|^2+|H_-|^2 \right) .
    \label{gbdsrho}
\end{eqnarray}
The decay rate for the process $\bar B_d\to D^{*+}\rho^-$ with two vector 
mesons in the final state has a non-trivial angular distribution given by 
\cite{KoGo79,CLEOconf}
\begin{eqnarray}\label{dang}
   &&\frac{d\Gamma}{d\cos\theta_1 d\cos\theta_2 d\phi}
    \sim \cos^2\theta_1\cos^2\theta_2 |H_0|^2+
    \frac{1}{4}\sin^2\theta_1\sin^2\theta_2 (|H_+|^2 + |H_-|^2)
    \nonumber\\
   &&\ \ +\frac{1}{4}\sin 2\theta_1 \sin 2\theta_2
    \left[\cos\phi\,{\rm Re}(H^*_0H_++H^*_0H_-)-\sin\phi\,
    {\rm Im}(H^*_0H_+-H^*_0H_-)\right]\nonumber\\
   &&\ \ +\frac{1}{2}\sin^2\theta_1 \sin^2\theta_2
    \left(\cos 2\phi\, {\rm Re}H^*_+ H_- +
    \sin 2\phi\, {\rm Im}H^*_+ H_-\right) .
\end{eqnarray}
Here $\theta_1$ is the angle between the direction of flight of the 
decaying $D^*$ meson and the daughter particle $D^0$, measured in the 
$D^*$ rest frame, $\theta_2$ is the angle between the direction of
flight of the decaying $\rho$ meson and the daughter particle $\pi^-$, 
measured in the $\rho$ rest frame, and $\phi$ is the 
angle between the decay planes of $D^*$ and $\rho$ in the rest frame 
of the $B$ meson. The variables $\cos\theta_{1,2}$ are to be 
integrated from $-1$ to $1$ and $\phi$ from $0$ to $2\pi$. Note that 
in the heavy-quark limit only the first term (proportional to $|H_0|^2$)
remains. Likewise, this term dominates in the total decay rate in 
(\ref{gbdsrho}).

The CLEO collaboration has studied the angular distribution in 
$\bar B_d\to D^{*+}\rho^-$ decays \cite{CLEOconf}. A particular focus 
of this investigation was the search for non-trivial complex phases in 
the helicity amplitudes. Such phases arise from strong final-state 
interactions and can manifest themselves in the terms proportional to
$\sin 2\theta_1\,\sin 2\theta_2\, \sin\phi$ and
$\sin^2\theta_1\,\sin^2\theta_2\, \sin 2\phi$ in the angular
distribution (\ref{dang}). From our discussion above it follows that
these interference terms are power suppressed in the heavy-quark 
limit. We expect them to be small and will not discuss them further 
here. (Experimentally, one finds $|H_+/H_0|\approx 0.15$ and
$|H_-/H_0|\approx 0.3$ with large uncertainties \cite{CLEOconf}.)

The various $B\to D^{(*)}$ form factors entering the expressions 
for the decay rates in (\ref{gbdpi})-(\ref{gbdsrho}) can be determined 
by combining experimental data on semi-leptonic decays with theoretical 
relations derived using heavy-quark effective theory \cite{IW89,NeSt97}. 
Since we work to leading order in $\Lambda_{\rm QCD}/m_b$, it is consistent 
to set the light meson masses to zero and evaluate these form factors at
$q^2=0$. In this case the kinematic relations
\begin{equation}\label{kinerela}
   F_0(0) = F_+(0) , \qquad
   (m_B+m_{D^*}) A_1(0) - (m_B-m_{D^*}) A_2(0) = 2 m_{D^*} A_0(0) ,
\end{equation}
the second of which implies 
\begin{equation}
   H_0\Big|_{m_\rho^2=0} = a_1(D^*\rho)\,f_\rho\,
    (m_B^2-m_{D^*}^2) A_0(0) ,
\end{equation}
allow us to express the two $\bar B_d\to D^+ L^-$ rates in terms
of $F_+(0)$, and the two $\bar B_d\to D^{*+} L^-$ rates in terms of
$A_0(0)$. Heavy-quark symmetry implies that these two form factors
are equal to within a few percent \cite{review}. Below we adopt the 
common value $F_+(0)=A_0(0)=0.6$. All our predictions for decay rates 
will be proportional to the square of this number.

\boldmath
\subsection{Meson distribution amplitudes and predictions for $a_1$}
\unboldmath

Let us now discuss in more detail the ingredients required for the
numerical analysis of the coefficients $a_1(D^{(*)} L)$. The Wilson 
coefficients $C_i$ in the effective weak Hamiltonian depend on the 
choice of the scale $\mu$ as well as on the value of the strong 
coupling $\alpha_s$, for which we take $\alpha_s(m_Z)=0.118$ and 
two-loop evolution down to a low scale $\mu\sim m_b$. To study the 
residual dependence of the results, which remains because the 
perturbation series are truncated at next-to-leading order, we vary 
$\mu$ between $m_b/2$ and $2m_b$. The hard-scattering kernels depend
on the ratio of the heavy-quark masses, for which we take 
$z=m_c/m_b=0.30\pm 0.05$.

\begin{table}
\caption{\label{tab:wfnint}
Numerical values for the integrals $\int^1_0 du\,F(u,z)\,\Phi_L(u)$ (upper 
portion) and $\int^1_0 du\,F(u,-z)\,\Phi_L(u)$ (lower portion) obtained 
including the first two Gegenbauer moments.}
\begin{center}
\begin{tabular}{|c|c|c|c|}
\hline
&&&\\[-0.4cm]
$z$ & Leading term & Coefficient of $\alpha_1^L$ &
 Coefficient of $\alpha_2^L$ \\
&&&\\[-0.4cm]
\hline
&&&\\[-0.4cm]
0.25 & $-8.41-9.51i$ & $5.92-12.19i$ & $-1.33+0.36i$ \\
0.30 & $-8.79-9.09i$ & $5.78-12.71i$ & $-1.19+0.58i$ \\
0.35 & $-9.13-8.59i$ & $5.60-13.21i$ & $-1.00+0.73i$ \\
&&&\\[-0.4cm]
\hline
&&&\\[-0.4cm]
0.25 & $-8.45-6.56i$ & $6.72-10.73i$ & $-0.38+0.93i$ \\
0.30 & $-8.37-5.99i$ & $6.83-11.49i$ & $-0.21+0.85i$ \\
0.35 & $-8.24-5.44i$ & $6.81-12.29i$ & $-0.08+0.75i$ \\[0.1cm]
\hline
\end{tabular}
\end{center}
\end{table}

Hadronic uncertainties enter the analysis also through the 
parameterizations used for the meson light-cone distribution 
amplitudes. It is convenient and conventional to expand the 
distribution amplitudes in Gegenbauer polynomials:
\begin{equation}\label{gpol}
   \Phi_L(u) = 6u(1-u) \left[ 1 + \sum_{n=1}^\infty 
   \alpha_n^L(\mu)\,C_n^{3/2}(2u-1) \right] ,
\end{equation}
where $C_1^{3/2}(x)=3x$, $C_2^{3/2}(x)=\frac32(5x^2-1)$, etc. Then 
the Gegenbauer moments $\alpha_n^L(\mu)$ are multiplicatively renormalized. 
The scale dependence of these quantities would, however, enter the 
results for the coefficients only at order $\alpha_s^2$, which is 
beyond the accuracy of our calculation. We assume that the leading-twist 
distribution amplitudes are close to the asymptotic form and thus 
truncate the expansion at $n=2$. However, it would be straightforward 
to account for higher-order terms if desired.
For the asymptotic form of the distribution amplitude, 
$\Phi_L(u)=6u(1-u)$, the integral in (\ref{a1dpi}) yields
\begin{eqnarray}\label{fintas}
   &&\int^1_0 du\,F(u,z)\,\Phi_L(u) 
    = 3\ln z^2 - 7 \nonumber\\
   &&\quad \mbox{}+ \Bigg[ \frac{6z(1-2z)}{(1-z)^2(1+z)^3}
    \left( \frac{\pi^2}{6} - \mbox{Li}_2(z^2) \right) 
    - \frac{3(2-3z+2z^2+z^3)}{(1-z)(1+z)^2} \ln(1-z^2) \nonumber\\
   &&\qquad\hspace*{0.3cm} \mbox{}+ \frac{4-17z+20z^2+5z^3}{2(1-z)(1+z)^2}
    + \{ z\to 1/z\} \Bigg] ,
\end{eqnarray}
and the corresponding result with the function $F(u,-z)$ is obtained
by replacing $z\to -z$. More generally, a numerical integration with 
a distribution amplitude 
expanded in Gegenbauer polynomials yields the results
collected in Table~\ref{tab:wfnint}. We observe that the first two
Gegenbauer polynomials in the expansion of the light-cone distribution
amplitudes give contributions of similar magnitude, whereas the
second moment gives rise to much smaller effects. This tendency 
persists in higher orders. For our numerical discussion, it is a safe
approximation to truncate the expansion after the first non-trivial 
moment. The dependence of the results on the value of the quark mass 
ratio $z=m_c/m_b$ is mild and can be neglected for all practical 
purposes. We also note that the difference of the convolutions with the
kernels for a pseudoscalar $D$ and vector $D^*$ meson are numerically
very small. This observation is, however, specific to the case of
$B\to D^{(*)} L$ decays and should not be generalized to other
decay modes.

\begin{table}
\caption{\label{tab:a1dpi}
The QCD coefficients $a_1(D^{(*)} L)$ at next-to-leading order for 
three different values of the renormalization scale $\mu$. The 
leading-order values are shown for comparison.}
\begin{center}
\begin{tabular}{|c|ccc|}
\hline
&&&\\[-0.4cm]
& $\mu=m_b/2$ & $\mu=m_b$ & $\mu=2 m_b$ \\
&&&\\[-0.4cm]
\hline
&&&\\[-0.4cm]
$a_1(D L)$ & $1.074+0.037i$ & $1.055+0.020i$ & $1.038+0.011i$ \\
 & $-(0.024-0.052i)\,\alpha_1^L$ & $-(0.013-0.028i)\,\alpha_1^L$ &
 $-(0.007-0.015i)\,\alpha_1^L$ \\
&&&\\[-0.4cm]
$a_1(D^* L)$ & $1.072+0.024i$ & $1.054+0.013i$ & $1.037+0.007i$ \\
 & $-(0.028-0.047i)\,\alpha_1^L$ & $-(0.015-0.025i)\,\alpha_1^L$ &
 $-(0.008-0.014i)\,\alpha_1^L$ \\
&&&\\[-0.4cm]
$a^{\rm LO}_1$ & $1.049$ & $1.025$ & $1.011$ \\[0.1cm]
\hline 
\end{tabular}
\end{center}
\end{table}

Next we evaluate the complete results for the parameters $a_1$ at
next-to-leading order, and to leading power in $\Lambda_{\rm QCD}/m_b$. 
We set $z=m_c/m_b=0.3$. Varying $z$ between 0.25 and 0.35 would change
the results by less than 0.5\%. The results are shown in 
Table~\ref{tab:a1dpi}. Note that the contribution proportional to the
second Gegenbauer moment $\alpha_2^L$ has coefficients of order 0.2\% 
or less. There is now increasing evidence that the leading-twist 
light-cone distribution amplitudes of light mesons are close to their 
asymptotic form, and that the Gegenbauer moments $\alpha_n^L$ take 
values at most of order 1 in magnitude. It then follows that the
contributions proportional to $\alpha_2^L$ can be safely neglected.
The contributions associated with $\alpha_1^L$ are present only for the
strange mesons $K$ and $K^*$, but not for $\pi$ and $\rho$.  Moreover,
the imaginary parts of the coefficients contribute to their modulus
only at order $\alpha_s^2$, which is beyond the accuracy of our 
analysis. To summarize, we thus obtain
\begin{eqnarray}\label{a1mods}
   |a_1(D L)| &=& 1.055_{-0.017}^{+0.019}
    - (0.013_{-0.006}^{+0.011})\alpha_1^L , \\
\label{a1star}
   |a_1(D^* L)| &=& 1.054_{-0.017}^{+0.018}
    - (0.015_{-0.007}^{+0.013})\alpha_1^L , 
\end{eqnarray}
where the quoted errors reflect the perturbative uncertainty due to 
the scale ambiguity (and the negligible dependence on the value of the 
ratio of quark masses and higher Gegenbauer moments), but not the 
effects of power-suppressed corrections. These will be estimated later.
It is evident that within theoretical uncertainties there is no 
significant difference between the two $a_1$ parameters, and moreover
there is only a very small sensitivity to the differences between
strange and non-strange mesons (assuming that $|\alpha_1^{K^{(*)}}|<1$).
In our numerical analysis below we take the fixed value $|a_1|=1.05$ 
for all decay modes.

\subsection{Tests of factorization}

The main lesson from the previous discussion is that corrections to
naive factorization in the class-I decays $\bar B_d\to D^{(*)+} 
L^-$ are very small. The reason is that these effects are governed 
by a small Wilson coefficient and, moreover, are colour 
suppressed by a factor $1/N_c^2$. For these decays, the most important
implications of the QCD factorization formula are to restore the
renormalization-group invariance of the theoretical predictions, and
to provide a theoretical justification for why naive 
factorization works so well. On the other hand, given the theoretical
uncertainties arising, e.g., from unknown power-suppressed corrections,
there is clearly no hope to confront the extremely small predictions
for non-universal (process-dependent) ``non-factorizable'' corrections
with experimental data. Rather, what we may do is ask whether data 
supports the prediction of a quasi-universal parameter $|a_1|\simeq 1.05$
in these decays. If this is indeed the case, it would support the 
usefulness of the heavy-quark limit in analyzing non-leptonic decay 
amplitudes. If, on the other hand, we were to find large non-universal
effects, this would point towards the existence of sizeable power
corrections to our predictions.

We will see that with present experimental errors the data are in good
agreement with our prediction of a quasi universal $a_1$ parameter. 
However, a reduction of the experimental 
uncertainties to the percent level would
be very desirable for obtaining a more conclusive picture.

\subsubsection{Ratios of non-leptonic decay rates}

We start by reconsidering the ratios of non-leptonic rates in 
(\ref{dpdsp}) and (\ref{drdp}). The calculable perturbative corrections 
to naive factorization are below the percent level. In the comparison
of $B\to D\pi$ and $B\to D^*\pi$ decays one is sensitive
to the difference of the values of the two $a_1$ parameters shown
in (\ref{a1mods}) and (\ref{a1star}) evaluated for $\alpha_1^L=0$. This 
difference is at most few times $10^{-3}$. Likewise, in the comparison 
of $B\to D\pi$ and $B\to D\rho$ decays one is sensitive
to the difference in the light-cone distribution amplitudes of the
pion and the $\rho$ meson, which start at the second Gegenbauer moment
$\alpha_2^L$. These effects are suppressed even more strongly. 

From the explicit expressions for the decay rates in 
(\ref{gbdpi})-(\ref{gbdrho}) it follows that
\begin{eqnarray}
   \frac{\Gamma(\bar B_d\to D^+\pi^-)}{\Gamma(\bar B_d\to D^{*+}\pi^-)}
   &=& \frac{(m_B^2-m_D^2)^2|\vec q\,|_{D\pi}}{4m_B^2|\vec q\,|_{D^*\pi}^3}
    \left( \frac{F_0(m_\pi^2)}{A_0(m_\pi^2)} \right)^2
    \left| \frac{a_1(D\pi)}{a_1(D^*\pi)} \right|^2 , \\
   \frac{\Gamma(\bar B_d\to D^+\rho^-)}{\Gamma(\bar B_d\to D^+\pi^-)}
   &=& \frac{4m_B^2|\vec q\,|_{D\rho}^3}{(m_B^2-m_D^2)^2|\vec q\,|_{D\pi}}
    \,\frac{f_\rho^2}{f_\pi^2}\,
    \left( \frac{F_+(m_\rho^2)}{F_0(m_\pi^2)} \right)^2
    \left| \frac{a_1(D\rho)}{a_1(D\pi)} \right|^2 .
\end{eqnarray}
Using the experimental values for the branching ratios reported by the
CLEO Collaboration in \cite{CLEO9701} we find (taking into account a 
correlation between some systematic errors in the second case)
\begin{eqnarray}
   \left| \frac{a_1(D\pi)}{a_1(D^*\pi)} \right|\,
   \frac{F_0(m_\pi^2)}{A_0(m_\pi^2)}
   &=& 1.00\pm 0.11 , \\
   \left| \frac{a_1(D\rho)}{a_1(D\pi)} \right|\,
   \frac{F_+(m_\rho^2)}{F_0(m_\pi^2)}
   &=& 1.16\pm 0.11 .
\end{eqnarray}
Within errors, there is thus no evidence for any deviations from 
naive factorization.

\subsubsection{Ratios of non-leptonic and semi-leptonic decay rates}

Our next-to-leading order results for the quantities 
$a_1(D^{(*)} L)$ allow us to make theoretical predictions which are
not restricted to ratios of hadronic decay rates. A particularly clean
test of these predictions, which is essentially free of hadronic 
uncertainties, is obtained by relating the $\bar B_d\to D^{(*)+} L^-$ 
decay rates to the differential semi-leptonic 
$\bar B_d\to D^{(*)+}\,l^-\nu$ decay rate evaluated at $q^2=m_L^2$ 
\cite{Bj89}. In this way the parameters $|a_1|$ can be measured directly.
One obtains
\begin{equation}\label{tfrpi}
   R_L^{(*)} = \frac{\Gamma(\bar B_d\to D^{(*)+} L^-)}
    {d\Gamma(\bar B_d\to D^{(*)+} l^-\bar\nu)/dq^2\big|_{q^2=m^2_L}}
   = 6\pi^2 |V_{ud}|^2 f^2_L\,|a_1(D^{(*)} L)|^2\,X^{(*)}_L ,
\end{equation}
where $X_\rho=X_\rho^*=1$ for a vector meson (because the production
of the lepton pair via a $V-A$ current in semi-leptonic decays is
kinematically equivalent to that of a vector meson with momentum $q$), 
whereas $X_\pi$ and $X_\pi^*$ deviate from 1 only by (calculable) terms 
of order $m_\pi^2/m_B^2$, which numerically are below the 1\% level
\cite{NeSt97}. We emphasize that with our results for $a_1$ in (\ref{a1dpi})
and (\ref{a1dspi}), Eq.~(\ref{tfrpi}) becomes a prediction based on first 
principles of QCD. This is to be contrasted with the usual 
interpretation of this formula, where $a_1$ plays the role of a 
phenomenological parameter that is fitted from data.

The most accurate test of factorization is at present possible for the 
class-I processes $\bar B_d\to D^{*+}L^-$, because the differential 
semi-leptonic decay rate in $B\to D^*$ transitions has been 
measured as a function of $q^2$ with good accuracy. The results of 
such an analysis, performed using the most recent CLEO data, have been 
reported in \cite{Rodr97}. One finds
\begin{eqnarray}\label{a1exp}
   R_\pi^* = (1.13\pm 0.15)\,\mbox{GeV}^2
   \quad &\Rightarrow& \quad
    |a_1(D^*\pi)| = 1.08 \pm 0.07 , \\
   R_\rho^* = (2.94\pm 0.54)\,\mbox{GeV}^2
   \quad &\Rightarrow& \quad
    |a_1(D^*\rho)| = 1.09\pm 0.10 , \\
   R_{a_1}^* = (3.45\pm 0.69)\,\mbox{GeV}^2
   \quad &\Rightarrow& \quad
    |a_1(D^* a_1)| = 1.08\pm 0.11 .
\end{eqnarray}
This is consistent with our theoretical result in (\ref{a1star}). In
particular, the data show no evidence for large power corrections to 
our predictions obtained at leading order in $\Lambda_{\rm QCD}/m_b$. 
However, a further improvement in the experimental accuracy would be 
desirable in order to become sensitive to process-dependent, 
non-factorizable effects.

\subsection{Predictions for class-I decay amplitudes}

We now consider a larger set of class-I decays of the form 
$\bar B_d\to D^{(*)+} L^-$, all of which are governed by the transition 
operator (\ref{heffa1}). In Table~\ref{tab:10decays} we compare
the QCD factorization 
predictions with experimental data. As previously we work in the 
heavy-quark limit, i.e.\ our predictions are model independent up to 
corrections suppressed by at least one power of $\Lambda_{\rm QCD}/m_b$. 
We keep the light meson masses in the phase space factors in 
(\ref{gbdpi})-(\ref{gbdsrho}), but we neglect them in the form factors,
i.e.\ we relate the various form factors to each other using the 
kinematic relations in (\ref{kinerela}).

\begin{table}
\caption{\label{tab:10decays}
Model-independent predictions for the branching ratios (in units of
$10^{-3}$) of class-I, non-leptonic $\bar B_d\to D^{(*)+} L^-$ decays 
in the heavy-quark limit. All predictions are in units of 
$(V_{cb}/0.04)^2\times(|a_1|/1.05)^2\times(\tau_{B_d}/1.56\,\mbox{ps})$.
The last two columns show the experimental results reported by the CLEO
Collaboration \protect\cite{CLEO9701}, and by the Particle Data Group 
\protect\cite{PDG}.}
\begin{center}
\begin{tabular}{|l|c|cc|}
\hline
&&&\\[-0.4cm]
Decay mode & Theory (HQL) & CLEO data & PDG98~ \\
&&&\\[-0.4cm]
\hline
&&&\\[-0.4cm]
$\bar B_d\to D^+\pi^-$   & 3.27 & $2.50\pm 0.40$ & $3.0\pm 0.4$ \\
$\bar B_d\to D^+ K^-$    & 0.25  & --- & --- \\
$\bar B_d\to D^+\rho^-$  & 7.64  & $7.89\pm 1.39$ & $7.9\pm 1.4$ \\
$\bar B_d\to D^+ K^{*-}$ & 0.39  & --- & --- \\
$\bar B_d\to D^+ a_1^-$  & 7.76  & $8.34\pm 1.66$ & $6.0\pm 3.3$ \\
 & $\times[F_+(0)/0.6]^2$ & & \\
&&&\\[-0.4cm]
\hline
&&&\\[-0.4cm]
$\bar B_d\to D^{*+}\pi^-$   & 3.05  & $2.34\pm 0.32$ & $2.8\pm 0.2$ \\
$\bar B_d\to D^{*+} K^-$    & 0.22  & --- & --- \\
$\bar B_d\to D^{*+}\rho^-$  & 7.59  & $7.34\pm 1.00$ & $6.7\pm 3.3$ \\
$\bar B_d\to D^{*+} K^{*-}$ & 0.40 & --- & --- \\
$\bar B_d\to D^{*+} a_1^-$  & 8.53 & $11.57\pm 2.02$ & $13.0\pm 2.7$ \\
 & $\times[A_0(0)/0.6]^2$ & & \\[0.1cm]
\hline 
\end{tabular}
\end{center}
\end{table}

The results show good agreement within the experimental errors, which
are still rather large. It would be desirable to reduce these
errors to the percent level. Note that we have not attempted to adjust 
the semi-leptonic form factors $F_+(0)$ and $A_0(0)$ so as to obtain
a best fit to the data. In this context we stress that the fact that
with $F_+(0)=A_0(0)=0.6$ our predictions for the 
$\bar B_d\to D^{(*)+}\pi^-$ branching ratios come out higher than the
central experimental results must {\em not\/} be taken as evidence
against QCD factorization in the heavy-quark limit. On the contrary, 
we have seen earlier in (\ref{a1exp}) that the value of $|a_1(D^*\pi)|$ 
extracted in a form-factor independent way is in very good agreement 
with our theoretical result.

We take the observation that, within errors, the experimental data on
class-I decays into heavy-light final states show good agreement with 
our predictions obtained in the heavy-quark limit as (weak) evidence 
that in these decays there are no unexpectedly large power corrections.
We will now address the important question of the size of power 
corrections theoretically. To this end we provide rough estimates of
two sources of power-suppressed effects: weak annihilation and spectator
interactions. We stress that a complete account of power corrections
to the heavy-quark limit cannot be performed in a systematic 
way, since these effects are no longer dominated by hard
gluon exchange. In other words, factorization breaks down beyond 
leading power. We believe that the estimates presented here are both
instructive and realistic. Yet, it is important to keep in mind that 
there are other sources of power corrections, e.g., contributions from 
higher Fock states in the light-cone expansion of meson wave functions,
which we will not address here. 

To obtain an estimate of the power corrections we adopt the following, 
heuristic procedure. We treat the charm quark as light compared to the 
large scale provided by the mass of the decaying $b$ quark ($m_c\ll
m_b$ and $m_c$ fixed as $m_b\to\infty$)
and use a light-cone projection similar to that of the pion also for 
the $D$ meson. In addition we assume that $m_c$ is still large compared 
to $\Lambda_{\rm QCD}$. We implement this by using a highly asymmetric
$D$-meson wave function, which is strongly peaked at a light-quark 
momentum fraction of order $\Lambda_{\rm QCD}/m_D$. This guarantees 
correct power counting for the heavy-light final states we are 
interested in and allows us to obtain simple, semi-quantitative 
estimates. As discussed in Sect.~\ref{subsec:annihilation}
there are four annihilation diagrams with single gluon exchange (see 
Fig.~\ref{fig9} (a)-(d)). The first two diagrams are ``factorizable'' 
and their contributions vanish because of current conservation in the 
limit $m_c\to 0$. For non-zero $m_c$ they therefore carry an additional 
suppression factor $m^2_D/m^2_B\approx 0.1$. Moreover, their 
contributions to the decay amplitude are suppressed by small Wilson 
coefficients. Diagrams (a) and (b) can therefore safely be neglected.
From the non-factorizable diagrams (c) and (d) in Fig.~\ref{fig9}, 
the one with the gluon attached to the $b$ quark turns out to be strongly
suppressed numerically, giving a contribution of less than $1\%$ of the 
leading class-I amplitude. We are thus left with diagram (d), in which 
the gluon couples to the light quark in the $B$ meson. This mechanism 
gives the dominant annihilation contribution. (Note that by deforming 
the light spectator-quark line one can redraw this diagram in such a way 
that it can be interpreted as a final-state rescattering process.) 

Adopting a common notation, we parameterize the annihilation contribution 
to the $\bar B_d\to D^+\pi^-$ decay amplitude in terms of a 
(power-suppressed) amplitude $A$ such that 
\begin{equation}
   {\cal A}(\bar B_d\to D^+\pi^-) = T + A ,
\end{equation}
where $T$ is the ``tree topology'', which contains the dominant 
factorizable contribution. A straightforward calculation using the 
approximations discussed above shows that the contribution of diagram (d) 
is (to leading order) independent of the momentum fraction $\xi$ of the 
light quark inside the $B$ meson:
\begin{equation}\label{wa9d}
   A \sim f_\pi f_D f_B \int du\,\frac{\Phi_\pi(u)}{u}
   \int dv\,\frac{\Phi_D(v)}{\bar v^2} 
   \simeq 3 f_\pi f_D f_B\,\int dv\,\frac{\Phi_D(v)}{\bar v^2} .
\end{equation}
The $B$-meson wave function simply integrates to $f_B$, and the 
integral over the pion distribution amplitude can be performed using
the asymptotic form of the wave function. We take $\Phi_D(v)$ in the 
form of (\ref{gpol}) with the coefficients $\alpha^D_1=0.8$ and 
$\alpha^D_2=0.4$ ($\alpha^D_i=0$, $i>2$). With this ansatz $\Phi_D(v)$ 
is strongly peaked at $\bar v\sim\Lambda_{\rm QCD}/m_D$. The integral 
over $\Phi_D(v)$ in (\ref{wa9d}) is divergent at $v=1$, and we
regulate it by introducing a cut-off such that $v\le 1-\Lambda/m_B$,
where $\Lambda\approx 0.3$\,GeV. Then 
$\int dv\,\Phi_D(v)/\bar v^2\approx 34$. Evidently, the proper value 
of $\Lambda$ is largely unknown and our estimate will be correspondingly
uncertain. Nevertheless, this exercise will give us an idea of the 
magnitude of the effect. For the ratio of the annihilation amplitude to 
the leading, factorizable contribution we obtain
\begin{equation}\label{ws1}
   \frac{A}{T} \simeq \frac{2\pi\alpha_s}{3}\,
   \frac{C_+ + C_-}{2C_+ + C_-}\,
   \frac{f_D f_B}{F_0(0)\,m_B^2}\int dv\,\frac{\Phi_D(v)}{\bar v^2}
   \approx 0.04 .
\end{equation}
We have evaluated the Wilson coefficients at $\mu=m_b$ and used 
$f_D=0.2$\,GeV, $f_B=0.18$\,GeV, $F_0(0)=0.6$, and $\alpha_s=0.4$.
This value of the strong coupling constant reflects that the typical 
virtuality of the gluon propagator in the annihilation graph is of 
order $\Lambda_{\rm QCD} m_B$. We conclude that the annihilation 
contribution is a correction of a few percent, which is what one would
expect for a generic power correction to the heavy-quark limit. 
Taking into account that $f_B\sim\Lambda_{\rm QCD}^{3/2} m_B^{-1/2}$,
$F_0(0)\sim\Lambda_{\rm QCD}^{3/2} m_B^{-3/2}$ and 
$f_D\sim\Lambda_{\rm QCD}$, we observe that in the heavy-quark limit 
the ratio $A/T$ indeed scales as $\Lambda_{\rm QCD}/m_b$, exhibiting 
the expected linear power suppression. (Recall that we consider the 
$D$ meson as a {\em light\/} meson for this heuristic analysis of 
power corrections.)

Using the same approach we may also derive a numerical estimate for 
the non-factorizable spectator interaction in $\bar B_d\to D^+\pi^-$ 
decays, discussed in Sect.~\ref{subsec:hardspec}. Adapting the 
corresponding result derived in \cite{BBNS99} for the spectator
interaction in $\bar B_d\to\pi^+\pi^-$ decays we find
\begin{equation}\label{estnfs}
   \frac{T_{\rm spec}}{T_{\rm lead}}
   \simeq \frac{2\pi\alpha_s}{3}\,
   \frac{C_+ - C_-}{2C_+ + C_-}\,
   \frac{f_D f_B}{F_0(0)\,m_B^2}\frac{m_B}{\lambda_B}
   \int dv\,\frac{\Phi_D(v)}{\bar v} \approx -0.03 ,
\end{equation}
where the hadronic parameter $\lambda_B=O(\Lambda_{\rm QCD})$ is 
defined as $\int_0^1(d\xi/\xi)\,\Phi_B(\xi)\equiv m_B/\lambda_B$.
For the numerical estimate we have assumed that 
$\lambda_B\approx 0.3$\,GeV. With the same model for $\Phi_D(v)$ as 
above we have $\int dv\,\Phi_D(v)/\bar v\approx 6.6$, where the integral 
is now convergent. The result (\ref{estnfs}) exhibits again the 
expected power suppression in the heavy-quark limit, and the numerical 
size of the effect is at the few percent level.

We conclude from this discussion that the typical size of power
corrections to the heavy-quark limit in class-I decays of $B$ mesons
into heavy-light final states is at the level of 10\% or less, and
thus our prediction for the near universality of the parameters $a_1$
governing these decay modes appears robust. 

\subsection{Remarks on class-II and class-III decay amplitudes}

In the class-I decays $\bar B_d\to D^{(*)+} L^-$ considered above, 
the flavour quantum numbers of the final-state mesons ensure that only
the light meson $L$ can be produced by the $(\bar d u)$ current 
contained in the operators of the effective weak Hamiltonian in 
(\ref{heff18}). The QCD factorization formula then predicts that the
corresponding decay amplitudes are factorizable in the heavy-quark 
limit. The formula also predicts that other topologies, in which
the heavy charm meson would be created by a $(\bar c u)$ current,
are power suppressed. To study these topologies we now consider
decays with a neutral charm meson in the final state. In the 
class-II decays $\bar B_d\to D^{(*)0} L^0$ the only possible topology
is to have the charm meson as the emission particle, whereas for the 
class-III decays $B^-\to D^{(*)0} L^-$ both final-state mesons can
be the emission particle. The factorization formula predicts that
in the heavy-quark limit class-II decay amplitudes are power suppressed 
with respect to the corresponding class-I amplitude, whereas class-III
amplitudes should be equal to the corresponding class-I amplitudes
up to power corrections. 

It is convenient to introduce two common parameterizations of the 
decay amplitudes, one in terms of isospin amplitudes $A_{1/2}$ and 
$A_{3/2}$ referring to the isospin of the final-state particles, and 
one in terms of flavour topologies ($T$ for ``tree topology'', $C$ for
``colour suppressed tree topology'', and $A$ for ``annihilation 
topology''). Taking the decays $B\to D\pi$ as an example, we have
\begin{eqnarray}\label{asw}
   {\cal A}(\bar B_d\to D^+\pi^-)
   &=& \sqrt{\frac13} A_{3/2} + \sqrt{\frac23} A_{1/2}
    = T + A , \label{asw+-} \\
   \sqrt2\,{\cal A}(\bar B_d\to D^0\pi^0) 
   &=& \sqrt{\frac43} A_{3/2} - \sqrt{\frac23} A_{1/2}
    = C - A , \label{asw00} \\
   {\cal A}(B^-\to D^0\pi^-) &=& \sqrt3 A_{3/2} = T + C . \label{asw0-}
\end{eqnarray}
A similar decomposition holds for the other $B\to D^{(*)} L$ 
decay modes. Note that isospin symmetry of the strong interactions 
implies that the class-III amplitude is a linear combination of the 
class-I and class-II amplitudes. (In the case of final states containing
two vector mesons, this statement applies separately for each of the 
three helicity amplitudes.) In other words, there are only two 
independent amplitudes, which can be taken to be $A_{1/2}$ and 
$A_{3/2}$, or $(T+A)$ and $(C-A)$. These amplitudes are complex due to
strong-interaction phases from final-state interactions. Only the 
relative phase of the two independent amplitudes is an observable. We 
define $\delta$ to be the relative phase of $A_{1/2}$ and $A_{3/2}$,
and $\delta_{TC}$ the relative phase of $(T+A)$ and $(C-A)$. The QCD 
factorization formula implies that
\begin{eqnarray}
   \frac{A_{1/2}}{\sqrt2\,A_{3/2}}
   &=& 1 + O(\Lambda_{\rm QCD}/m_b) , \qquad
    \delta = O(\Lambda_{\rm QCD}/m_b) , \\
\label{delTC}
   \frac{C-A}{T+A}
   &=& O(\Lambda_{\rm QCD}/m_b) , \hspace{1.54cm}
    \delta_{TC} = O(1) .
\end{eqnarray}
In the remainder of this section, we will explore to what extent 
these predictions are supported by data.

\begin{table}
\caption{\label{tab:brbdpi}
CLEO data on the branching ratios for the decays $B\to D^{(*)} L$ 
in units of $10^{-3}$ \protect\cite{CLEO9701,CLEOsupp}. Upper limits 
are at 90\% confidence level. See text for the definition of the 
quantities $\delta$ and ${\cal R}$.}
\begin{center}
\begin{tabular}{|c|cccc|}
\hline
&&&&\\[-0.4cm]
 & $B\to D\pi$ & $B\to D\rho$ & $B\to D^*\pi$ & $B\to D^*\rho$ \\
&&&&\\[-0.4cm]
\hline
&&&&\\[-0.4cm]
Class-I~~ ($D^{(*)+} L^-$) & $2.50\pm 0.40$ & $ 7.89\pm 1.39$
 & $2.34\pm 0.32$ & $7.34\pm 1.00$ \\
Class-II~~ ($D^{(*)0} L^0$) & $<0.12$ & $<0.39$ & $<0.44$ & $<0.56$ \\
Class-III ($D^{(*)0} L^-$) & $4.73\pm 0.44$ & $9.20\pm 1.11$
 & $3.92\pm 0.63$ & $12.77\pm 1.94$ \\
&&&&\\[-0.4cm]
\hline
&&&&\\[-0.4cm]
$\delta$ & $<22^\circ$ & $<30^\circ$ & $<57^\circ$ & $<31^\circ$ \\
${\cal R}$ & $1.34\pm 0.13$ & $1.05\pm 0.12$ & $1.26\pm 0.14$ 
 & $1.28\pm 0.13$ \\[0.1cm]
\hline 
\end{tabular}
\end{center}
\end{table}

In Table~\ref{tab:brbdpi} we show the experimental results for the
various $B\to D^{(*)} L$ branching ratios reported by the CLEO
Collaboration \cite{CLEO9701,CLEOsupp}. We first note that no evidence
has been seen for any of the class-II decays, in accordance with our 
prediction that these decays are suppressed with respect to the class-I
modes in the heavy-quark limit. Below we will investigate in more 
detail how this suppression is realized. The fourth line in the table
shows upper limits on the strong-interaction phase difference $\delta$ 
between the two isospin amplitudes. These bounds follow from the relation 
\cite{NeSt97}
\begin{equation}
   \sin^2\!\delta < \frac92\,\frac{\tau(B^-)}{\tau(B_d)}\,
   \frac{\mbox{Br}(\bar B_d\to D^0\pi^0)}
        {\mbox{Br}(B^-\to D^0\pi^-)} ,
\end{equation}
where we use $\tau(B^-)=1.65$\,ps and $\tau(B_d)=1.56$\,ps for the 
$B$-meson lifetimes \cite{PDG}. The strongest bound arises in the
decays $B\to D\pi$, where the strong-interaction phase is bound
to be less than $22^\circ$. This confirms our prediction that the
phase $\delta$ is suppressed in the heavy-quark limit.

Let us now study the suppression of the class-II amplitudes in 
more detail. We have already mentioned in Sect.~\ref{subsec:diffcharm}
that the observed smallness of class-II amplitudes is more a reflection
of colour suppression than power suppression. This is already apparent
in the naive factorization approximation, because the appropriate
ratios of meson decay constants and semi-leptonic form factors exhibit
large deviations from their expected scaling laws in the heavy-quark 
limit, see (\ref{scalingviolations}). Indeed, it is obvious from 
Table~\ref{tab:brbdpi} that there are significant differences between
the class-I and class-III amplitudes, indicating that some 
power-suppressed contributions are not negligible. In the last line in 
the table we show the experimental values of the quantity
\begin{equation}
   {\cal R} = \left|  
   \frac{{\cal A}(B^-\to D^{(*)0} L^-)}
          {{\cal A}(\bar B_d\to D^{(*)+} L^-)} \right|
   = \sqrt{ \frac{\tau(B_d)}{\tau(B^-)}\,
    \frac{\mbox{Br}(B^-\to D^{(*)0} L^-)}
         {\mbox{Br}(\bar B_d\to D^{(*)+} L^-)} } ,
\end{equation}
which parameterizes the magnitude of power-suppressed effects at the
level of the decay amplitudes. If we ignore the decays 
$B\to D^*\rho$ with two vector mesons in the final state, which
are more complicated because of the presence of different helicity
amplitudes, then the ratio ${\cal R}$ is given by
\begin{equation}\label{Rdef}
   {\cal R} = \left| 1 + \frac{C-A}{T+A} \right|
   = \left| 1 + x\,\frac{a_2}{a_1} \right| ,
\end{equation}
where $a_1$ are the QCD parameters entering the transition operator in 
(\ref{heffa1}), and
\begin{equation}
   a_2 = \frac{N_c+1}{2 N_c}\,C_+ - \frac{N_c-1}{2 N_c}\,C_-
   + \mbox{``non-factorizable corrections''} 
\end{equation}
are the corresponding parameters describing the deviations from 
naive factorization in the class-II decays (see e.g.\ 
\cite{NeSt97}). All the quantities in (\ref{Rdef}) depend on the
nature of the final-state mesons. In particular, the parameters
\begin{eqnarray}
   x(D\pi)
   &=& \frac{(m_B^2-m_\pi^2)\,f_D\,F_0^{B\to\pi}(m_D^2)}
            {(m_B^2-m_D^2)\,f_\pi\,F_0^{B\to D}(m_\pi^2)} 
    \approx 0.9 , \\
   x(D\rho)
   &=& \frac{f_D\,A_0^{B\to\rho}(m_D^2)}
            {f_\rho\,F_+^{B\to D}(m_\rho^2)} 
    \approx 0.5 , \\
   x(D^*\pi)
   &=& \frac{f_{D^*} F_+^{B\to\pi}(m_{D^*}^2)}
            {f_\pi\,A_0^{B\to D^*}(m_\pi^2)} 
    \approx 0.9 ,
\end{eqnarray}
account for the ratios of decay constants and form factors entering
in the naive factorization approximation. For the numerical
estimates we have assumed that the ratios of heavy-to-light over
heavy-to-heavy form factors are approximately equal to 0.5, and we have 
taken $f_D=0.2$\,GeV and $f_{D^*}=0.23$\,GeV for the charm meson decay 
constants. Note that in (\ref{Rdef}) it is the quantities $x$ that are 
formally power suppressed (by a factor of order 
$\Lambda_{\rm QCD}^2/m_b^2$) in the heavy-quark limit, 
not the ratios $a_2/a_1$. For the final states containing a pion the 
power suppression is clearly not operative, mainly due to the fact that 
the pion decay constant $f_\pi$ is much smaller than the quantity 
$(f_D\sqrt{m_D})^{2/3}\approx 0.42$\,GeV. To reproduce the experimental
values of the ratios ${\cal R}$ shown in Table~\ref{tab:brbdpi}
requires values of $a_2/a_1$ of order 0.1--0.4 (with large 
uncertainties), which is consistent with the fact that these ratios 
are of order $1/N_c$ in the large-$N_c$ limit, i.e.\ they are colour 
suppressed. 

The QCD factorization formula (\ref{fff}) allows us to compute the
coefficients $a_1$ in the heavy-quark limit, but it does {\em not\/}
allow us to compute the corresponding parameters $a_2$ in class-II 
decays. The reason is that in class-II decays the emission particle is
a heavy charm meson, and hence the mechanism of colour transparency,
which was essential for the proof of factorization, is not operative. 
For a rough estimate of $a_2$ in $B\to\pi D$ decays we consider 
as previously the limit
in which the charm meson is treated as a light meson ($m_c\ll m_b$), 
however with a highly asymmetric distribution amplitude. In this 
limit we can adapt our results for the class-II amplitude in 
$B\to\pi\pi$ decays derived in \cite{BBNS99}, with the only
modification that the hard-scattering kernel must be generalized to
the case where the leading-twist light-cone distribution amplitude
of the emission meson is not symmetric. We find that
\begin{eqnarray}
   a_2 &\simeq& \frac{N_c+1}{2 N_c}\,\bar C_+(\mu)
    - \frac{N_c-1}{2 N_c}\,\bar C_-(\mu) \nonumber\\
   &&\mbox{}+ \frac{\alpha_s}{4\pi}\,\frac{C_F}{2 N_c}\,
    [\bar C_+(\mu)+\bar C_-(\mu)] \left( -6\ln\frac{\mu^2}{m_b^2}
    + f_I + f_{II} \right) ,
\end{eqnarray}
where
\begin{eqnarray}
   f_I &=& \int_0^1 dv\,\Phi_D(v) \left[
    \ln^2\!\bar v + \ln\bar v + \frac{\pi^2}{3} - 6
    + i\pi (2\ln\bar v - 3) + O(\bar v) \right] , \\
   f_{II} &=& \frac{12\pi^2}{N_c}\,
    \frac{f_\pi f_B}{F_0^{B\to\pi}(m_D^2)\,m_B^2}\,
    \frac{m_B}{\lambda_B}\int dv \frac{\Phi_D(v)}{\bar v}.
\end{eqnarray}
The contribution from $f_{II}$ describes the hard, non-factorizable 
spectator interaction. Note that this term involves 
$\int dv\,\Phi_D(v)/\bar v$, which can be sizeable but remains 
constant in the heavy-quark limit implied here ($m_b\to\infty$ with  
$m_c$ constant). Using the same numerical inputs as previously, we find 
that $f_{II}\approx 13$ and $f_I\approx-1-19i$. In writing the 
hard-scattering kernel for $f_I$ we have only kept the leading terms in 
$\bar v$, in accordance with the strongly asymmetric shape of 
$\Phi_D(v)$. Note the large imaginary 
part arising from the ``non-factorizable'' vertex corrections with a gluon 
exchange between the final-state quarks. Combining all contributions, and 
taking $\mu=m_b$ for the renormalization scale, we find
\begin{equation}\label{a2est}
   a_2 \approx 0.25\,e^{-i 41^\circ} ,
\end{equation}
which is significantly larger in magnitude than the leading-order 
result $a_2^{\rm LO}\approx 0.12$ corresponding to naive 
factorization. We hasten to add that our estimate (\ref{a2est}) 
should not be taken too seriously since it is most likely not a good
approximation to treat the charm meson as a light meson. Nevertheless,
we find it remarkable that in this idealized limit one obtains indeed
a very significant correction to naive factorization, which gives
the right order of magnitude for the modulus of $a_2$ and, at the same 
time, a large strong-interaction phase. For completeness, we note that 
the value for $a_2$ in (\ref{a2est}) would imply a strong-interaction 
phase difference $\delta\approx 10^\circ$ between the two isospin amplitudes 
$A_{1/2}$ and $A_{3/2}$ in $B\to D\pi$ decays, and hence is not in 
conflict with the experimental upper bound on the phase $\delta$ given in 
Table~\ref{tab:brbdpi}. The phase $\delta_{TC}$, on the other hand, is to 
leading order simply given by the phase of $a_2$ and is indeed large, in 
accordance with (\ref{delTC}).


\section{Comparison with previous approaches}
\label{sec:comparison}

The theoretical understanding of non-leptonic weak decays has always 
been a challenge for theorists. Because of the complexity of the
corresponding hadronic matrix elements progress in this field has been 
very slow, and most of the phenomenological work was based on simple 
models and assumptions. In the present work we have presented, 
for the first time, a consistent theoretical framework allowing us 
to perform a systematic, model-independent study of a large class of 
two-body $B$ decays in the heavy-quark limit. For the particular case 
of $B$ decays into a heavy-light final state such as $D\pi$, a 
factorization formula of the form (\ref{bdpi18}) has been used previously 
by Politzer and Wise \cite{PW91}. Although in their work no attempt is 
made to prove factorization, the underlying physical motivation for 
their approach was the same as in our case. The extension of the 
factorization formula to a wider class of decay modes, including in
particular those with two light mesons in the final state, is however 
non-trivial and has been presented here and in 
\cite{BBNS99} for the first time.

We will now set our approach in perspective with previous attempts to 
tackle the problem of non-leptonic decays. These can be 
grouped into three classes: phenomenological models, dynamical 
approaches, and methods based on classifications in terms of flavour 
topologies or Wick contractions. The first class 
consists of various formulations and generalizations of the naive 
factorization hypothesis, which typically introduce a small set of 
phenomenological parameters in order to parameterize important 
non-factorizable effects. No attempt is made to calculate these 
parameters from a fundamental theory. The second class consists of 
several different approaches aiming at a dynamical understanding of 
non-leptonic weak decays starting from QCD and making a controlled set 
of approximations. We will briefly discuss the large-$N_c$ expansion, 
lattice field theory, QCD sum rules, large-energy effective theories, 
and hard-scattering approaches in this category. We also comment on 
previous treatments of final-state rescattering phases. The third class 
of approaches aims at a convenient parameterization of non-leptonic 
amplitudes rather than at a dynamical calculation. To this end, the 
amplitudes are decomposed into invariant subamplitudes, which are 
either associated with certain flavour topologies and classified 
according to their transformation properties under isospin or SU(3) 
flavour symmetries, or chosen to correspond to certain Wick 
contractions of operators in the effective weak Hamiltonian, defined 
in a renormalization-scheme invariant way. Apart from flavour symmetry 
relations, these amplitudes are treated as phenomenological parameters
to be determined from experiment.

\subsection{Phenomenological approaches}
\label{subsec:pheno}

Here we summarize different formulations and generalizations of the
concept of naive factorization in non-leptonic $B$ decays, taking as
an example the decays $\bar{B}_d\to D^+\pi^-$ and $\bar{B}_d\to D^0\pi^0$.
The non-factorizable effects in these decays can be parameterized in
terms of quantities $a_1$ and $a_2$, respectively. The QCD 
factorization formula applies only in the first case and leads to 
a calculable expression for $a_1$ given in (\ref{a1dpi}), 
which is valid to leading power in $\Lambda_{\rm QCD}/m_b$. 

We have already mentioned that for class-I decays into heavy-light 
final states (and all decays into two light mesons) the naive 
factorization model \cite{Schw64,Feyn65,Stec70}, in which all 
non-factorizable gluon exchanges are ignored, is contained as the 
leading term in our approach. In this model the parameters
$a_1=C_1(\mu)+C_2(\mu)/N_c$ and $a_2=C_2(\mu)+C_1(\mu)/N_c$ carry a 
renormalization-scale and -scheme dependence, which remains uncompensated 
because the factorized amplitudes multiplying these quantities are scale 
and scheme independent. (In this section we
adopt the standard parameterization of the effective weak Hamiltonian. 
The coefficients $C_0$ and $C_8$ of the singlet-octet basis used in 
Sects.~\protect\ref{oneloop}-\protect\ref{bdpi} are related to
the standard coefficients by $C_0=C_1+C_2/N_c$ and $C_8=2 C_2$.) 
Without knowing that ``non-factorizable'' 
corrections are actually dominated by hard scattering the scale $\mu$ 
could be of order $\Lambda_{\rm QCD}$, in which case the scale-dependent 
terms would not be suppressed by any small parameter. This indicates 
that an important aspect of the physics (i.e.\ ``non-factorizable'' 
exchanges) is missing in this model. 

Several phenomenological recipes have been proposed for fixing this 
deficiency. Typically, they aim at parameterizing (rather than 
ignoring) the dominant part of the non-factorizable corrections by 
introducing a small number of phenomenological parameters. To maintain 
predictive power, it is assumed that these parameters are universal 
(i.e.\ process independent) for classes of decays sharing similar 
kinematics. This treatment is known as the ``generalized factorization 
hypothesis''. The first proposal in this direction was the basis of 
the Bauer-Stech-Wirbel model for non-leptonic decays \cite{BSW}, in 
which one sets $a_1=C_1(m_b)+\xi\,C_2(m_b)$ and 
$a_2=C_2(m_b)+\xi\,C_1(m_b)$ with the quantity $\xi$ of order $1/N_c$ 
treated as a free parameter. A phenomenological analysis of charm meson 
decays indicated that setting $\xi\approx 0$ provided a successful 
description of two-body $D$ decays. This observation found
theoretical support in the framework of a systematic
$1/N_c$ expansion \cite{BGR86}.
However, it was soon realized that 
the ``rule of discarding the $1/N_c$ terms'' would not work in $B$ 
decays \cite{Xu92,Dean93}. An equivalent formulation uses the notion 
of an ``effective number of colours'', $\xi\to 1/N_c^{\rm eff}$, where 
typically $N_c^{\rm eff}$ is varied between 2 and infinity 
\cite{Nceff}. To some extent, the generalized factorization ansatz 
was motivated by the large-$N_c$ counting rules of QCD, which show 
that non-factorizable effects are of the same order as the $1/N_c$ 
terms kept in the naive factorization approach. Hence, it was natural 
to replace these terms by a more flexible parameterization. However,
using the same parameter $\xi$ in the expressions for $a_1$ and 
$a_2$, and assuming that $\xi$ should be the same in $K$, $D$ and $B$ 
decays, was an oversimplification lacking any theoretical 
justification. 

A refined version of generalized factorization has been proposed,
in which non-factorizable contributions are classified according to 
their behaviour in the large-$N_c$ limit \cite{NeSt97}. For $B$ decays
one finds the simple relations $a_1=1+O(1/N_c^2)$ and 
$a_2=C_2(m_b)+\xi\,C_1(m_b)+O(1/N_c^3)$, where 
$\xi=1/N_c+\varepsilon_8(m_b)$ is related to a non-factorizable 
colour-octet matrix element of order $1/N_c$, which in general is a
process-dependent quantity \cite{DGSu80,Chen94,Soa95,KaAl}. In the 
next step, the colour-transparency argument \cite{Bj89} was invoked 
to argue that for renormalization scales of order $m_b$ the process 
dependence of this matrix element is expected to be a small effect 
\cite{NeSt97}. This led to a successful phenomenological description of
a large class of Cabibbo-favoured two-body $B$ decays. However, it 
remained unclear if (and why) a similar treatment should work for
more complicated, rare decay processes, in which penguin operators
play an important role. In the literature, it has often been assumed
that the same effective parameter $N_c^{\rm eff}$ can be used to 
account for non-factorizable contributions to the matrix elements of 
all operators in the effective weak Hamiltonian \cite{AlGr98,AKL98}, 
or that two parameters $N_c^{\rm eff}(LL)$ and $N_c^{\rm eff}(LR)$, 
referring to operators with chiral structure $(V-A)\otimes(V-A)$ or 
$(V-A)\otimes(V+A)$ respectively, would suffice to account for these 
effects \cite{Nceff}. In the present paper and in our previous work 
\cite{BBNS99}, we have shown that even at leading power in 
$\Lambda_{\rm QCD}/m_b$ the ``non-factorizable'' effects in rare $B$ decays 
have a richer structure than assumed in these analyses.

Because of the renormalization-scale and -scheme dependence of the
Wilson coefficients $C_1(\mu)$ and $C_2(\mu)$, the parameters $\xi$
and $N_c^{\rm eff}$ introduced in generalized factorization are
unphysical quantities, which carry a scale and scheme dependence 
in such a way that the resulting expressions for the quantities $a_1$
and $a_2$ are renormalization-group invariant. One must therefore
be careful when trying to give physical significance to the values
extracted for $\xi$ and $N_c^{\rm eff}$ from a fit to data 
\cite{Bur95}. The parameters $a_1$ and $a_2$, however, are physical 
by definition. In many recent phenomenological analyses based on 
generalized factorization the authors have tried to avoid the 
problems of renormalization-group dependence of the Wilson 
coefficients by using so-called ``effective, scheme-independent 
Wilson coefficients'' $C_i^{\rm eff}$ 
\cite{Kram,Nceff,AlGr98,AKL98,KaAl}. 
These coefficients are related to the original ones by an equation 
of the form
\begin{equation}\label{sec7:Cieff}
   C_i^{\rm eff} = C_i(\mu) + \frac{\alpha_s(\mu)}{4\pi} \left(
   \gamma_V^T \ln\frac{m_b}{\mu} + r_V^T \right)_{ij} C_j(\mu),
\end{equation}
where the matrix $(\ldots)_{ij}$ contains the ultraviolet
logarithms and certain process-independent parts of the vertex-correction 
diagrams of the operators in the effective weak Hamiltonian 
\cite{Bur92}. (In addition, for $i\ne 1,2$ the penguin contractions 
of the local operators were evaluated in perturbation theory and their 
contributions absorbed into the definition of the effective 
coefficients.) These matrices are chosen in such a way that the 
resulting expressions for $C_i^{\rm eff}$ are formally scale and scheme 
independent. It is important to realize, however, that such a treatment
does not achieve an improvement of the accuracy of naive or generalized
factorization in a parametric way. Non-trivial, process-dependent 
corrections of the same order as the extra terms in (\ref{sec7:Cieff}) 
are neglected. In practice, the definition of the effective 
coefficients is nothing but the choice of a particular renormalization
scheme $V$, defined such that $\mu=m_b$ and $r_V=0$. It has also
been pointed out that, in general, the values of the effective 
coefficients depend on the gauge and infrared regulator, and as such
are unphysical \cite{Bur95}. All of these shortcomings are resolved in 
our approach, where all ``non-factorizable'' terms of leading power are 
retained.

\subsection{Dynamical approaches}
\label{subsec:dynamic}

Because of the complexity of non-leptonic weak decays, dynamical
calculations starting from first principles of QCD have not been
very successful so far in producing useful predictions
for the decay amplitudes, or even in providing a semi-quantitative
understanding of the hadronic dynamics involved in these processes. 
Indeed, most of the approaches face difficult
conceptual problems, which cannot be overcome in a straightforward
way by increasing the level of technical sophistication. 

The factorization formula established in the present work changes 
this situation in that it provides a systematic basis for 
a discussion of most non-leptonic $B$ 
decays in a well-defined approximation given by the heavy-quark limit, 
i.e.\ an expansion in powers of $\Lambda_{\rm QCD}/m_b$. In the 
following we summarize earlier dynamical approaches and compare them 
to our results obtained in the heavy-quark limit where appropriate.

\boldmath
\subsubsection{Large-$N_c$ expansion}
\unboldmath

An expansion around the limit of a large number of colours is an 
important theoretical tool in the study of non-perturbative properties 
of QCD, which in particular has led to insights into the
dynamics of hadronic weak decays. Factorization of non-leptonic decay
amplitudes becomes exact in the large-$N_c$ limit, and hence an
expansion in powers of $1/N_c$ provides a natural framework in which
to discuss the structure of non-factorizable corrections \cite{BGR86}. 
For kaon decays, detailed calculations of non-factorizable effects
have been performed by combining the $1/N_c$ expansion with methods of 
chiral perturbation theory \cite{kaon_largeN}. An important outcome of
this analysis was that at subleading order in $1/N_c$ there are, in 
general, two types of contributions: $1/N_c$ terms present
in naive factorization (which result from a colour Fierz-reordering of 
the operators in the effective Hamiltonian), and non-factorizable 
effects that are genuinely non-perturbative. As a consequence, naive
factorization, where the second type of contribution is neglected, 
cannot be justified theoretically beyond the large-$N_c$ limit. This
observation initiated attempts to generalize the naive factorization
approach by treating the terms of order $1/N_c$ as phenomenological
parameters \cite{BSW}. 

The large-$N_c$ counting rules were useful also in the analysis of 
non-leptonic $B$ decays, despite the fact that chiral perturbation 
theory does not apply in this case. For so-called class-I and class-II
decays governed by phenomenological parameters $a_1\sim 1$ and 
$a_2\sim 1/N_c$, respectively, one can show that
\begin{equation}
   a_1 = C_1(m_b) + O(1/N_c^2) \,, \qquad
   a_2 = C_2(m_b) + \xi\,C_1(m_b) + O(1/N_c^3),
\end{equation}
where $\xi=1/N_c+\varepsilon_8(m_b)$ is a non-perturbative, 
process-dependent hadronic matrix element of a colour octet-octet
operator \cite{NeSt97}. In addition to the arguments based on the 
heavy-quark limit presented in this paper, this discussion shows
that non-factorizable corrections in class-I decays are generally 
suppressed by two powers of $1/N_c$. The situation is, however, 
different for other decays, where such a suppression does not persist.
For instance, in class-II decays non-factorizable corrections have the
same $1/N_c$ scaling as the leading factorizable contributions. On
the other hand, our discussion based on the heavy-quark limit still 
applies to class-II decays, provided there are two light mesons in the
final state.

\subsubsection{Lattice field theory}

The evaluation of the matrix elements corresponding to exclusive
non-leptonic $B$ decays represents a major challenge for lattice field
theory. No results have been obtained up to now and new theoretical
ideas have to be developed and tested before amplitudes which are
sufficiently precise to be phenomenologically useful can be
computed. Lattice determinations of matrix elements traditionally
follow from computations of correlation functions of two or more local
operators, separated by large time distances in order to isolate the
lightest hadrons with the required quantum numbers. In general,
therefore, energy is not conserved in such correlation functions. For
example, if we consider a decay of a $B$ meson at rest into two hadrons,
then the lowest energy final state is the one in which the two hadrons
are also both at rest.  The correlation functions at large time
separations are therefore dominated by the unphysical process of a
$B$ meson decaying into two hadrons, all at rest.

For kaon decays the use of chiral perturbation theory allows one to
estimate the physical $K\to\pi\pi$ amplitude from the computed value
of the matrix element obtained with all three particles at rest (for a
theoretical introduction and references to the original literature see
e.g.\ \cite{lattice_kaons}, and for a review of recent numerical
results see \cite{Kuramashi}). For $B$ decays this is clearly not
applicable.  Moreover for $K\to\pi\pi$ decays the momenta of the 
final-state pions are sufficiently small that one can hope to compute 
the matrix elements corresponding to physical kinematics in the
foreseeable future (see for example~\cite{LL00} for a discussion of
the applications of finite-volume techniques to kaon decays).

When, as in physical decay amplitudes, the two final-state hadrons
have non-zero momenta, final-state interactions are present
and the corresponding scattering phases need to be evaluated.  There
is considerable effort currently being devoted to developing efficient
techniques for the computation of scattering phases in lattice
simulations (which are performed in Euclidean space so that the
dependence of the correlation functions on the scattering phases is
different than in Minkowski space), with a realistic expectation of
success, at least for kaon decays~\cite{CFMS96,LMS}.

It is an intriguing question whether the results of this paper might
be potentially helpful for lattice computations of the amplitudes for
exclusive $B$ meson non-leptonic decays. Since we have shown above that
the strong phases in leading order of the heavy-quark expansion are
perturbative, it should be investigated whether rescattering effects
are sufficiently small to enable the extraction of the amplitude into
final-state hadrons with non-zero momenta. At least for those
$B$ decays in which penguin diagrams do not contribute significantly,
this becomes a realistic and interesting possibility (although
extrapolations in the mass of the $b$ quark would have to be
performed). This can be checked by studying the time behaviour of the
correlation functions to see if it is indeed given by energies
corresponding to hadrons with the expected non-zero momentum. For
decays in which penguin diagrams contribute significantly there remain
many technical difficulties. These are being studied intensively for
$K\to\pi\pi$ decays, and, before turning to $B$ decays, one needs first
to establish that these kaon decays can be controlled. 

\subsubsection{QCD sum rules}

QCD sum rules provide a powerful field-theoretic approach with which
to study the properties of hadronic bound states, incorporating
essential non-perturbative features of QCD, such as chiral symmetry
breaking, vacuum condensates, unitarity and dispersion relations. They
have been used extensively to compute the masses, decay constants,
form factors and other strong-interaction couplings of mesons and
baryons \cite{SRreview}. In many areas, sum rules have been
established as a serious competitor to lattice gauge theory
computations.

QCD sum rules have also been applied to the difficult problem of 
non-leptonic weak decays. The first such applications were presented 
in the pioneering work by Blok and Shifman dealing with decays of 
charm mesons \cite{BlSh87}. Later, the same authors studied 
non-factorizable effects in $B\to D\pi$ decays and identified 
a long-distance, non-factorizable contribution to the decay amplitude
which shows a tendency to reduce the $1/N_c$ terms arising in naive 
factorization \cite{Blok93}. In terms of the parameter $\xi$ introduced 
earlier in this section, they found that 
$\xi=1/N_c-x\cdot 3\lambda_2/(4\pi^2f_\pi^2)$, where 
$\lambda_2\approx 0.12$\,GeV$^2$ is determined from the $B$-$B^*$ mass 
splitting, and $x\approx 1$ is a parameter of the model. Note that the 
presence of such a non-perturbative term at leading power would 
contradict the factorization formula (\ref{fff}), according to which 
$\xi$ is calculable up to corrections of order $\Lambda_{\rm QCD}/m_b$. 
(With the definition of $\xi$ above, its value can be determined from 
(\ref{a1dpi}).) The resolution is that in \cite{Blok93} the authors 
worked in a special kinematic regime, where the pion energy $E_\pi$ in 
the $B$-meson rest frame is assumed to stay of order $\Lambda_{\rm QCD}$ 
as $m_b\to\infty$. However, this would require that 
$m_b-m_c=O(\Lambda_{\rm QCD})$ in the heavy-quark limit. This scaling 
is different from the one we assumed in the derivation of the 
factorization formula, which crucially relies on having 
$E_\pi\gg\Lambda_{\rm QCD}$. 

Khodjamirian and R\"uckl have applied QCD sum rules to the study 
of non-factor\-iz\-able effects in the decays $B\to J/\psi\,K_S$, 
first using the conventional approach based on three-point vacuum 
correlation functions \cite{Khod98a}, and more recently using the 
method of light-cone sum rules \cite{Khod98}. Earlier, Halperin had
applied light-cone sum rules to estimate soft non-factorizable gluon 
exchanges in the colour-suppressed decay $\bar B_d\to D^0\pi^0$ 
\cite{Halp95}.

A problem common to all QCD sum-rule calculations of non-leptonic
decay amplitudes is that, because of the technical complexity, up to
now ``non-factorizable'' corrections to the amplitudes arising from
hard gluon exchange have not been included. These are, however, the
leading ``non-factorizable'' effects in the heavy-quark limit. The
non-factorizable contributions which have been included are those due
to vacuum condensates or higher-dimensional form factors involving
gluon fields, which correspond to formally subleading corrections in
the heavy-quark limit. As in the lattice calculations described above,
in addition to technical problems, QCD sum rule applications to
non-leptonic decays also face conceptual limitations. In particular,
it will not be possible to obtain a realistic description of
final-state interactions if the projection on the external hadron
states is, as is usually the case, performed using an ad hoc 
continuum subtraction.

We also mention that the sum rule technique has not yet been 
applied to heavy-light final states in which the light meson is 
emitted and energetic, as well as to decays into two light particles, 
because the current that couples to the emitted meson cannot be 
expanded in a series of local operators in these cases. These are
exactly the cases for which the theoretical description discussed in
this paper is most useful.

\subsubsection{Large-energy effective theories}

We have seen that the physical principle of colour transparency 
plays an important role in our approach, as it implies a systematic 
cancellation of soft divergences in the ``non-factorizable'' diagrams. 
The notion that colour transparency would imply an approximate
factorization (in the sense that $a_1$ is close to 1) in energetic
two-body $B$ decays in which the emission particle is a light
meson was introduced by Bjorken \cite{Bj89} and subsequently used
to argue in favour of an approximate universality of the parameters 
$a_i$ in energetic $B$ decays \cite{NeSt97}. Dugan and Grinstein 
made a first step towards formalizing the concept of colour 
transparency by introducing a ``large-energy effective theory'' 
(LEET) to describe the soft interactions of gluons with a pair of 
fast-moving quarks inside a pion \cite{DG91}. The effective theory was 
derived in analogy with heavy-quark effective theory by considering 
the Feynman rule for the gluon-quark coupling in the soft limit 
$k^\mu=O(\Lambda_{\rm QCD})$, $k\cdot p_\pi=O(\Lambda_{\rm QCD} m_b)$,
in much the same way as we did in Sect.~\ref{oneloopcancel}. 
However, since collinear gluon exchanges provide another source of 
infrared singularities, the decoupling of soft gluons is not sufficient 
to establish factorization.

The LEET gives, essentially, an operator description of what is often 
called the eikonal approximation for the coupling of soft particles 
to energetic ones. But since the eikonal approximation does not 
apply to hard-collinear lines, it cannot be universally used in 
infrared factorization proofs for hard processes. This is a very 
general feature of all factorization theorems in QCD for processes  
which involve (nearly) massless, hard particles. It raises the 
important conceptual question of 
whether it is possible to perform a consistent matching of QCD onto 
the LEET, in other words, whether the LEET correctly represents the 
long-distance dynamics of QCD in $B$ decays into at least one light 
particle. Due to collinear singularities, the answer to this question 
is negative. (In a somewhat different language, this point was 
already discussed in \cite{Agli92}.) In the particular situation of 
non-leptonic $B$ decays, we should distinguish collinear singularities
which cancel in non-leptonic decay amplitudes 
from those which do not. For example, the non-factorizable
collinear singularities cancel, and this is crucial for the validity
of (\ref{fff}), but because of their cancellation they do not 
invalidate a description provided by the LEET. On the other hand, 
as seen explicitly in Sect.~\ref{allorders}, factorizable collinear
singularities do not cancel. This is perfectly consistent with 
(\ref{fff}), since these singularities can be factorized into the 
light-cone distribution amplitudes. However, one cannot introduce 
light-cone distribution amplitudes in the context of the 
LEET effective Lagrangian, and therefore the LEET provides 
no means for absorbing
these collinear singularities into non-perturbative parameters. This
is a particular example of the statement that the LEET is not the
correct low-energy theory for non-leptonic $B$ decays.  

Let us be more specific and compare our result for the ``non-factorizable'' 
corrections to the $\bar B_d\to D^+\pi^-$ decay amplitude with the 
corresponding results obtained by Dugan and Grinstein in \cite{DG91}. 
Using the LEET, these authors have resummed large logarithms in two 
different kinematic regimes:
$m_b\gsim m_c\gg E_\pi$ (case 1), and $m_b\gsim E_\pi\gg m_c$ (case 2). 
Because of the kinematic relation $1-2 E_\pi/m_b\approx (m_c/m_b)^2$ it 
follows that $m_c/m_b=O(1)$ in case 1, whereas $E_\pi/m_b=O(1)$ in case 
2. Hence, for consistency the ratios $\alpha_s(m_c)/\alpha_s(m_b)$ 
(case 1) or $\alpha_s(E_\pi)/\alpha_s(m_b)$ (case 2) must be set to 1 
in the leading-logarithmic approximation. With these replacements the 
results of Dugan and Grinstein precisely correspond to the result of 
naive factorization with the Wilson coefficients evaluated at a scale
of order $m_b$. In the present work, we have developed a general 
formalism that allows us to calculate in a systematic way the leading
non-trivial corrections to this picture.

Aglietti and Corb\`o have argued that the correct way of dealing with 
the collinear singularities missed in the Dugan-Grinstein approach is 
to consider, instead of the exclusive decay $B\to D\pi$, a 
semi-inclusive process such as $B\to D+\mbox{jet}$ \cite{Corb98}. 
Then the collinear singularities may cancel by virtue of the KLN theorem, 
and the LEET can be applied to prove the factorization of the soft 
contributions. Here we have shown that in the heavy-quark limit a 
stronger form of factorization holds even for the 
exclusive process $B\to D\pi$. The reason for this is the 
cancellation of collinear singularities in the sum of all 
``non-factorizable'' diagrams. 

Another interesting analysis related to our work is a study of 
$B\to D\pi$ decays by Donoghue and Petrov \cite{DoPe96}, in which 
they calculate the non-factorizable one-gluon corrections in a 
background gluon field. Their ansatz is equivalent to the calculation 
of soft non-factorizable one-gluon contributions performed in 
Sect.~\ref{snf}. They find that this contribution vanishes exactly. 
This contradicts our result (\ref{o8snf}), which shows that the
resulting  
contribution is power suppressed in $\Lambda_{\rm QCD}/m_b$ but not 
vanishing. The origin of this discrepancy is that the Lorentz 
decomposition of a certain matrix element in Eq.~(22) of \cite{DoPe96} 
misses a term proportional to the $\epsilon$-tensor, which gives rise 
to the non-vanishing result.

\subsubsection{Hard-scattering approaches}

Methods familiar from the hard-scattering approach play an important 
role in our analysis. Yet, as we have emphasized the presence of a 
soft spectator quark in the $B$ meson prevents us from describing its 
weak decays exclusively using the language of perturbative QCD. In 
fact, the semi-leptonic form factors governing $B\to M$ decays receive 
dominant soft contributions by power counting. This is evident for 
heavy-to-heavy transitions, where the dominance of the soft 
contribution is the basis for the validity of an approximate 
heavy-quark symmetry and leads to the construction of the heavy-quark 
effective theory \cite{EIC90,GRI90,GEO90,review}. 
Only at very large recoil, such that 
$v\cdot v'\sim m_b/\Lambda_{\rm QCD}$, are the form factors dominated 
by hard gluon exchange \cite{NeGr96}. 

For heavy-to-light transitions hard and soft contributions
have the same power behaviour in $\Lambda_{\rm QCD}/m_b$, as discussed
in Sect.~\ref{formfactor}, so the soft
contribution is a leading effect again~\cite{CZ90,BBB98,Char98,FeKr99}. 
Contrary to familiar applications of the hard-scattering approach, 
such as to the pion form factor at large momentum transfer, 
the endpoint suppression of the light-cone distribution 
amplitude of the light meson is not sufficient to render the 
soft contribution power suppressed. 
(In fact, the hard gluon correction is suppressed by one power of 
$\alpha_s(\sqrt{\Lambda_{\rm QCD} m_b})$ relative to the soft one.) 
This discussion refers to counting powers only. It ignores the 
possibility 
that a resummation of Sudakov logarithms may suppress the soft 
contribution beyond naive power counting (see e.g.\ \cite{ASY94}). 
This possibility deserves further investigation. 
In this paper we have taken  
the point of view that Sudakov suppression is not sufficiently 
effective for realistic $B$ mesons and showed that a useful factorization 
formula holds even in the presence of soft contributions to the form 
factor. This also provides a common framework to discuss decays into
light-light final states and heavy-light final states, while the 
hard-scattering approach is never an option for heavy-light final states.  

Several authors have analyzed exclusive, semi-leptonic and non-leptonic 
$B$ decays using a perturbative hard-scattering approach 
\cite{SHB90,BuDo91,SiWy91,CaMi93,Ward1995,DaJK95,Li97,GLDu99,Mel99}.
The basic assumption in these studies is that non-leptonic $B$ 
decays are indeed dominated by hard gluon exchange, either because soft
exchanges are Sudakov suppressed, or because they are negligible for 
other dynamical reasons. A systematic formulation of this approach 
can be found in \cite{Li97,CLY99}, which adapts the 
``modified hard-scattering approach'' of \cite{BS89} to the case 
of non-leptonic $B$ decays. The decay amplitudes are expressed 
as a convolution of a hard-scattering amplitude,
containing a resummed Sudakov factor, and meson wave functions, e.g.\
\begin{equation}
   {\cal A}(B\to\pi\pi) = T *\Phi_B *\Phi_\pi *\Phi_\pi.
\end{equation}
The structure of this equation is similar to the hard 
spectator-interaction term in (\ref{fff}), because the hard-scattering 
approach {\em assumes\/} that the spectator quark in the $B$ meson 
{\em always\/} participates in a  hard interaction. However, 
here $T$ also contains a convolution in impact parameter space with
a Sudakov form factor and therefore takes a more complicated form
than in our approach. Because the hard subamplitudes are
evaluated with on-shell quark states, each term in the factorization
formula is separately gauge invariant (as in our approach). The
authors of \cite{CLY99} also observed the cancellation of the 
infrared double poles ($\sim 1/\epsilon_{\rm IR}^2$) in the sum of 
all ``non-factorizable'' diagrams.
In their scheme, the remaining single poles are absorbed into the 
definition of the $B$-meson wave function. 

Although this approach 
shares some similarities with our method, and is intrinsically 
self-consistent, it is important to note that it is based -- in
addition to the heavy-quark limit -- on the further assumption that
soft contributions to the $B\to M$ form factors are negligible. 
As mentioned in Sect.~\ref{formfactor} and above, the theoretical 
framework proposed in this work is more general. It does in fact
include the (modified) hard-scattering approach as the special case
when the form factors that appear as parameters in (\ref{fff}) are
assumed to be perturbatively calculable. This amounts to a 
slightly different power-counting scheme than the one adopted 
throughout our discussion, since for instance the leading-order
diagram in Fig.~\ref{fig4} is absent altogether. (It does not contain a
hard interaction with the spectator quark.) As a consequence the 
form-factor term in the factorization formula is suppressed relative 
to the hard spectator interaction. In particular, this means that naive
factorization is not recovered in any limit, because the
``non-factorizable'' hard spectator interaction is always as important as
the form-factor term.

\subsubsection{Models of final-state interactions}

Whereas final-state interactions may be of little importance in
Cabibbo-allowed $B$ decays into heavy-light final states such as 
$B\to D\pi$, their understanding is crucial for studies of CP 
violation in rare $B$ decays, such as decays into two light mesons. 
The reason is that interference of at least two contributions to the 
decay amplitude which differ in both their weak (CP-violating) and 
strong (rescattering) phases is necessary for observing a CP-violating 
rate asymmetry. Hence, in the study of direct CP-violation final-state
interactions are a crucial ingredient, and a theoretical handle on
the corresponding strong-interaction phases is of great importance. 
In the recent literature there have been numerous attempts to 
estimate these phases in decays such as $B\to\pi\pi$ or 
$B\to\pi K$, where they directly affect the determination of the 
angles $\alpha$ and $\gamma$ of the unitarity triangle 
\cite{BH96,Blok,Rob,Ge97,Ne97,Fa97,At97,CMB99}. Our approach provides 
a first systematic attempt to calculate these phases in a heavy-quark 
expansion.

We have stressed in Sect.~\ref{fsi} 
that the dominance of hard rescattering in the heavy-quark limit 
justifies the use of both a partonic and a hadronic language when discussing
final-state rescattering effects. However, the large number of 
intermediate states makes it intractable to observe systematic 
cancellations using a hadronic description. 
(An example of this is familiar from other 
applications of the heavy-quark 
expansion such as to the inclusive semi-leptonic decay width of a heavy 
quark. Here the leading term is given by the free quark decay, but 
the attempt to 
reproduce this obvious result by summing exclusive modes has been 
successful only in solvable two-dimensional toy models, but not in 
QCD \cite{GL97,Bigi99}.) 
In many phenomenological 
discussions of final-state interactions, it has been assumed that such 
cancellations are absent. It is then reasonable to
consider the size of rescattering effects for a subset of intermediate
states (such as the two-body states), assuming that this will 
provide a correct order-of-magnitude estimate for the total rescattering
effect. This strategy underlies all estimates of final-state phases
using dispersion relations and Regge phenomenology 
\cite{DGPS96,Fa97,CMB99}. Such approaches suggest that soft 
rescattering phases do not vanish in the heavy-quark limit. However, 
they also leave open the possibility of systematic cancellations.

In the present work, we have shown (implicitly) 
that systematic cancellations 
do indeed occur in the sum over all intermediate states. It is 
worth recalling that similar infrared cancellations are not uncommon 
for hard processes, such as $e^+ e^-\to\mbox{hadrons}$ at large 
center-of-mass energy. In a somewhat more remote context, 
cancellations among many individually large contributions from 
hadronic intermediate states are also known to occur for hadronic loop 
corrections to the Okubo-Zweig-Iizuka (OZI) rule \cite{GeIsgu}.
In our case, the underlying physical reason is that the 
{\em sum\/} over all states is accurately represented 
by a $q\bar{q}$ fluctuation in the emitted light meson 
of small transverse size of order $1/m_b$. Because the 
$q\bar{q}$ pair is small, the physical picture of rescattering is very 
different from elastic $\pi\pi$ scattering -- and hence the 
Regge phenomenology applied to $B$ decays  
is difficult to justify in the heavy-quark limit. We stress that this
important result of our analysis is not in conflict with the findings
of \cite{DGPS96} that {\em individual\/} intermediate states give
rise to large rescattering effects. However, we have identified a 
dynamical mechanism for systematic cancellations among {\em all\/} 
intermediate states that contribute to the decay. 
Because of these cancellations, the numerical
estimates for rescattering effects and final-state phases obtained
using Regge models are likely to overestimate the correct size of 
the effects.

An alternative proposal to deal with the large number of accessible
hadronic intermediate states in $B$ decays was made by Suzuki and 
Wolfenstein \cite{SW99}, who stress the importance of multi-channel
final-state interactions. The key element of their analysis is the
postulate of a randomness (under variation of $n$) of the relative 
strong phases between the weak decay matrix elements $M_{B\to n}$ and 
the elements $S_{nf}$ of the strong-interaction $S$ matrix, which 
connect an intermediate state $n$ to a particular final state $f$. 
The elastic contribution $S_{ff}$ is again estimated on the basis of 
pomeron exchange. Although the randomness assumption 
implies some degree of cancellations at a statistical level, it 
does not incorporate the QCD dynamics that results in 
systematic cancellations of the kind observed in the present 
work. 

\subsection{Classifications in terms of flavour topologies or
Wick contractions}

In the absence of reliable field-theoretic methods for calculating
non-leptonic weak decay amplitudes, strategies have been developed
to classify the various contributions to these amplitudes in a 
convenient way, and then to use symmetry arguments or ``plausible 
dynamical assumptions'' to derive relations between different decay 
processes. The most common classification scheme is based on SU(3)
flavour topologies, providing a catalogue of invariant amplitudes 
classified according to their transformation properties under 
SU(3) flavour symmetry \cite{Zepp81,SaWeSU3,Chau91,GRLH94}. Sometimes,
SU(3) subgroups such as isospin, U- or V-spin provide useful
classification schemes, too. Neglecting flavour-symmetry breaking, 
it is then possible to derive model-independent relations between
different decay amplitudes (e.g.\ isospin triangles or more complicated
constructions).

More recently, some authors have pointed out that the diagrammatic
approach, though perfectly correct, may not provide the most suitable
classification scheme, because the amplitudes defined there are in
general not renormalization-scheme invariant. It is possible, however,
to modify the approach in order to deal with this problem. To this end,
one classifies the various contributions to non-leptonic amplitudes
in terms of Wick contractions of the various operators in the effective 
weak Hamiltonian \cite{CFMS97,Rob}. This can be done in such a way 
that operators that mix under renormalization are grouped together. 
It has been argued that some Wick contractions that had previously
been neglected may play an important role in some cases. In particular,
the so-called ``charming penguins'', referring to penguin contractions
of four-quark operators containing a $c\bar c$ pair of quark fields,
may be of relevance for many rare $B$ decays \cite{CFMS97,BuFl95}. 
A complete classification of all possible Wick contractions to two-body 
hadronic $B$ decays can be found in \cite{BuSi98}.

\begin{table}
\caption{\label{tab:BScomparison}
Summary of the scaling properties of the various amplitudes contributing 
to $B\to D\pi$ and $B\to\pi\pi$ decays with respect to the large-$N_c$ 
and heavy-quark limits ($\Lambda\equiv\Lambda_{\rm QCD}$). The scaling 
laws for the leading hard (and computable) and soft (and incalculable) 
non-factorizable corrections to naive factorization are given separately. 
The scaling laws refer to the amplitude without a factor 
$G_F f_\pi m_B^2$.}
\vspace{0.2cm}
\begin{center}
\begin{tabular}{|l|ccccc|}
\hline
&&&&&\\[-0.35cm]
Topology & $E_1$ & $E_2$ & $A_2$ & $P_1$, $P_1^{\rm GIM}$ & 
 $P_3$, $P_3^{\rm GIM}$ \\
Large-$N_c$ counting & 1 & $1/N_c$ & $1/N_c$ & $1/N_c$ & $1/N_c^2$ \\[0.1cm]
\hline
&&&&&\\[-0.4cm]
$B\to D\pi$: & & & & & \\[0.1cm]
Hard non-fact.\ cor.\  & 1 & $(\Lambda/m_b)^2$ & $\Lambda/m_b$ & --- & ---
\\[0.1cm]
Soft non-fact.\ cor.\  & $\Lambda/m_b$  & $(\Lambda/m_b)^2$ &
$\Lambda/m_b$ & --- & --- \\[0.1cm]
\hline
&&&&&\\[-0.4cm]
$B\to\pi\pi$: & & & & & \\[0.1cm]
Hard non-fact.\ cor.\ &  $(\Lambda/m_b)^{3/2}$ & $(\Lambda/m_b)^{3/2}$ &
 $(\Lambda/m_b)^{5/2}$ & $(\Lambda/m_b)^{3/2}$ &
 $(\Lambda/m_b)^{5/2}$ \\[0.1cm]
Soft non-fact.\ cor.\ &$(\Lambda/m_b)^{5/2}$  &$(\Lambda/m_b)^{5/2}$  &
$(\Lambda/m_b)^{5/2}$  & $(\Lambda/m_b)^{5/2}$ & $(\Lambda/m_b)^{5/2}$
 \\[0.1cm]
\hline
\end{tabular}
\end{center}
\end{table}

In their most general form, such complete parameterizations are of a
limited use for phenomenological analyses, since they do not provide
any dynamical insight into the underlying strong-interaction phenomena
governing non-leptonic decays. For instance, in \cite{BuSi98} the 
authors define up to fourteen process-dependent phenomenological 
parameters for a single $B$ decay. Only when combined with a dynamical
approach such as the heavy-quark expansion presented in the present 
work, one can make statements about the relative importance of these
parameters. To illustrate this point, we summarize in 
Table~\ref{tab:BScomparison} the properties of the various amplitudes 
defined in terms of Wick contractions of the operators relevant to
$B\to D\pi$ and $B\to\pi\pi$ decays, as defined by Buras
and Silvestrini \cite{BuSi98}. In their notation, $E_1$ and $E_2$ are
``exchange amplitudes'', $A_2$ are ``annihilation amplitudes'', $P_1$
($P_3$) are ``penguin (annihilation) amplitudes'', and $P_1^{\rm GIM}$ 
($P_3^{\rm GIM}$) are ``GIM-penguin (annihilation) amplitudes''. 
There are many other 
amplitudes defined in this reference, which however are not relevant
to our discussion here. The table shows that in $B\to D\pi$
decays the exchange amplitude $E_1$ is parametrically enhanced with
respect to the other amplitudes, and that the ``non-factorizable'' 
contributions to this amplitude are calculable in perturbation theory. 
This allows us to obtain model-independent predictions for the 
$\bar{B}_d\to D^+\pi^-$ and $B^-\to D^0\pi^-$ decay amplitudes in the
heavy-quark limit. On the other hand, since only $E_2$ and $A_2$ 
contribute to the $\bar{B}_d\to D^0\pi^0$ decay amplitude, we do not
obtain a model-independent prediction in this case. In the case of
$B\to\pi\pi$ decays a significantly larger number of amplitudes
contributes. We find that the annihilation amplitudes 
$A_2$, $P_3$ and $P^{\rm GIM}_3$ are power 
suppressed, whereas all other amplitudes are of the same order in the
heavy-quark expansion and only receive calculable ``non-factorizable'' 
contributions at leading power. 


\section{Conclusion}
\label{conclusion}

With the recent commissioning of the $B$ factories and the planned emphasis 
on heavy-flavour physics in future collider experiments, the role of $B$ 
decays in providing fundamental tests of the standard model and potential 
signatures of new physics will continue to grow. In many cases the principal 
source of systematic uncertainty is a theoretical one, namely our inability 
to quantify the non-perturbative QCD effects present in these decays. This 
is true, in particular, for almost all measurements of CP violation at 
the $B$ factories. In this paper we have presented a rigorous framework for 
the evaluation of strong-interaction effects for a large class of exclusive, 
two-body non-leptonic decays of $B$ mesons. Our main results are contained 
in the factorization formula (\ref{fff}), which expresses the amplitudes 
for these decays in terms of experimentally measurable semi-leptonic form 
factors, hadronic light-cone distribution amplitudes, and hard-scattering 
functions that are calculable in perturbative QCD. For the first time, 
therefore, we have a well founded field-theoretic basis for phenomenological 
studies of exclusive hadronic $B$ decays, and a formal justification for 
the ideas of factorization. In this work we have focused on $B\to D\pi$ 
decays. A detailed discussion of $B$ decays into two light mesons will be 
presented in a forthcoming paper.

It is our belief that the factorization formula (\ref{fff}) will form the 
foundation for future phenomenological studies of non-leptonic two-body 
decays of $B$ mesons. We stress, however, that a considerable amount of 
conceptual work remains to be completed. This includes proving the validity 
of the factorization formula to all orders in perturbation theory. In 
particular, the two-loop proof for heavy-light final states presented in this 
paper must be extended to the more complicated case of $B$ decays into final
states containing two light mesons. Next, it will be important to investigate
better the limitations on the numerical precision of the factorization 
formula, which are valid in the formal heavy-quark limit. We have presented 
some preliminary estimates of power-suppressed effects in the present work, 
but a more complete analysis would be desirable. In particular, for rare $B$ 
decays into two light mesons it will be important to understand the role of 
chirally-enhanced power corrections. Finally, we mention that at present 
there are still large uncertainties associated with the description of the 
hard spectator interactions, which enter the factorization formula for $B$ 
decays into two light mesons. Some of these uncertainties are related to the 
fact that only little in known about the light-cone structure of heavy mesons 
and the properties of their wave functions. 

Theoretical investigations along these lines should be pursued with great
vigour. We are confident that, ultimately, this research will result in a
{\em theory\/} of non-leptonic $B$ decays, which should be as useful for this 
area of heavy-flavour physics as the large-$m_b$ limit and heavy-quark 
effective theory were for the phenomenology of semi-leptonic weak decays.

\vspace{0.6cm}
\noindent
{\bf Note added:} 
While this paper has been written, Ref.~\cite{Cha00} appeared, in which 
the matrix elements for $\bar{B}_d\to D^{(*)}\pi^-$ are also computed to 
next-to-leading order. The result for the symmetric part of the 
hard-scattering kernel disagrees with our result (\ref{delo8}) and 
(\ref{fxzsym}). (The asymmetric part of the kernel is not needed for the 
pion final state and has not been computed in \cite{Cha00}.) The correct 
result is obtained if the sign of the imaginary parts in $f_1(x)$ and 
$g_1(x)$ defined in \cite{Cha00} is inverted, and a term $-r^2 \ln r^2$ 
is added to the square bracket in the expression for $f_1(x)$.


\section*{Acknowledgements}

We would like to thank J.D.~Bjorken, V.M.~Braun, J.C.~Collins, J.F.~Donoghue,
T.~Feldmann, G.P.~Korchemsky, G.P.~Lepage, A.A.~Petrov and H.R.~Quinn for 
useful discussions on various aspects of this work.
M.N.~is supported in part by the National Science Foundation, and C.T.S.~by 
the Particle Physics and Astronomy Research Council (through grants
GR/L56329 and PPA/G/O/1998/00525).


\vfill\eject
 \end{document}